\documentclass[11pt,a4paper,openright,floatfix]{t}

\usepackage[dvips]{graphics}
\usepackage[dvips]{graphicx}
\usepackage{bm}
\usepackage{subfigure}
\usepackage{epsfig}
\usepackage{amsmath} 
\usepackage{bm}
\usepackage{chapterbib} 
\usepackage{amsfonts}
\usepackage{verbatim}
\usepackage{graphicx}
\usepackage{tabularx}
\begin{document}
\pagestyle{fancy}
\pagenumbering{roman}
\begin{spacing}{1.5} 
\end{spacing}

  \thispagestyle{empty}
\begin{center}
 {\bf \Large Finite field dependent BRST transformations and its applications to gauge field theories}
 \end{center}

\begin{figure}[htb]
\begin{center}
\includegraphics{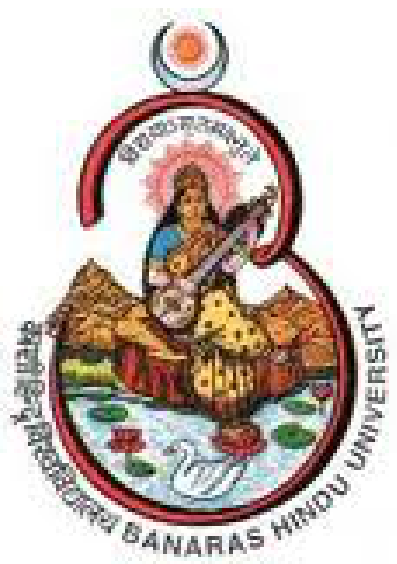}
\end{center}
\end{figure}

\begin{center}
A Thesis for the Degree of\\

\vspace{0.1in}

{ \large Doctor of Philosophy}\\

\vspace{0.1in}

{ in}\\

\vspace{0.in}

{ \large  Physics}\\

\vspace{0.1in}

{\large by}\\

\vspace{0.2in}

{\bf \large SUDHAKER UPADHYAY}\\

\vspace{0.2in}

{\it \large under the supervision of}\\

\vspace{0.2in}

{\bf \large Dr.  Bhabani Prasad Mandal}\\

\vspace{0.3in}

{ Nuclear and Particle Physics Section,\\
Centre of Advanced Study,\\
Department of Physics,\\
Banaras Hindu University,\\
Varanasi-221005.}
\end{center}
 
\newpage

$\left. \right.$

\vspace{1.9in}

\bigskip

\bigskip

\thispagestyle{empty}
\begin{center}
{\it {\Large \bf{ Dedicated to My Family} }} 
\end{center} 
\begin{acknowledgements}
\thispagestyle{empty}
\vspace{-1cm}
{\it I am indebted to my thesis supervisors, Dr. Bhabani Prasad Mandal, for introducing the subject to me and 
guiding me throughout my Ph.D. carrier.  Without his guidance and persistent  help this thesis would not 
have been possible. 

I express the great honor towards my family members, on whose consistent
encouragement and love I have relied throughout my life. I am indebted
to my parents Mr. Gauri Shanker Upadhyay and Mrs. Madhuri Devi, brother Deepu, and sisters Anita and Alka, 
for their loving care, blessings and patience
shown during the course of my research work and I hope to continue, in my own small way,
the noble missions to which they expect from me. I dedicate this thesis to them.

We gratefully acknowledge the financial support from University Grant Commission (UGC),
 Department of Science and Technology 
(DST), and Council for Scientific and Industrial Research (CSIR), New Delhi,  at various 
stages of my research carrier. 

I would like to thank our group mates, Sumit Kr. Rai, Ananya Ghatak, Rajesh Kr. Yadav, Brijesh Kr. Maurya
and Manoj Kr. Dwivedi, for their kind supports. I also take this opportunity to express my 
gratitude towards Dr. Raka D. Ray Mandal and loving brother Satyajay Mandal for their inspirational
and moral support.
I extend my sincere thanks to all my friends, specially  Manoj,  Sudhir, Charu, Anand 
and their family members, for kind help in many respects during the stay in Varanasi. 

It is pleasure to acknowledge all the faculty members, research scholars, laboratory staff members, of 
the high energy physics group and the DRC members for helping me throughout my research work.

Last but not the least I thank the almighty for bringing all the above people into my 
life and  thus guiding me as an invisible, silent but steady friend, mentor and guardian.}
\vspace{.35in}
\begin{flushright}
{\bf (Sudhaker Upadhyay)}
\end{flushright}
\end{acknowledgements}

\newpage
\thispagestyle{empty}
\begin{abstract}
 
The Becchi-Rouet-Stora and Tyutin (BRST) transformation  plays a crucial role in the quantization of gauge theories.  
The BRST transformation is also very important tool in characterizing  the various renormalizable field theoretic 
models. 
The generalization of the usual BRST transformation,  by making the infinitesimal global parameter finite and field 
dependent, is commonly known as the finite field dependent BRST (FFBRST) transformation.
 In this thesis,  we  have extended the FFBRST transformation
 in an auxiliary field formulation and have developed both on-shell and off-shell 
FF-anti-BRST transformations. The different aspects of such transformation are  studied in Batalin-Vilkovisky 
(BV) formulation.  FFBRST transformation has further been used to study the celebrated Gribov problem and to
 analyze   the constrained dynamics in gauge theories. A new finite field dependent symmetry (combination of 
FFBRST and FF-anti-BRST) transformation has been invented. 
The FFBRST transformation is shown useful in connection of first-class constrained
theory to that of second-class also.
 Further, we have applied the Batalin-Fradkin-Vilkovisky (BFV) technique 
to quantize a field theoretic model in the Hamiltonian framework. The Hodge de Rham theorem for differential geometry
has also been studied in such context.

\end{abstract}
\tableofcontents 


\chapter{Introduction}
\pagenumbering{arabic}
Due to the lack of hard empirical data, symmetry principles have been proved to be the most invaluable  
tools in describing physical phenomenon. Gauge field theories (based on the local gauge invariance 
of the Lagrangian density of the theories) have  enormous importance in describing all the fundamental
interactions of nature and play the central role in understanding the present state of the
art of modern particle physics. The standard  model of particle physics which describes strong, weak and
electromagnetic interactions on the same footing is a non-Abelian gauge theory (Yang-Mills theory) \cite{ym}. 

However, one faces various problems to develop the quantum version of such theories with
local gauge invariance consistently. 
In particular, the generating functional, $Z=\int {\cal D} A\ e^{i\int d^4 x{\cal L}}$,
for such gauge theories becomes ill-defined due to the over counting of physically equivalent gauge configuration.
This in turn leads to the ill-defined Green's functions of these theories. 
 Therefore, it is necessary to eliminate the redundant degrees of freedom from the functional 
 integral representation of the generating functional $Z$. This can be achieved by adding a gauge variant 
 term, called as gauge-fixing term, to the Lagrangian density ${\cal L}$ of the theory. 
The generating functional is made well-defined
 in the cost of the gauge symmetry. The gauge-fixing was achieved by adding an extra term 
 consisting of  arbitrary function of the gauge field and arbitrary gauge parameter. This of course solves the 
problem of 
 over counting but the physical theory now depends on arbitrary function of gauge field and/or arbitrary parameter
which is not desirable for any physical theory.
Faddeev-Popov (FP) resolved this problem by introducing unphysical ghost fields which are scalars but 
behave like Grassmannian \cite {fp}. These unphysical fields compensate the effect of arbitrary function and in term 
preserves 
the unitarity of the theory.  Various difficulties in different situations occur due to the  gauge non-invariance
of the theory; For 
example, the choices of the counter terms in the renormalization program in such theories 
are no more restricted to the gauge invariant terms as the gauge invariance is broken.

C. Becchi, A. Rouet and R. Stora and independently I. V.
Tyutin  came to resolve the situation by discovering  a new symmetry of the FP effective 
theory 
known as BRST symmetry \cite{brst, tyu}.  This BRST transformation  is characterized by 
(i) infinitesimal, 
(ii) global (i.e. does not depend on the space-time)  and (iii) anticommuting parameter.
Such BRST transformation leaves the effective action, including gauge-fixing and ghost parts, invariant
and is nilpotent in nature.
Sometimes  the nilpotency is proved using equation of motion of one or more fields then it is referred 
as on-shell nilpotent. However, BRST transformation can be made off-shell nilpotent by introducing
Nakanishi-Lautrup type auxiliary fields to the theory.  
 BRST symmetry is extremely useful in quantizing different gauge field theoretic models and the
 renormalization program is greatly facilitated by the use of such symmetry \cite{brst,tyu,ht,wei}.

To cover the wider class of gauge theories, including open or reducible gauge theories, a powerful technique was 
introduced by I. A. Batalin and G. A. Vilkovisky \cite{ht,wei,bv,bv1,bv2}, known as field/antifield (or BV) 
formulation.
The main idea of this formulation is to construct an extended action by introducing the antifields 
 for each fields in the theories.  
The antifields satisfy  the opposite  statistics corresponding to that of fields and have the ghost number equal to 
$-gh(\phi)-1$, where $gh(\phi)$ is the ghost number of the fields.
 However, the extended action satisfies the certain rich mathematical formula known as
quantum master equation which reflects the gauge symmetry in the zeroth order of antifields
and in the first order of antifields it reflects the nilpotency of BRST transformation. 
These extended theories work extremely well in the frame of gauge theories which  
are always endowed with first-class constraints in the language of 
Dirac's constraints analysis \cite{c,frad, ku, dir}.
 The systems with second-class constraints are quantized by converting these to a
first-class  theory in an extended phase space  \cite{brst,c,frad,ku,ste}. 
This procedure has
been  introduced by I. A. Batalin, E. S. Fradkin  and I. V. Tyutin \cite{bt,bf} 
and has been applied to the various models \cite{fik,kkk,fs,bs,wz}.
Another way of approaching the problem, which is very
different from the Dirac's method, is the BFV (due to I. A. Batalin, E. S. Fradkin and G. A. Vilkovisky)
quantization \cite{ht,bv0,hen}.  The main features of BFV approach are as follows: (I)
it does not require closure off-shell of the gauge algebra and therefore does not need an 
auxiliary fields, (II) this formalism relies on BRST transformation which is independent 
of gauge-fixing condition and (III) it is also applicable to the first order Lagrangian. Hence it
is more general than the strict Lagrangian approach.

In all these approaches of studying gauge theories the main ingredient is the underlying BRST symmetry
 of the FP effective theory. Therefore, any modification or reformulation or generalization of BRST
 transformation is extremely important in the study of fundamental interactions which are described by gauge theories.
 With various motivations and goals, BRST transformation has been generalized 
in many different ways. M. Lavelle
and D. Mcmullan had found a generalized BRST symmetry adjoint to usual
BRST symmetry in the case of QED which is
nonlocal and noncovariant \cite{liv}. The motivation behind the emergence of this
symmetry was to refine the characterization of physical states given by the
BRST charge.  Later, Z. Tang and D. Finkelstein  had
found another generalized BRST symmetry which is nonlocal but covariant \cite{tan}.
Such a BRST symmetry is not nilpotent generally and additional conditions are required  in auxiliary field 
formulation to make them nilpotent. H. S. Yang and B. H. Lee  had presented a local
and noncovariant BRST symmetry in the case of Abelian gauge theories \cite{yan}.
Finite field dependent BRST (FFBRST) transformation, where the parameter
is finite and field dependent  but still anticommuting in nature, is the most important among the  generalizations
of BRST symmetry which
 was developed by S. D. Joglekar and B. P. Mandal for the first time in 1995 \cite{jm}. 
They had shown that the usual infinitesimal, global BRST transformation can be 
 integrated out to construct the  FFBRST transformation   
\cite{jm}. The parameter in such a transformation is anticommuting, finite in nature, depends on the fields,
 and does not depend on space-time explicitly. FFBRST transformation is also the symmetry of 
the effective theories and maintains the on-shell nilpotency property.
Moreover, FFBRST transformation is capable of connecting the 
generating functionals of two different effective field theories with suitable choice of  
the finite field dependent parameters \cite{jm}.
For example, this transformation was used to connect the FP
effective action in Lorentz gauge with a gauge parameter $\lambda $ to
(i) the most general BRST/anti-BRST symmetric action in Lorentz gauge \cite{jm}, 
(ii) the FP effective action  in axial gauge \cite{sdj0,sdj2,sdj3,sdj4,sd}, (iii)  the 
FP effective action  in Coulomb gauge \cite{sdbp},
(iv) FP effective action  with another distinct gauge parameter 
$\lambda^\prime $ \cite{jm} and (v) the FP effective action in quadratic gauge \cite{jm}.
The FFBRST transformation was also used to connect the generating functionals corresponding
to different solutions of the quantum master equation  in field/antifield formulation \cite{subm}.
 The choice of the finite parameter  is crucial in connecting different
effective gauge theories by means of the FFBRST transformation. The path integral measure in the 
expression of generating functional is not invariant under FFBRST transformation. The 
nontrivial Jacobian of such FFBRST transformation is the source for new results.
The FFBRST formulation has many applications \cite{sdj0,sdj2,sdj3,sdj4,sd,sdj,rb,sdj1,etc,sudha1,sudha2,sudha3} on the 
gauge theories. A correct prescription for the poles in the gauge field propagators in 
noncovariant gauges has been derived by connecting effective theories in covariant gauges to 
the theories in noncovariant
 gauges by using FFBRST transformation \cite{sdj1}. The divergent energy integrals in the Coulomb gauge are 
regularized by modifying the time like propagator using FFBRST transformation \cite{sdbp}. The FFBRST transformation, 
which is discussed so far in literature, is only on-shell nilpotent \cite{jm, sdj0, sdj2,rb}.

In this thesis, we would like to address different issues of BRST transformation, its generalizations 
and applications to different gauge field theoretic models. We further
generalize the FFBRST transformation and find new applications. 
We develop the off-shell nilpotent FFBRST transformation by introducing
a Nakinishi-Lautrup type auxiliary field and show that such transformation is more elegant
to use in certain specific cases \cite{susk}. The anti-BRST transformation, where the role of ghost and antighost
fields are interchanged with some coefficients, does not play as fundamental
role as BRST symmetry itself but it is a useful tool in geometrical description \cite{boto} of
BRST transformation, in the investigation of perturbative renormalization \cite{gaba}.  We develop 
both the on-shell and off-shell nilpotent finite field dependent anti-BRST (FF-anti-BRST)
 transformations for the first time which play similar role as FFBRST
transformation \cite{susk}.

We study these transformations in the context of higher form gauge theories \cite{sm1}. 
The gauge theories of Abelian rank-2 antisymmetric tensor field play crucial role in studying the 
theory for classical strings \cite{a}, vortex motion in an irrotational, incompressible 
fluid \cite{c,b} and the dual formulation of the Abelian Higgs model \cite{d,e}. 
Abelian 
rank-2 antisymmetric tensor fields are also very useful in studying supergravity multiplets 
\cite{g}, excited states in superstring theories \cite{h,i} and anomaly cancellation in 
certain superstring theories. Geometrical aspects of Abelian rank-2 antisymmetric tensor 
fields are studied in a U(1) gauge theory in loop space. 
We extend the FFBRST formulation to study Abelian rank-2 tensor field theories. 
We establish the connection between different effective 2-form gauge theories using the FFBRST 
and FF-anti-BRST transformations. The FF-anti-BRST transformation
plays similar role to connect different effective theories.  
We further extend these FFBRST and FF-anti-BRST transformations to the field/antifield formulation
of 2-form gauge theory \cite{sm1}.

In non-Abelian gauge theories even after gauge-fixing the redundancy of gauge 
fields is not completely removed in certain gauges for large gauge fields (Gribov problem)
\cite{gri}. The  Yang-Mills (YM)  theories in those gauges contain so-called Gribov copies. 
Gribov copies play 
a crucial role in the infrared (IR) regime while it can be neglected in the perturbative 
ultraviolet (UV) regime \cite{gri, zwan, zwan2}.
This topic has become very exciting currently due to the fact that color confinement is closely 
related to the asymptotic behavior of the ghost and gluon propagators in deep IR regime
\cite{kon0}.
In order to resolve the Gribov problem, Gribov and Zwanziger (GZ) proposed a 
theory, which restricts the domain of integration in the functional integral within
 the first Gribov horizon \cite{zwan}. This restriction to first  Gribov horizon  is
achieved by adding a nonlocal term, commonly known as horizon term, to the YM action \cite
{zwan, zwan2, zwan1}. But the YM action restricted
in Gribov region (i.e. GZ action) does not exhibit the usual BRST invariance, due to the
presence of the nonlocal horizon term \cite{sore1}. 
The famous Kugo-Ojima (KO) criterion for color confinement \cite{ku} is based on the assumption of an 
exact BRST invariance of YM theory in the manifestly covariant gauge.
Recently, a nilpotent BRST transformation which leaves the GZ 
action invariant has been obtained and  
can be applied to KO analysis of the GZ theory \cite{sor}.
The BRST symmetry in presence of the Gribov horizon has great applicability in order to 
solve the nonperturbative features of confining YM theories \cite{dud, fuj},
where the soft breaking of the BRST symmetry exhibited by the GZ action is converted 
into an exact invariance \cite{sor1}.  We  consider FFBRST transformation  in Euclidean 
space to show the mapping between the generating functional of GZ theory to that of YM theory \cite{sudbm}. 
Such a mapping is also shown to exists in field/antifield formulation of GZ theory \cite{sudbpm}.

So far we have seen that  FFBRST and FF-anti-BRST transformations 
are symmetry of the effective action  but do not leave the generating functionals invariant. The Jacobians
of the path integral measure in the expression of generating functional are not unity as
these transformations are  finite in nature. We address the important question whether 
it is possible to develop a finite nilpotent symmetry for both effective action as well as the generating
functional? In search of answer to this question, we propose the finite version of 
 mixed BRST (combination of BRST and anti-BRST) transformation.
Such a finite mixed BRST (FFMBRST) transformation is shown to be nilpotent as well as
 the symmetry of both the effective action and the
generating functional \cite{epjc}. Our  results are established  with the help of
several explicit examples. This formulation is further extended to field/antifield formulation \cite{epjc}. 

In another problem, we study the systems with constraints in the framework of FFBRST transformation.
 The gauge variant model  for the single self-dual 
chiral boson in (1+1) dimensions (2D) is well known example of the second-class 
theory \cite{pp,pp1,kk,fj,vo,cg,ggk}. This model was made gauge invariant by adding the
Wess-Zumino (WZ) term  and had been studied using  BFV
formulation \cite{sud,sub}. Such a model is very useful in the study of certain string theoretic 
models \cite{ms} and plays a crucial role in the study of quantum Hall effect \cite{wen}.
The Proca model in (1+3) dimensions (4D) 
for massive spin 1 vector field  also is another example of a system with the second-class constraint as  
the gauge symmetry is broken by the mass term of the theory. However, 
Stueckelberg  converted 
this theory to a first-class theory  by introducing  a scalar 
field \cite{stuc1,stuc2,stuc3}. Such a gauge invariant description for massive spin 1
field has many applications in  gauge field theories as well as in string theories \cite
{al,mr,r}. We establish the connection between the generating functionals for the first-class theories
and the generating functionals for the second-class theories using FFBRST transformation \cite{aop}.
 The generating functional of the Proca model is obtained
from the generating functional of the Stueckelberg theory for massive spin 1 vector field using
FFBRST transformation with appropriate choice of the finite parameter. In the other example we relate the generating 
functionals for the
gauge invariant and the gauge variant theory for self-dual chiral boson by constructing suitable FFBRST
transformation. Thus, the complicated nonlocal Dirac bracket analysis in the study of the
second-class theories is avoided in our formulation. 

In a different problem, we study the analogy between the conserved
charges of different BRST and co-BRST symmetry transformations with
exterior and co-exterior derivatives \cite{sud}. In the BRST formulation of gauge theories  one requires that
 the physical subspace of total Hilbert space of states contains only those states that 
are annihilated by the nilpotent and conserved BRST charge $Q_b$ i.e. 
$Q_b\left|phys\right>=0$ \cite{dir,sund}.
The nilpotency  of the BRST charge ($Q^2_b = 0$)
and the physicality criterion  ($Q_b\left|phys \right>= 0$) are the two essential ingredients
 of BRST quantization. In the language of differential geometry  defined on compact,
 orientable
 Riemannian manifold, the cohomological aspects of BRST charge is realized in a simple, 
elegant manner. The nilpotent BRST charge is connected with  exterior derivative 
(de Rham cohomological operator $d = dx^\mu\partial_\mu,$ with $ d^2 =
0$)\cite{egu,nis,nis1,kala,hol,hol1,arn,hari,sr,sr1}. It has been found that the co-BRST 
transformation which is also the symmetry of the action and leaves the gauge-fixing 
part of the action invariant separately. The conserved charge corresponding to the co-BRST 
transformation is shown to be analog to the co-exterior derivative 
($\delta = \pm \ast d \ast,$ with $ \delta^2 =
0$)\cite{hari}. 

This thesis
is divided into following nine chapters. The detail contents of these chapters are given below. 

General introductions to i) BRST transformation, its generalizations and applications, ii) 
basic techniques of field/antifield formulation and BFV formulation are 
presented in chapter one. We brief the contents of the different chapters of the thesis.

In Chapter two, we provide the different mathematical techniques which will be used to construct the thesis.
In particular FFBRST transformation, BV formulation and BFV technique are discussed in brief.

In chapter three,
we formulate the FFBRST transformation in auxiliary field formulation to make 
it off-shell nilpotent \cite{susk}. We consider several examples to demonstrate that 
the  off-shell nilpotent FFBRST transformation also leads to similar results in 
connecting the different generating functionals. 
  We further construct the finite field dependent anti-BRST (FF-anti-BRST) 
transformation  analogous to 
the FFBRST transformation  by integrating infinitesimal anti-BRST transformation.
 By considering several choices of finite field dependent 
parameter in 
FF-anti-BRST transformation we show that FF-anti-BRST transformation also 
plays the same role as FFBRST transformation in connecting different effective theories but with
different parameters. Finally, 
 we consider 
the formulation of FF-anti-BRST transformation in auxiliary field formulation also to make it off-shell nilpotent.

In chapter four, we study the quantization of the Abelian rank-2 antisymmetric tensor 
field by using the FFBRST transformation \cite{sm1}. We  show  that it is possible to 
construct the Abelian rank-2 tensor field theory in noncovariant gauges by using 
FFBRST transformation. In particular, we show that the generating functional for Abelian 
rank-2 tensor field theory in covariant gauges transformed to the generating functional for 
the same theory in a noncovariant gauges for a particular choice of the finite parameter in 
FFBRST transformation. The new results arise from the nontrivial Jacobian of the 
path integral measure under such finite BRST transformation. The 
connections between the theories in two different noncovariant gauge, namely axial 
gauge and 
Coulomb gauge are also established explicitly. Further, we consider 
the field/antifield formulation of Abelian rank-2 tensor field theory by introducing 
the antifield $\phi^\star $ corresponding to each field $\phi$ with opposite statistics
to study the role of FFBRST transformation in such formulation. We show that 
the FFBRST
transformation 
changes the generating functional corresponding to one gauge-fixed fermion to the generating 
functional to another gauge-fixed fermion. Thus, the FFBRST transformation connects the different 
solutions of 
the master equation in field/antifield formulation. We show this by considering explicit 
examples.

Chapter five  is devoted for the construction of FFBRST transformation in Euclidean space 
to study the  GZ theory \cite{sudbm, sudbpm}. 
By constructing an appropriate finite field dependent parameter we show that such 
FFBRST transformation relates the generating functional for GZ theory to the generating 
functional in YM theory. Thus, we are able to connect the theories with and without Gribov copies.

In chapter six, we construct the finite version of MBRST transformation having finite and field 
dependent 
parameters \cite{epjc}. The usual  FFBRST  and  FF-anti-BRST  transformations 
are the symmetry transformations of  the effective action only but do not leave the generating functional 
invariant as the path integral measure in the definition of generating functional transforms in a nontrivial 
manner \cite{jm,susk}. 
Unlike the usual FFBRST and FF-anti-BRST
transformations, this finite field dependent MBRST (FFMBRST) transformation  is shown to be  the symmetry of the 
both effective 
action as well as the generating functional of the theory. We construct the  finite parameters in the FFMBRST 
transformation in 
such a way that the Jacobian contribution due to FFBRST part compensates the same due to FF-anti-BRST 
part. 
Thus, we are able to construct the finite nilpotent transformation which leaves the generating 
functional as well as the effective action of the theory invariant. We further show that the effect of FFMBRST 
transformation is equivalent to the effect of successive operations of FFBRST and FF-anti-BRST transformations. 
Our results are supported by several explicit examples. First of all 
we consider the gauge  invariant model for single 
self-dual chiral boson in (1+1) dimensions \cite{pp,pp1,sud} to show our results.
(1+3) dimensional Abelian as well as  non-Abelian  YM theory in the Curci-Ferrari-Delbourgo-Jarvis (CFDJ) 
gauge 
\cite{cf, cf1, del} are also considered to demonstrate the above
finite nilpotent symmetry.
 To study the 
role of FFMBRST transformation in field/antifield formulation we consider the same 
 three  simple models in BV formulation. We show that the FFMBRST transformation does  not 
change  the generating functional written in terms of extended quantum action in BV formulation. Hence
the FFMBRST 
transformation leaves  the different solutions of 
the quantum master equation in field/antifield formulation invariant. 

In chapter seven, we study  the theories  with constraints in the context of FFBRST transformation \cite{aop}.
Theories with first-class constraints are shown to be related to the 
theories with second-class constraints through FFBRST transformation. The generating 
functional of Stueckelberg theory for massive spin 1 vector field is related to the 
generating functional of Proca model for the same theory  through FFBRST and FF-anti-BRST transformations. 
Similar relationship is also established between the gauge invariant and gauge variant 
models for single self-dual chiral boson  through FFBRST and FF-anti-BRST formalism. 

In chapter eight, we present BFV formulation for the model of single self-dual 
chiral boson \cite{sud}.
 Along with the usual nilpotent BRST symmetry,  anti-BRST, co-BRST and anti-co-BRST symmetries  
are investigated in this framework.
 The generators of all these continuous 
symmetry transformations
are shown to obey the algebra of de Rham cohomological operators of differential geometry.
The Hodge decomposition theorem in the quantum Hilbert space of states is also discussed. 
We show that the classical theory for a self-dual chiral boson is a field theoretic model for 
Hodge theory.

The last chapter is devoted for summary and conclusion.
\label{Chap:Intro}

\chapter{The mathematical basis}
The aim of this  chapter  is to provide the basic techniques and mathematical tools to prepare   
the necessary background relevant to this thesis. In particular, we briefly outline the
basic ideas of  the on-shell finite field dependent BRST transformation, BV formulation of
gauge theories and BFV technique. We start with on-shell FFBRST transformation in the next section.
\section{ On-shell finite field dependent BRST (FFBRST) transformation}
We begin with the {\it on-shell} FFBRST formulation of pure gauge theories \cite
{jm}. 
The usual BRST transformation for the generic fields $\phi$ of an effective theory is  defined compactly as
\begin{eqnarray}
\delta_b \phi  =s_b \phi \  \Lambda,
\end{eqnarray}
where $s_b \phi$ is the BRST variation of the fields with infinitesimal, anticommuting and
global parameter $\Lambda$.  Such transformation is on-shell nilpotent, i.e. $s_b^2 =0$, with the use of 
some equation of motion for fields and leaves the FP effective action invariant. It was observed by Joglekar and
Mandal \cite{jm} that $\Lambda$ needs neither to be infinitesimal, nor to be field-independent 
to maintain the symmetry of the FP effective action of the theory as long as it does not depend explicitly on 
space-time.
They made it infinitesimally field dependent and then integrated the infinitesimal field dependent BRST 
transformation to construct FFBRST transformation 
which preserves the same form as  
\begin{eqnarray}
\delta_b \phi =s_b \phi \ \Theta_b [\phi ],
\end{eqnarray}
where $\Theta_b[\phi ]$ is an $x$-independent functional of fields $\phi $.

We briefly mention the important steps to construct FFBRST transformation. We start 
with the fields, $ \phi (x, \kappa)$, which are made to depend on  some parameter, $\kappa: 0\le \kappa \le 1$, in 
such a manner that $\phi(x, \kappa =0 ) = \phi(x) $ is the initial field and
 $ \phi(x, \kappa=1) = \phi ^\prime
(x)$ is the transformed field. The infinitesimal parameter $\Lambda$ in the BRST transformation is made field 
dependent and hence the BRST transformation can be written as
\begin{equation}
\frac{ d}{d \kappa}\phi(x, \kappa ) = s_b \phi (x, \kappa )\
\Theta_b^\prime [\phi(x,\kappa )],
\label{ibr}
\end{equation}
where $\Theta_b^\prime $ is an infinitesimal field dependent parameter. By integrating these equations 
from $ \kappa=0$ to $\kappa=1$, it has been shown \cite{jm} 
that the $\phi^\prime ( x) $ are related to $\phi(x)
$ by the FFBRST transformation  as
\begin{equation}
\phi^\prime (x) = \phi(x) + s_b \phi (x)\ \Theta_b [\phi(x)],
\label{fbrs}
\end{equation}
where $ \Theta_b [\phi(x)] $ is obtained from $\Theta_b^\prime [\phi(x)]$
through the relation \cite{jm} 
\begin{equation}
\Theta_b [\phi(x, \kappa)] = \Theta_b^\prime [\phi(x, 0)] \frac{ \exp f[\phi(x, 0)]
-1}{f[\phi(x, 0)]},
\label{80}
\end{equation}
and $f$ is given by $ f= \sum_{i} \frac{ \delta \Theta_b^\prime (x)}{\delta
\phi_i(x)} s_b \phi_i(x). $
This transformation is nilpotent and symmetry of the  effective action. The generating functional, 
 defined as
\begin{equation}
Z=\int [{\cal D}\phi ]\ e^{i S_{eff}},\label{generator}
\end{equation} 
is not invariant under such FFBRST transformation as the Jacobian in the above expression is not invariant
under it. 
Under FFBRST transformation
Jacobian changes as  
\begin{eqnarray}
{\cal D}\phi =J[\phi(\kappa)] {\cal D}\phi(\kappa). \label{jac}
\end{eqnarray}               
It has been shown \cite{jm} that under certain condition this nontrivial Jacobian can be replaced (within the 
functional integral) as
\begin{equation}
J[\phi(\kappa)] \rightarrow e^{iS_1[\phi(\kappa)]},
\end{equation}
where $S_1[\phi(\kappa)]$ is some local functional of $\phi (x)$. The condition
 for existence of $S_1$ is
\begin{eqnarray}
\int [{\cal D}\phi] \left[\frac{1}{J}\frac{d J}{d\kappa} -i\frac{dS_1}{d\kappa}\right] \exp {i[S_{eff}+S_1]}=0.
\label{mcond}
\end{eqnarray}
Thus,
\begin{equation}
Z\left(=\int [{\cal D} \phi]\ e^{i S_{eff}}\right)\stackrel{FFBRST}{----\longrightarrow }Z' \left(=\int [{\cal D}
\phi] 
\ e^{i [S_{eff}(\phi)+S_1 (\phi)]}\right).
\end{equation} 
$S_1[\phi]$ depends on the finite field dependent parameter. Therefore, the generating functional corresponding
to the two different effective theories can be related through FFBRST transformation with 
appropriate choices of finite parameters.  The FFBRST transformation has also been used to solve many of the long 
outstanding problems in quantum field theory \cite{sdj0,sdj2,sdj3,sdj4,sd,sm1,sdj,rb,sdj1,etc}.
For example, the gauge field propagators in noncovariant gauges contain singularities on the 
real momentum axis \cite{sdj3}. Proper prescriptions for these singularities in gauge field propagators have been 
found by using 
FFBRST transformation \cite{sdbp}.

\subsection{Evaluation of Jacobian}
Due to the finiteness of FFBRST transformation the Jacobian is not unity and hence it is important 
to consider the Jacobian contribution.
 In this subsection we present the general method to evaluate the nontrivial Jacobian
of path integral measure for FFBRST transformation. Here we utilize the fact that
FFBRST transformation can be written as a succession of infinitesimal transformation given in Eq. (\ref{ibr}).
Now, one defines the path integral measure  as 
\begin{eqnarray}
{\cal D}\phi =J(\kappa){\cal D}\phi (\kappa) =J(\kappa+ d\kappa){\cal D}\phi (\kappa +d\kappa).
 \end{eqnarray}
 Since the transformation $\phi(\kappa)$ to $\phi (\kappa +d\kappa)$ is an infinitesimal one, then
 the equation reduces to
 \begin{eqnarray}
 \frac{J(\kappa)}{J(\kappa+ d\kappa)} =\int d^4x\sum_\phi \pm \frac{\delta \phi(x, \kappa +d\kappa)}{\delta \phi(x, \kappa)},
 \end{eqnarray}
 where $\Sigma_\phi$ sums over all fields in the measure and $\pm$ refers to whether $\phi$ is bosonic or fermionic
 field. 
 Using the Taylor's expansion in the above equation, the expression for infinitesimal change in Jacobian 
is obtained as follows:
 \begin{equation}
 \frac{1}{J(\kappa)}\frac{dJ(\kappa)}{d\kappa}= -\int d^4x\sum_\phi  \left[\pm s_b\phi \frac{\delta \Theta_b'[\phi(x,\kappa)]}{
\delta\phi(x,\kappa)}\right].\label{jaceva}
\end{equation}
 The nontrivial Jacobian is the source of new results in the FFBRST formulation.  
 \section{Batalin-Vilkovisky (BV) formalism}
The BV  formulation (also known as field/antifield formulation) is a powerful  technique in the Lagrangian
framework to deal  
with more general gauge theories. This method is applicable to gauge theories 
with both reducible/open as well as irreducible/close algebras \cite{wei, bv, bv1}. The basic idea in this approach
is to introduce the so-called antifield ($\phi^\star$)  for  each field ($\phi$) in the theory.
The antifields satisfy opposite statistics with ghost number $-gh(\phi)-1$, where $gh(\phi)$ is ghost number of 
field $\phi$.

The effective action $S_{eff}$ is then extended with the antifields as
\begin{equation}
S_{eff}[\phi, \phi^\star ]=I[\phi ]+(s_b \phi)\phi^\star,
\end{equation}
where $I[\phi ]$ is gauge invariant action.
The antifields $\phi^\star$ are obtained from gauge-fixing fermion $\Psi$ as
\begin{equation}
\phi^\star =\frac{\delta\Psi}{\delta \phi}.
\end{equation}
The extended effective action is then written in terms of $\Psi$ as
\begin{equation}
S_{eff}[\phi ]=I[\phi ]+s_b \Psi.
\end{equation}
 The effective action $S_{eff}$
 satisfies certain rich mathematical relation which is known as `quantum master equation' as follows: 
\begin{equation}
(S_{eff},S_{eff})-2i\Delta S_{eff}=0,\ \mbox{at}\  \phi^\star =\frac{\delta\Psi}{\delta \phi},
\end{equation}
where the antibracket of effective action, $(S_{eff},S_{eff})$, is defined by
\begin{eqnarray}
(S_{eff},S_{eff})=\frac{\delta_r S_{eff}}{\delta\phi}\frac{\delta_l S_{eff}}{\delta\phi^\star}-\frac{\delta_r 
S_{eff}}{\delta\phi^\star}
\frac{\delta_l S_{eff}}{\delta\phi },
\end{eqnarray} 
and  $\Delta$ is defined with left and right differentials ($ \delta_l $ and $ \delta_r$ respectively) as
\begin{equation}
\Delta =\frac{\delta_r}{\delta\phi^\star}\frac{\delta_l}{\delta\phi}.
\end{equation}
Usually it is easy to construct an action that satisfies the `classical
master equation'
\begin{equation}
 (S_{eff},S_{eff})=0.
\end{equation}
The generating function given in Eq. (\ref{generator}) can also be written in compact form as
\begin{equation}
Z = \int [{\cal D}\phi] \exp{\left [iW_{\Psi}(\phi,\phi^\star)\right]},
\end{equation}
where $W_{\Psi}(\phi,\phi^\star)$ is an extended action satisfying following 
`quantum master equation' \cite{wei}:
\begin{equation}
\Delta e^{iW_\Psi[\phi, \phi^\star ]} =0.\label{2mq}
\end{equation}
The quantum master equation in the zeroth order of antifields gives the condition of gauge invariance.
On the other hand it reflects the nilpotency of BRST transformation in the first order
of antifields. We will be using this formulation in the framework of FFBRST
formulation in different contexts. 
\section{Batalin-Fradkin-Vilkovisky (BFV) formulation} 
We briefly mention the BFV formalism \cite{ht,bv0}  which is applicable
for the general theories with first-class constraints. This formalism
is developed on  an extended phase space with finite number of 
canonically conjugate variables. The basic features of this approach are as follows:
i) it does not
require closure off-shell of the gauge algebra and therefore does not need an
auxiliary field, ii) heavily relies on BRST transformation which is independent
of the gauge condition and iii) it is even applicable to Lagrangian
which are not quadratic in velocities and hence is more general than the strict
Lagrangian approach. 
The action, in terms of canonical Hamiltonian density ${\cal H}_0$ 
and first-class constraints, $\Omega_a(a = 1, 2, ..., m)$,
 can be written in this formalism as
\begin{eqnarray}
S=\int d^4x (p^\mu \dot q_\mu -{\cal H}_0 -\lambda^a\Omega^a),
\end{eqnarray}
where ($q_\mu, p^\mu$) are the canonical variables and $\lambda^a$ are the Lagrange multiplier associated with
first-class constraints. In this
method, the Lagrange multipliers $\lambda^a$ are treated as the
canonical variables. Therefore, one introduces its canonical  conjugate momenta $p^a$,
 where $p^a$ further imposes new constraints such that the dynamics of the theory must not be changed.
In order to make the extended theory to be consistent with the initial theory
a pair  of canonically conjugate anticommuting ghost coordinate and momenta $(c^a, \pi^a)$ is introduced for each 
constraint. These canonically conjugate ghosts satisfy 
the following 
anticommutation relation: 
\begin{equation}
\{c^a ({\bf x}, t), \pi^b({\bf y}, t)\} =-i\delta^{ab}\delta^3 ({\bf x-y}),
\end{equation}
where $c_a$ and $\pi_a$ have ghost number $1$ and $-1$, respectively.
The generating functional for this extended theory is then defined as
\begin{eqnarray}
Z_{\Psi}=\int [{\cal D}\phi] e^{iS_{eff}[\phi]},
\end{eqnarray} 
where $[{\cal D}\phi]$ is the path integral measure and the effective action $S_{eff}$ is
\begin{eqnarray}
S_{eff}=\int d^4x (p^\mu\dot q_\mu +p^a\dot\lambda^a  +\pi^a \dot c^a - {\cal H}_\Psi ).
\end{eqnarray}
${\cal H}_\Psi$ is the extended Hamiltonian  and is written as
\begin{eqnarray}
{\cal H}_\Psi ={\cal H}_0 +\{Q_b, \Psi\},
\end{eqnarray}
where $\Psi$ is the gauge-fixing fermion and $Q_b$ is the nilpotent BRST charge which has the 
following 
general form:
\begin{equation}
Q_b =c^a \Omega^a +\frac{1}{2} \pi^a f^{abc} c^b c^c,
\end{equation}
where the $f^{abc}$
is a structure constant. 

The BRST symmetry transformation for fields $\phi$ can be calculated with BRST charge $Q_b$ using the relation
\begin{equation}
s_b \phi= -[\phi, Q_b]_\pm,\label{anticom}
\end{equation}
where $ +$  and $ -$ signs, respectively, denote anticommutator and commutator for the fermionic and bosonic nature 
of fields 
$\phi$.

In this chapter we have provided the basic mathematical techniques which are relevant for the later part of the 
thesis.
In  the next chapter we would like to deal with the off-shell FFBRST and FF-anti-BRST transformations.

\label{Chap:chapter2}

\chapter{Off-shell nilpotent FFBRST transformation }
The FFBRST transformation which have been discussed in the previous chapter is on-shell nilpotent.  
In this chapter we extend this formulation using auxiliary  fields to make FFBRST transformation off-shell nilpotent
\cite{susk}.
Several explicit examples of off-shell nilpotent FFBRST transformation are considered. We further construct
both on-shell and off-shell nilpotent FF-anti-BRST transformations and show that such transformations
also play the similar role. 
\section{FFBRST transformation for Faddeev-Popov (FP) effective theory: short survey}
Let us now briefly review the FFBRST formulation, particularly, in case of  FP  effective theory as
an example of pure gauge theories \cite{jm}.
In this case the Jacobian of the path integral measure changes the generating functional corresponding to FP 
effective theory to the generating functional for a different effective theory.               
The meaning of these field transformations is as follows. We consider the vacuum expectation value of a gauge 
invariant functional $G[A]$ in some effective theory,
\begin{equation}
<< G[A]>> \equiv \int [{\cal D} \phi ] \ G[A] \exp( iS_{eff}[\phi]),
\label{90}
\end{equation}
where $ \phi$ is generic notation for all fields and the FP effective action is defined as 
\begin{equation}
S_{eff} = S_0 + S_{gf}+S_g. 
\label{sf}
\end{equation}
Here, $S_0$  is the pure YM action\footnote{Here we adopt the following notations throughout the thesis.
The $d$ dimensional metric tensor in Minkowski space-time is defined  as $g_{\mu\nu} = g^{\mu\nu} = \mbox{diag} (+1,-1,-1,-1,...)$ and the quantity 
 $X_\mu X^\mu =X\cdot X$ is a Lorentz scalar. We always use the small letters (a,b,c,...) in superscript
 to denote the   group indices.  This should not be confused with the small letters (b, ab, d, ad,...) used in
 subscript which denote the different forms of BRST transformations (e.g. BRST, anti-BRST, co-BRST, anti-co-BRST,...). }

\begin{equation}
S_0= \int d^4x \left [-\frac{1}{4}F^{a\mu\nu}F_{\mu\nu}^a\right ],
 \label{s0}
\end{equation} and the gauge-fixing and ghost part of the effective action in Lorentz gauge are
given as
\begin{eqnarray}
S_{gf} &=& -\frac{1}{2\lambda}\int d^4x (\partial \cdot{A^a})^2 \nonumber,\\
S_g &=& -\int d^4x\left [ \bar{c}^a\partial^\mu
D^{ab}_\mu c^b \right ].
\label{s12}
\end{eqnarray}
The covariant derivative is defined as $ D^{ab}_\mu [A] \equiv \delta ^{ab}\partial
_\mu + g f^{abc}A^c_\mu $. 
                                                                                
Now we perform the FFBRST transformation $\phi\rightarrow \phi^\prime
 $ given by Eq. (\ref{fbrs}). Then we have 
\begin{equation}
<<G[A]>>=  <<G[A^\prime ]>> = \int [{\cal D} \phi ^\prime ]
J[\phi^\prime ] G[A^\prime ] \exp(iS^F_{eff}[\phi^\prime ]),
\label{def}
\end{equation}
on account of BRST invariance of $S_{eff}$ and the  gauge invariance of $G[A]$.
Here $J[\phi^\prime ]$ is the Jacobian associated with FFBRST transformation and is defined
as
\begin{equation}
{\cal D} \phi = {\cal D} \phi(\kappa) J[\phi(\kappa) ].
\label{ojac}
\end{equation}
Note that unlike the usual infinitesimal BRST transformation, the Jacobian
for FFBRST is not unity. In fact, this nontrivial Jacobian
is the source of the new results in this formulation. 
As shown in Ref. \cite{jm} for the special case $G[A]=1$, the Jacobian
$ J[\phi(\kappa) ]$ can always be replaced by
$ e^{iS_1[\phi(\kappa) ]}$,  where $S_1[\phi(\kappa)]$ is some {\it local}
functional of the fields and can be added to the action at $\kappa =1$,
\begin{equation}
S_{eff}[\phi^\prime ] +S_1[\phi^\prime ] = S_{eff}^\prime
[\phi^\prime].
\label{s1}
\end{equation}
Thus, the FFBRST transformation changes the FP effective action of the theory \cite{sdj1}.
\section{ Off-shell nilpotent FFBRST transformation }
In this section, we intend to 
generalize the FFBRST 
transformation in an auxiliary field formulation. We only mention the necessary steps of the FFBRST 
formulation in presence of  auxiliary field. For simplicity, we consider the case 
of pure 
YM theory described by the effective action in Lorentz gauge 
\begin{equation}
 S^L_{eff}= \int d^4x\left [-\frac{1}{4}F^{a \mu \nu }F^{a}_{\mu \nu } +\frac{\lambda 
}{2}(B^a) ^2 - B^{a}
\partial \cdot{A^a }-\bar{c}^a \partial^{\mu}D^{
ab}_\mu  c^b \right ]. \label{seff}
\end{equation} 
This effective action is invariant under an off-shell nilpotent usual BRST transformation
with infinitesimal parameter. 
Following the procedure outlined in the previous chapter, it is straightforward to construct FFBRST transformation 
under 
which the above $S^L_{eff}$ remains 
invariant.
The transformation is as follows:
\begin{eqnarray}
A^{a}_\mu &\rightarrow & A^a_\mu + D_\mu^{ab}c^b \  \Theta_b
(A,c,\bar{c},B),\nonumber \\
c^a &\rightarrow & c^a -\frac{g}{2}f^{abc}c^b c^c \ \Theta_b(A,
 c, \bar{c},B), \nonumber \\
\bar{c}^a &\rightarrow & \bar{c}^a +B^a \ \Theta_b(A,c,\bar{c},B),\nonumber \\
B^a &\rightarrow & B^a.
\label{ffb}
\end{eqnarray}
The finite parameter, $\Theta_b (A,c,\bar c,B)$ depends also on the auxiliary field $B$ and hence 
the nontrivial modification arises in the calculation of Jacobian for this FFBRST in 
an auxiliary field formulation. The Jacobian is defined as
\begin{eqnarray}
{\cal D}A(x){\cal D}c(x){\cal D}\bar{c}(x){\cal D}B(x) &=&J(x,k) {\cal D}A(x,k)c{\cal D}(x,k){
\cal D}\bar{c}(x,k){\cal D}B(x,k) \nonumber \\
&=& J(k+dk){\cal D}A(k+dk)Dc(k+dk){\cal D}\bar{c}(k+dk)\nonumber\\
&&{\cal D}B(k+dk).\label{jac1}
\end{eqnarray}
The transformation from $\phi(k)$ to $\phi(k+dk)$ is an infinitesimal one and one has, for 
its Jacobian
\begin{equation}
\frac{J(k)}{J(k+dk)}=\int d^4x\sum_{\phi }\pm\frac{\delta{\phi(x,k+dk)}}{\delta{\phi(x,k)}},
\end{equation}
where $\sum_{\phi }$ sums over all the fields in the measure 
$A_\mu^a,c^a,\bar{c}^a,B^a$ and the $\pm$ sign refers to the cases of fields $\phi$ being
bosonic or fermionic in nature.
We evaluate the right hand side as
\begin{eqnarray}
&& \int d^4x \sum_{a,\mu}\left [ \frac{\delta A_\mu^a(x,k+dk)}{\delta A_
\mu^a(x,k)} - \frac{\delta c^a(x,k+dk)}{\delta c^a(x,k)} - \frac{\delta\bar{c}^
a(x,k+dk)}{\delta\bar{c}^a(x,k)}\right.\nonumber\\
& + &\left.\frac{\delta B^a(x,k+dk)}{\delta B^a(x,k)}
 \right ],
\end{eqnarray}
dropping those terms which do not contribute on account of the antisymmetry of structure 
constant. We calculate the infinitesimal change in Jacobian,  as mentioned in \cite{jm}, as\footnote{The small 
letter (b) used in the subscript of Eq. (3.13),
 which denotes BRST, should not be confused with the group index 
which is always written in the superscript.}
\begin{equation}
\frac{1}{J(k)}\frac{dJ(k)}{dk}= - \int d^4x\left[(s_b A^a _{\mu})\frac{\delta 
\Theta_b^\prime}{\delta A_\mu ^a}-(s_b c^{a})\frac{\delta \Theta_b^\prime}
{\delta c^a} -(s_b \bar{c}^a) \frac{\delta \Theta_b^\prime}{\delta \bar{c}^a}
+(s_b B^a)\frac
{\delta \Theta_b^\prime}{\delta B^a} \right ].\label{jc1}
\end{equation}

Further, it can be shown that the Jacobian in Eq. ({\ref
{jac1}}) can be expressed as 
$e^{iS_1[\phi]}$ if it satisfies the  condition given in Eq. (\ref{mcond})

Now, we consider different choices of the parameter $\Theta_b^\prime$ (which is 
related to $\Theta_b$ through the relation in Eq. (\ref{80})) to show the connection between different pairs of 
effective theories. 
\subsection{ Connecting YM theory in Lorentz gauge to the same theory in axial gauge}
To show the connection between YM theories in Lorentz gauge and axial gauge 
we start with the Lorentz gauge YM theory in the auxiliary field formulation, described by the effective action 
given in Eq. (\ref{seff}) and
 choose the infinitesimal field dependent parameter as
\begin{equation} 
\Theta_b^\prime =i\int d^4x \; {\bar{c}}^a \left [\gamma_1\lambda B^a + \gamma_2 
\left (\partial \cdot A^a -\eta \cdot A^a \right )\right ],\label{tlab}
\end{equation}
where $\gamma_1$, $\gamma_2$ are arbitrary constants and $\lambda $ is a gauge parameter.
 Using 
Eq. (\ref{jc1}), we calculate the change in the Jacobian of such 
transformation as
\begin{equation}
\frac{1}{J}\frac{dJ}{dk} =i \int d^4x \left [ \gamma_1 \lambda {(B^a)}^2 +\gamma_2 B^
a \left (\partial \cdot A^a -\eta \cdot A^a \right )+\gamma _2 {\bar{c}}^a
 \left( M^{ab}c^b -{\tilde {M}}^{ab}c^b \right )\right ],
\end{equation}
where $ M^{ab}\equiv\partial\cdot D^{ab}$ and $\tilde M^{ab}\equiv\eta\cdot D^{ab}$.
We further make an ansatz for $S_1$ in this case as
\begin{eqnarray}
S_1\left [\phi(\kappa), \kappa \right ]&=& \int d^4x\left[\xi_1 (\kappa){(B^a)}^2 +\xi_2 (\kappa ) B^
a \partial\cdot A^a +\xi_3 (\kappa) B^a \eta \cdot A^a \right .\nonumber\\
&+&\left .\xi_4 (\kappa ){\bar{c}}^a M^{ab}c^b +\xi_5 (\kappa){\bar{c}}^a{\tilde{M}}^{ab}c^
b\right]. \label{s1a}
\end{eqnarray}
The constants $\xi_i(\kappa)$  depend on $\kappa$ explicitly and satisfies the 
following initial condition:
\begin{equation}
\xi_i(\kappa=0)=0. \label{xcond}
\end{equation}
Using Eq. (\ref{ibr}) we calculate
\begin{eqnarray}
\frac{dS_1}{d\kappa}&=&\int d^4x\left[\frac{d\xi_1}{d\kappa}{(B^a)}^2+\frac{d\xi_2}{d\kappa} B^a \partial 
\cdot A^
a+\frac{d\xi_3}{d\kappa} B^a \eta\cdot A^a +\frac{d\xi_4}{d\kappa}\bar{c}^a M^{ab}c^b  
\right.\nonumber\\
&+&\left.  
\frac{d\xi_5}{d\kappa} \bar{c}^a\tilde{M}^{ab}c^b+\xi_2 B^a M^{ab}c^b \Theta_b^\prime +\xi_3 B^
a \tilde{M}^{ab}c^b\Theta_b^\prime
 -\xi_4 B^a M^{ab}c^b \Theta_b^\prime \right.\nonumber\\
&-&\left.\xi_5 B^
a \tilde{M}^{ab}c^b\Theta_b^\prime\right].\label{ds1}
\end{eqnarray}
From the condition mentioned in Eq. (\ref{mcond}), we obtain
\begin{eqnarray}
& &\int[{\cal{D}}\phi]  \; e^{i \left (S^L_{eff}+S_1 \right )} \left 
\{ M^{ab}c^b\Theta_b ^\prime[B^a (\xi _2-\xi _4)]+ {\tilde{M}}^{ab}c^b\Theta_b^\prime[B^
a(\xi_3 -\xi_5)]\right. \nonumber\\
&+ &\left.{(B^a)}^2\left(\frac{d\xi_1}{d\kappa}-\gamma _1\lambda \right) +B^a\partial\cdot A^
a\left(\frac{d\xi_2}{d\kappa}-\gamma_2\right) +B^a \eta\cdot A^a \left( \frac{d\xi_3}{d\kappa
} + \gamma _2\right)
\right.\nonumber\\
&+&\left.{\bar{c}}^a M^{ab} c^b \left(\frac{d\xi_4}{d\kappa}-\gamma_2 \right) +{\bar{c}}^a{
\tilde{M}}^{ab}c^b\left(
\frac{d\xi_5}{d\kappa}+\gamma_2 \right)\right \} = 0. \label{cond}
\end{eqnarray}
The last two terms in the integrand of Eq. (\ref{cond}) depend on $\bar c$ in a local 
fashion. The contribution of these terms   vanish by antighost equation of motion
 \cite{jm, sdj1}
\begin{equation}
\int {\cal D}\bar c^a \frac{\delta}{\delta \bar c^a}e^{i(S_{eff}+S_1)}=0.
\end{equation} 
This can only happen if the ratio of coefficients of the two terms is identical to the ratio 
of coefficients of $\bar c^{a}M^{ab}c^{b}$ and $\bar c^{a} \tilde M^{ab}c^{b}$ in $S^L_{eff}+S_1$.
This requires that
\begin{equation}
\frac{{d\xi_4}/{d\kappa}-\gamma_2}{\xi_4 -1} =\frac{{d\xi_5}/{d\kappa}+ \gamma_2}{\xi_5}. \label{orc1}
\end{equation}
The nonlocal $\Theta_b^\prime$ dependent terms are canceled by converting them to local terms 
using antighost equation of motion \cite{sdj1}. This can only work if the two $\Theta_b^\prime$ dependent terms
 in a certain manner, depending again on the ratio of coefficients of $\bar{c}^a 
M^{ab}c^b$ and $ \bar{c}^a \tilde{M}^{ab}c^b$ in terms in $ S^L_{eff}+S_1$. This requires that 
\begin{equation}
\frac{\xi_2 -\xi_4}{\xi_4 -1}=\frac{\xi_3-\xi_5}{\xi_5}.\label{rc2}
\end{equation}
When the above two equations (\ref{orc1}) and (\ref{rc2}) are satisfied, the nonlocal 
$\Theta_b^\prime$ dependent terms get converted to local terms. The coefficients of local terms 
${(B^a)}^2$, $ B^a \partial\cdot A^a$, $B^a \eta\cdot A^a$, $\bar{c}^a
M^{ab}c^b$, and $\bar{c}^a \tilde{M}^{ab}c^b $, independently, vanish and are giving rise to following 
differential equations, respectively:
\begin{eqnarray}
&&\frac{d\xi_1}{d\kappa} -\gamma_1\lambda +\gamma_1\lambda(\xi_2-\xi_4 ) + \gamma_1 \lambda(\xi_3-\xi_5) =0, 
\nonumber\\
&&\frac{d\xi_2}{d\kappa} -\gamma_2 +\gamma_2 (\xi_2-\xi_4 ) + \gamma_2 (\xi_3-\xi_5) =0, \nonumber\\
&&\frac{d\xi_3}{d\kappa} +\gamma_2 -\gamma_2 (\xi_2-\xi_4 ) - \gamma_2 (\xi_3-\xi_5) =0, \nonumber\\
&&\frac{d\xi_4}{d\kappa} -\gamma_2 =0, \ \ \ \frac{d\xi_5}{d\kappa} +\gamma_2 =0. 
\end{eqnarray}
The above equations can be solved for various $\xi_i(\kappa)$ using the boundary conditions 
given by Eq. (\ref{xcond}) and the solutions ($\gamma_2 =1$) are given as
\begin{eqnarray}
\xi_1 &=& \gamma_1\lambda\kappa, \ \
\xi_2  =  \kappa, \ \
\xi_3  =  -\kappa, \ \
\xi_4 =\kappa, \ \
\xi_5  = -\kappa. \label{sol}
\end{eqnarray}
Putting the above values in Eq. (\ref{s1a}), we get
\begin{equation}
S_1=\gamma_1\lambda\kappa{(B^a)}^2 +\kappa B^a\partial \cdot A^a -\kappa B^
a\eta \cdot A^a +\kappa \bar{c}^a M^{ab}c^b -\kappa \bar{c}^a \tilde{M}^{ab}c^
b. 
\end{equation}
FFBRST transformation in Eq. (\ref{ffb}) with the parameter given in Eq. (\ref{tlab}) connects the 
generating functional in Lorentz gauge,
\begin{equation}
 Z_L=\int [{\cal{D}}\phi ]\  e^{iS_{eff}^L},\label{gf}
\end{equation}
 to the
generating functional corresponding to the effective action
\begin{eqnarray}
S^\prime_{eff}&=&S_{eff}^L+S_1(\kappa=1)=  \int d^4x \left[-\frac {1}{4}F_{\mu\nu}^a F^{a\mu\nu} +\frac{
\zeta}{2}{(B^a)}^2-B^
a\eta\cdot A^a\right.\nonumber\\ 
 &-&\left.\bar{c}^a \tilde{M}^{ab}c^b \right]
 =  S_{eff}^{\prime A},
\end{eqnarray}
where $S_{eff}^{\prime A}$ is nothing but FP effective action in axial gauge with the gauge parameter
$\zeta=(2\gamma_1+1)\lambda$.

\subsection{Relating theories in Coulomb gauge and Lorentz gauge}
We again  start with Lorentz gauge theory given in Eq. (\ref{seff}) and choose another parameter 
\begin{equation}
\Theta_b^\prime = i\int d^4x\ \bar{c}^a\left[ \gamma_1\lambda B^a +\gamma_2\left( 
\partial \cdot A^a -\partial^j A_j^a \right ) \right ];\  j=1,2,3, \label{tcg}
\end{equation}
to show the connection with theory in the Coulomb gauge.
The change in the Jacobian due to this FFBRST transformation is calculated using Eq. (\ref{jc1}) as
\begin{equation}
\frac{1}{J}\frac{dJ}{dk} =i \int d^4x \left [ \gamma_1 \lambda {(B^a)}^2 +\gamma_2 B^
a\left (\partial \cdot A^a - \partial^j A_j^a \right )+\gamma _2 {\bar{c}}^a
 \left( M^{ab}c^b -{\tilde {M'}}^{ab}c^b \right )\right ],
\end{equation}
where ${\tilde M'}  ={\partial^j} D_j^{ab}$.
We try the following ansatz for $S_1$ for this case:
\begin{eqnarray}
S_1 &=&\int d^4x \left[ \xi_1 (\kappa ){(B^a)}^2 + \xi_2 (\kappa) B^a \partial \cdot
 A^a + \xi_3 (\kappa )B^a \partial^j A_j^a \right .\nonumber\\
&+&\left .\xi_4 (\kappa) \bar{c}^a (Mc)^a + \xi_5 (\kappa) \bar{c}^a (\tilde M^\prime c)^
a\right ]. \label{s1cg}
\end{eqnarray}
Now, using the condition Eq. (\ref{mcond}) for replacing the Jacobian as $ e^{iS_1}$ and following the similar 
procedure as discussed in the previous case, we obtain exactly same solutions as given in Eq. (\ref{sol}) for the 
coefficients $\xi_i$. Putting these solutions in Eq. (\ref{s1cg}) we obtain
\begin{equation}
S_1 =\int d^4x \left [\gamma_1 \lambda \kappa {(B^a)}^2 + \kappa B^a\partial \cdot 
A^a -\kappa B^a\partial^j A^a_j +\kappa{\bar c}^a (Mc)^a-\kappa{
\bar c}^a(\tilde M^\prime c)^a\right ].
\end{equation}
The transformed effective action 
\begin{eqnarray}
S^\prime_{eff} &=&S^L_{eff} +S_1(\kappa =1), \nonumber\\
&=&\int d^4x \left[-\frac{1}{4}F^{a \mu \nu }F^{a}_{\mu 
\nu}+\frac{\zeta}{2}{(B^a)}^2-B^a\partial ^j A^a_j 
-\bar c^a(\tilde M^\prime c)^a \right], 
\end{eqnarray}
which is the FP effective action in Coulomb gauge with gauge parameter $\zeta $.

Thus, FFBRST transformation 
with parameter given in equation (\ref{tcg}) connects the generating functional
 for YM theories in Lorentz gauge to the generating functional for the same theory in Coulomb gauge.
\subsection{FFBRST transformation to link FP effective action in Lorentz gauge to quadratic gauge}
Next we consider theories in quadratic gauges which are often useful in doing
calculations \cite{thoo}.
The effective action in quadratic gauge in terms of auxiliary field can be written as
\begin{eqnarray}
S^Q_{eff} &=&\int d^4x \left[ -\frac{1}{4}F^a_{\mu\nu}F^{a\mu\nu} +\frac{\lambda}{2}{(B^
a)}^2 - B^a \left ( \partial\cdot A^a +d^{abc}A_\mu^b A^{
\mu c} \right )\right. \nonumber\\
 &-& \left.\bar{c}^a\partial^\mu {(D_\mu c)}^a -2d^{abc}\bar{c}^
a A^{\mu c}{(D_\mu c)}^b\right ],\label{qg}
\end{eqnarray}
where $d^{abc}$ is structure constant symmetric in $b$ and $c$. 
This effective action is invariant under the FFBRST transformation mentioned in Eq. (\ref{ffb}).

For this case, we start with the following choice of the infinitesimal field dependent parameter:
\begin{equation}
\Theta_b^\prime = i\int d^4x\ \bar{c}^a \left [ \gamma_1\lambda B^a +\gamma_2 d^{
abc} A_\mu^b A^{\mu c} \right ]. \label{qeff}
\end{equation}
We calculate the Jacobian change as
\begin{equation}
\frac{1}{J}\frac{dJ}{d\kappa}=i\int d^4x \left [\gamma_1\lambda {(B^a)}^2 + \gamma_2 B^a 
d^{abc}A_\mu^b A^{\mu c} + 2\gamma_2 d^{abc}\bar{c}^a
 (D_\mu c)^b A^{\mu c} \right ].
\end{equation}
We make an ansatz for $S_1$ for this case as
\begin{equation}
S_1=\int d^4x \left [\xi_1(\kappa) {(B^a)}^2 +\xi_2 (\kappa) B^a d^{
abc}A_\mu^b A^{\mu c} +\xi_3 (\kappa) d^{abc}\bar{c}^
a (D_\mu c)^b A^{\mu c} \right ].
\end{equation}
The unknown coefficients $\xi_i$ are determined by using the condition in equation  (\ref{mcond}) and the initial 
condition in equation (\ref{xcond}), to get
\begin{equation}
S_1(\kappa =1)=\gamma_1\lambda {(B^a )}^2-B^a d^{abc}A^b _\mu A^{
\mu c}-2d^{abc}\bar c^a{(D_\mu c)}^b A^{\mu c},
\end{equation}
and $ S^\prime_{eff} =S^L_{eff} +S_1(\kappa =1)= S^{\prime Q}_{eff} $, 
which is effective action in quadratic gauge as given in Eq. (\ref{qg}) with gauge parameter $\zeta $.

Thus, the FFBRST transformation with parameter given in equation (\ref{qeff}) connects 
Lorentz gauge theory to the theory for 
quadratic gauge.
\subsection{FP action to the most general BRST/anti-BRST invariant action}
The most general BRST/anti-BRST invariant action for YM theories in Lorentz gauge is given as \cite{bath}
\begin{eqnarray}
S_{eff}^{AB} &=& \int d^4x \left[-\frac{1}{4}F^{a\mu \nu }F^{a}_{\mu 
\nu} - \frac{(\partial\cdot A^a)^2}{2\lambda}  + \partial ^\mu\bar{c}^a D_\mu^{ab} c^b
+\frac{\alpha}{2} gf^{abc }\partial\cdot A^a\bar{c}^b c^
c \right. \nonumber 
\\ &-&\left.\frac{1}{8}\alpha (1-\frac{1}{2}\alpha )\lambda g^2 f^{abc}\bar{c}^
b\bar{c}^c f^{alm}c^l c^m \right ].
\end{eqnarray}
This effective action has the following global symmetries.\\
BRST:
\begin{eqnarray}
\delta_b A_\mu ^a &=&(D_\mu c)^a  \Lambda, \ \ 
\delta_b c^a =  -\frac{1}{2} g f^{abc}c^b c^c \Lambda,\nonumber\\ 
\delta_b\bar{c}^a &=& \left(\frac{\partial\cdot A^a}{\lambda} -\frac{1}{2}a g f^{
abc} \bar{c}^b c^c \right) \Lambda. 
\end{eqnarray}
Anti-BRST:
\begin{eqnarray}
\delta_{ab} A_\mu ^a &=&(D_\mu \bar c)^a \Lambda, \ \
\delta_{ab} \bar c^a =-\frac{1}{2} g f^{abc }\bar c^b \bar c^c \Lambda,\nonumber\\
\delta_{ab}{c}^a &=&-\left( \frac{\partial\cdot A^a}{\lambda}+(1-\frac{1}{2}\alpha) g f^{
abc} \bar{c}^b c^c \right) \Lambda.\label{mostanti}
\end{eqnarray}
The above most general BRST/anti-BRST effective action can be re-expressed in the auxiliary  
field formulation as, 
\begin{eqnarray}
S_{eff}^{AB} &=& \int d^4x \left[-\frac{1}{4}F^{a \mu \nu }F^{a}_{\mu 
\nu}+\frac{\lambda}{2}(B^a)^2 - B^a(\partial\cdot A^a-\frac{\alpha g \lambda}{2}f^{
abc}\bar{c}^b c^c)\right. \nonumber\\
&+&\left. \partial ^\mu\bar{c}^aD_\mu^{ab} c^b -\frac{1}{8}\alpha\lambda g^2f^{abc}\bar{c}^
b \bar{c}^c f^{alm}c^l c^m\right].\label{seffab}
\end{eqnarray}
The off-shell nilpotent, global BRST/anti-BRST symmetries for this effective action are 
given as\\
BRST:
\begin{eqnarray}
\delta_b A_\mu ^a &=&(D_\mu c)^a \ \Lambda, \ \ \ \
\delta_b c^a = -\frac{1}{2} g f^{abc }c^b c^c\; \Lambda, 
\nonumber \\
\delta_b\bar{c}^a &=& B^a\ \Lambda,\ \ \ \ \ \
\delta_b B^a = 0.
\end{eqnarray}
Anti-BRST:
\begin{eqnarray}
\delta_{ab}A^a_\mu &=&{(D_\mu \bar c)}^a  \ \Lambda, \ \ \ \ \
\delta_{ab}\bar c^a =-\frac{1}{2}gf^{abc }\bar c^b \bar c^c
\ \Lambda,\nonumber\\
\delta_{ab} c^a &=&{(-B^a -gf^{abc}\bar c^b c^c )}\  
\Lambda,\ \ \
\delta_{ab} B^a =-gf^{abc}B^b \bar c^c \ \Lambda.
\label{antb}
\end{eqnarray}
To obtain the generating functional corresponding to this theory, we apply the FFBRST transformation with the finite 
field parameter obtainable from
\begin{equation}
\Theta_b^\prime =i\int d^4x\ \bar{c}^a\left [ \gamma_1 \lambda  B^a +\gamma_2 f^{abc}\bar{c}^
b c^c \right ],
\end{equation}
on the generating functional given in Eq. (\ref{gf}).
Using Eq. (\ref{jc1}), change in Jacobian can be calculated as
\begin{equation}
\frac{1}{J}\frac{dJ}{d\kappa}=i\int d^4x \left[ \gamma_1\lambda {(B^a)}^2 +2 \gamma_2 f^{
abc}B^a\bar{c}^b c^c -\frac{g}{2} \gamma_2 f^{abc}
\bar{c}^b \bar{c}^c f^{alm}c^l c^m \right ].
\end{equation}
We further make an ansatz for $S_1$ as
\begin{equation}
S_1=\int d^4x \left[ \xi_1 (\kappa ) {(B^a)}^2 +\xi_2 (\kappa )B^a f^{abc}\bar{c}^b c^
c + \xi_3 (\kappa )f^{abc}\bar{c}^b\bar{c}^c f^{alm}c^l c^
m \right ].
\end{equation}
The condition given in equation (\ref{mcond}), 
then be written for this case as
\begin{eqnarray}
&&\int [{\cal{D}}\phi]  \; e^{i \left (S^L_{eff}+S_1 \right )} \left 
[ \left (\frac{d\xi_1}{d\kappa} -\gamma_1\lambda \right )(B^a)^2  
+  \left(\frac{d\xi_2}{d\kappa}-2\gamma_2\right) B^a f^{abc}\bar{c}^b c^
c \right.  \nonumber \\
&+&\left.\left (\frac{d\xi_3}{d\kappa} +\frac{g}{2}\gamma_2 \right ) f^{abc}
\bar{c}^b \bar{c}^c f^{alm}c^l c^m  
 -  \left ( \frac{g}{2} \xi_2 
+2 \xi_3 \right )f^{abc} B^b\bar{c}^c f^{alm}c^l c^
m \Theta_b' \right ]\nonumber\\
&=&0.\label{mgcond} 
\end{eqnarray}
We look for a special solution corresponding to the condition
\begin{equation}
\frac{g}{2} \xi_2 
+2 \xi_3 =0.\label{sp}
\end{equation}
Comparing the different coefficients, we get the following differential equations for $\xi_i (\kappa )$:
\begin{eqnarray}
\frac{d\xi_1}{d\kappa} -\gamma_1\lambda = 0, \ \
\frac{d\xi_2}{d\kappa}-2\gamma_2 = 0,\ \
\frac{d\xi_3}{d\kappa} +\frac{g}{2}\gamma_2 = 0.
\end{eqnarray}
Solutions of the above equations subjected to the initial condition given in Eq. (\ref{xcond}) are, 
\begin{eqnarray}
\xi_1  = \gamma_1 \lambda \kappa,  \ \
\xi_2  =  2\gamma_2 \kappa, \ \
\xi_3  =  -\frac{g}{2} \gamma_2\kappa.\label{soln}
\end{eqnarray}
These solutions are consistent with condition in Eq. (\ref{sp}). Since $\gamma_2$ is arbitrary, we choose 
$\gamma_2=\frac{1}{4} \alpha g \zeta$ 
to get
\begin{eqnarray}
S^L_{eff}&+&S_1 ( \kappa =1 )= \int d^4x \left[-\frac {1}{4}F_{\mu\nu}^a F^{a\mu\nu}+\frac{\zeta}{2}(B^
a)^2 
- B^a(\partial\cdot A^a \right. \nonumber\\ 
&-&\left.\frac{\alpha g \zeta}{2}f^{
abc}\bar{c}^b c^c)
+ \partial ^\mu\bar{c}^a D_\mu^{ab} c^b -\frac{1}{8}\alpha\zeta g^2f^{abc}\bar{c}^
b \bar{c}^c f^{alm}c^l c^m \right]\nonumber\\
& \equiv & S^{\prime AB}_{eff}, \label{mga}
\end{eqnarray}
which is the same effective action as mentioned in Eq. (\ref{seffab}), where $\zeta$ is the 
gauge parameter.

Thus, even in the auxiliary field formulation the different generating functionals corresponding to the different 
effective theories are connected through off-shell nilpotent FFBRST transformation with different choices of the 
finite parameter which also depends on the auxiliary field. 
\section{Finite field dependent anti-BRST (FF-anti-BRST) formulation}
In this section, we construct the FF-anti-BRST transformation analogous to FFBRST transformation. For simplicity, we 
consider the pure YM theory in Lorentz gauge described by the effective action in Eq. (\ref{sf}) which is invariant 
under the following on-shell anti-BRST transformation:
\begin{eqnarray}
\delta_{ab} A_\mu ^a &=&(D_\mu \bar{c})^a\; \Lambda , \nonumber \\
\delta_{ab}\bar c^a &=& -\frac{1}{2} g f^{abc }\bar{c}^b \bar{c}^c
\;\Lambda , \nonumber \\
\delta_{ab}{c}^a &=&\left( -\frac{\partial\cdot A^a}{\lambda} - g 
f^{abc} \bar{c}^b c^c \right) \;\Lambda, \label{abt} 
\end{eqnarray}
where $\Lambda$ is infinitesimal, anti commuting and global parameter. This anti-BRST transformation is special
case $(\alpha =0)$ of general anti-BRST transformation given in Eq. (\ref{mostanti}).
Following the procedure similar to the construction of FFBRST transformation as outlined in the 
chapter 2,
we can easily construct the FF-anti-BRST transformation for the pure YM theory as
\begin{eqnarray}
\delta_{ab} A_\mu ^a &=&(D_\mu \bar{c})^a \;\Theta_{ab} , \nonumber \\
\delta_{ab}\bar c^a &=& -\frac{1}{2} g f^{abc }\bar{c}^b \bar{c}^c 
\;\Theta_{ab}, \nonumber \\
\delta_{ab}{c}^a &=&\left( -\frac{\partial\cdot A^a}{\lambda} - g 
f^{abc} \bar{c}^b c^c \right)\;\Theta_{ab}, \label{anti}
\end{eqnarray}
where $ \Theta_{ab} (A,c,\bar c)$ is finite, field dependent and anticommuting parameter.

 In FF-anti-BRST formulation, 
 the infinitesimal change in fields $\phi$  is written as
\begin{equation}
\frac{d}{d \kappa}\phi(x, \kappa ) = s_{ab}\phi (x, \kappa )\
\Theta_{ab}^ \prime [\phi(x,\kappa )],
\label{ibro}
\end{equation}
where $s_{ab}\phi$ is anti-BRST variation of fields $\phi$ and 
$\Theta_{ab}^\prime $ is an infinitesimal field dependent parameter  which is related with 
finite field dependent parameter $ \Theta_{ab} (A,c,\bar c)$ as
 \begin{equation}
\Theta_{ab} [\phi(x, \kappa)] = \Theta_{ab}^\prime [\phi(x, 0)] \frac{ \exp g[\phi(x, 0)]
-1}{g[\phi(x, 0)]},
\label{81}
\end{equation}
and $g$ is given by $ g= \sum_{i} \frac{ \delta \Theta_{ab}^\prime (x)}{\delta
\phi_i(x)} s_{ab} \phi_i(x).$

In FF-anti-BRST formulation, we use the following relation
to calculate the infinitesimal change in Jacobian:
\begin{equation}
 \frac{1}{J(\kappa)}\frac{dJ(\kappa)}{d\kappa}= -\int d^4x\sum_\phi  \left[\pm s_{ab}\phi \frac{\delta \Theta_{ab}'
[\phi(x,\kappa)]}{
\delta\phi(x,\kappa)}\right].\label{jaceva1}
\end{equation}
  
Now, we would like to investigate the role of such transformation by considering different infinitesimal
 field dependent parameters $ \Theta_{ab}^\prime(A,c,\bar c)$
\subsection{ FF-anti-BRST transformation to change the gauge parameter $\lambda$ }
We consider a very simple example to  show that a simple 
FF-anti-BRST transformation can transfer the generating functional corresponding to YM effective action in Lorentz 
gauge with a gauge parameter $\lambda$ to the generating functional corresponding to same effective action with a 
different gauge parameter $\lambda^\prime$.
We start with the Lorentz gauge effective action given in Eq. ({\ref {seff}) with the gauge 
parameter $\lambda$ and consider
\begin{equation}
\Theta_{ab}^\prime = -i\gamma \int {d^4x\ { c^a (x,\kappa)}\partial\cdot A^a (x,\kappa)},\label{tllb}
\end{equation}
with $\gamma$ as arbitrary parameter.   
 
Using Eq. (\ref{jaceva1}), we calculate the infinitesimal change in Jacobian as
\begin{equation}
\frac{1}{J}\frac{dJ}{d\kappa}=i\gamma \int d^4x\ \left[\frac{(\partial\cdot A^a )^2}{\lambda} +\bar c^a M^{ab}c^
b \right].
\end{equation}
We choose
\begin{equation}
S_1=\xi (\kappa )\int d^4x\ \frac{(\partial\cdot A^a )^2}{\lambda}.
\end{equation}
The condition, for replacing the Jacobian of the FF-anti-BRST transformation with parameter given in 
Eq. (\ref{tllb}) as $e^{iS_1}$, is given in Eq. (\ref{mcond}) and is calculated as
\begin{eqnarray}
 \int [{\cal{D}}\phi] \; e^{i \left (S^L_{eff}+S_1 \right )}  \left 
[\frac{(\partial\cdot A^a )^2}{\lambda}(\xi^\prime -\gamma )+2\xi\frac{\partial\cdot A^a}{\lambda}M^{ab}\bar c^
b \Theta_{ab}^\prime  
-\bar c^a Mc^a\right] 
 = 0.\label{xy}
\end{eqnarray}
The last term of above equation gives no contribution due to dimensional regularization  and we can substitute \cite
{jm}
\begin{equation}
\int d^4x\ \frac{\partial\cdot A^a}{\lambda}M^{ab}\bar c^b \Theta_{ab}^\prime\longrightarrow \gamma\frac{(\partial\cdot 
A^a )^2}{\lambda}.
\end{equation}  

Thus, the L.H.S. of Eq. (\ref{xy}) is vanish iff
\begin{equation}
\xi^\prime -\gamma +2\xi\gamma =0.
\end{equation}
We solve this equation subjected to the initial condition given in Eq. (\ref{xcond}) to obtain
\begin{equation}
\xi =\frac{1}{2}(1-e^{-2\gamma\kappa}).
\end{equation}
Thus, at $\kappa=1$ the extra term in the net effective action from the Jacobian is 
\begin{equation}
S_1 =\frac{1}{2}(1-e^{-2\gamma})\int d^4x\ \frac{(\partial\cdot A^a )^2}{\lambda}.
\end{equation}
The new effective action becomes $S^\prime_{eff}=S^L_{eff}+S_1 $. In this case, 
\begin{equation}
S^L_{eff}+S_1=\int d^4x\ \left[-\frac {1}{4}F_{\mu\nu}^a F^{a\mu\nu} -\frac{1}{2\lambda^\prime}{(\partial\cdot A^
a)}^2 
-\bar c^a {M}^{ab}{c}^b \right] ,
\end{equation}
which is effective action in Lorentz gauge with gauge parameter $\lambda^\prime =\lambda/e^{-2\gamma}$.
Thus, the FF-anti-BRST transformation in Eq. (\ref{anti}) with parameter given in Eq. (\ref{tllb}) connects two 
effective theories which differ only by a gauge parameter.

\subsection{ Lorentz gauge  to axial gauge theory }
In this subsection we show that FF-anti-BRST transformation plays exactly same role
of FFBRST transformation in connecting different effective theories. For this purpose
we consider same pair of theories which were connected by FFBRST transformation.  
The effective action in Lorentz gauge given in Eq. (\ref{sf}) can be written as 
\begin{equation}
S^L_{eff}= \int d^4x \left[-\frac {1}{4}F_{\mu\nu}^a F^{a\mu\nu} -\frac{1}{2\lambda}({\partial\cdot A^
a})^2+ c^a M^{ab}\bar{c}^b -g f^{
abc}\bar{c}^b c^c{(\partial \cdot A)}^a \right],
\end{equation}
where we have interchanged the position of $c, \bar c$ in the ghost term for the seek of convenience.
This action is invariant under anti-BRST transformation given in Eq. (\ref{abt}).
Similarly, the effective action in axial gauge can be written as 
\begin{equation}
S^A_{eff} = \int d^4x \left[-\frac {1}{4}F_{\mu\nu}^a F^{a\mu\nu}-\frac{1}{2\lambda}{(\eta\cdot A^a)
}^2 +c^a\tilde{M}\bar{c}^a -g f^{
abc}\bar{c}^b c^c {(\eta\cdot A)}^a \right],
\end{equation}
which is invariant under the following anti-BRST symmetry transformation:
\begin{eqnarray}
\delta_b A_\mu ^a =(D_\mu \bar{c})^a \;\Lambda , \ 
\delta_b c^a =-\frac{1}{2} g f^{abc }\bar{c}^b \bar{c}^c
\;\Lambda , 
 \
\delta_b\bar{c}^a =\left( -\frac{\eta\cdot A^a}{\lambda} - g 
f^{abc} \bar{c}^b c^c \right)\; \Lambda. 
\end{eqnarray}
Now, we show that the generating functionals corresponding to these two effective action are related through
 FF-anti-BRST transformation.

To show the connection, we choose 
\begin{equation}
\Theta_{ab}^\prime =-i\gamma \int d^4x\ c^a \left (\partial\cdot A^a -\eta \cdot A^a
\right ).
\end{equation}
We calculate the change in Jacobian corresponding to this FF-anti-BRST transformation using Eq. (\ref{jaceva1}) as
\begin{eqnarray}
\frac{1}{J}\frac{dJ}{d\kappa}&=& i\gamma \int d^4x \left [ \frac{1}{\lambda}{(\partial\cdot A^
a)}^2-\frac{1}{\lambda}(\partial\cdot A^a)(\eta\cdot A^a) +gf^{
abc}\bar{c}^b c^c (\partial\cdot A^a)\right. \nonumber \\
&-&\left. gf^{abc}\bar{c}^b c^c(\eta\cdot A^a )-c^a M^{ab}\bar{c}^b+c^a \tilde
{M}^{ab}\bar{c}^b \right ].
\end{eqnarray}
We make an ansatz for $S_1$ as the following: 
\begin{eqnarray}
S_1&=&\int d^4x\left[\xi_1 (\kappa)(\partial\cdot A^a )^2+\xi_2 (\kappa)(\eta\cdot A^a )^2+\xi_3 (
\kappa)(\partial\cdot A^
a )(\eta\cdot A^a )+\right. \nonumber\\
&&\left.\xi_4 (\kappa)\left (c^a M^{ab}\bar c^b - gf^{abc}
\bar c^b c^c \partial\cdot A^a \right ) 
+\xi_5 (\kappa)\left ( c^a \tilde M^{ab}\bar c^b - gf^{abc}\bar c^b c^c
 \eta\cdot A^a\right )\right],
\end{eqnarray}
where $\xi_i(\kappa)$ are parameters to be determined.
The condition mentioned in Eq. (\ref{mcond} ) to replace the Jacobian as $e^{iS_1} $ for this case is
\begin{eqnarray}
&&\int [{\cal{D}}\phi]  \; e^{i \left (S^L_{eff}+S_1 \right )} \left 
[ \left ( M^{ab}\bar{c}^b - gf^{abc}\bar{c}^b \partial\cdot A^
c\right )\Theta_{ab}^\prime \left \{ \partial\cdot A^a
\left( 2\xi_1 +\frac{\xi_4}{
\lambda}\right )\right.\right. \nonumber \\ 
 &+&\left. \left.  \eta\cdot A^a\xi_3\right \}
+ \left (\tilde{M}^{ab}\bar{c}^b -gf^{abc}\bar{c}^b\eta\cdot A^
c\right ) \Theta_{ab}^\prime \left \{(\partial\cdot A^a)
 \left( \xi_3 +\frac{\xi_5}{
\lambda}\right)\right. \right. \nonumber\\
&+&\left.\left.2\xi_2(\eta\cdot A^a)\right \} +\left(\frac{d\xi_1}{d\kappa} -\frac{\gamma}{\lambda}\right ){(
\partial\cdot A^a)}^2 +
\frac{d\xi_2}{d\kappa} {(\eta \cdot A^a)}^2\right. \nonumber\\
&+&\left. (\partial\cdot A^a)(\eta\cdot A^a) 
\left(\frac{d\xi_3}{d\kappa}+\frac{\gamma}{\lambda}\right) +\left(\frac{d\xi_4}{d\kappa}+\gamma\right) c^a M^{ab} 
\bar{c}^b +\left( 
\frac{d\xi_5}{d\kappa}-\gamma \right) c^a\tilde{M}^{ab}\bar{c}^b  \right. \nonumber\\
&-&\left.  gf^{abc}\bar{c}^
b c^c(\partial\cdot A^
a)\left(\frac{d\xi_4}{d\kappa}+\gamma\right)
-g f^{abc}\bar{c}^b c^c(\eta\cdot A^a) 
\left(\frac{d\xi_5}{d\kappa}-\gamma\right) \right ]=0. \label{rto}
\end{eqnarray}
The last four terms in the integrand of Eq. (\ref{rto}) are dependent on $c$ in a local 
fashion. The contribution of these terms can possibly vanish by ghost equation of motion 
\cite{jm, sdj1}. 
\begin{equation}
\int {\cal D} c^a \frac{\delta}{\delta c^a}e^{i(S_{eff}+S_1)}=0.\label{gh}
\end{equation} 
This can only happen if the ratio of coefficients of the four terms is identical to the ratio 
of coefficients of $\bar c^{a }M^{ab}c^{ b}$ and $\bar c^{a} \tilde M^{ab}c^{b}$ in $S^L_{eff}+S_1$.
This requires that
\begin{equation}
\frac{{d\xi_4}/{d\kappa}+\gamma}{\xi_4 +1} =\frac{{d\xi_5}/{d\kappa}- \gamma}{\xi_5}. \label{rc1}
\end{equation}
The nonlocal $\Theta_{ab}^\prime$ dependent terms are cancelled by converting them to local terms 
using ghost equation of motion \cite{sdj1}. This occurs only if the two $\Theta_{ab}^\prime$ dependent terms
 combine in a 
certain manner, depending again on the ratio of coefficients of $\bar{c}^a 
M^{ab}c^b$ and $ \bar{c}^a \tilde{M}^{ab}c^b$ in  terms in $ S^L_{eff}+S_1$. i.e. 
\begin{eqnarray}
\frac{2\xi_1 +{\xi_4}/{\lambda}}{\xi_4 +1}&=&\frac{\xi_3 +{\xi_5}/{\lambda}}{\xi_5},\nonumber\\
\frac{\xi_3}{\xi_4 +1}&=&\frac{2\xi_2}{\xi_5}, \nonumber\\
\frac{{d\xi_4}/{d\kappa} +\gamma}{\xi_4 +\gamma}&=&\frac{{d\xi_5}/{d\kappa} -\gamma}{\xi_5}.
\end{eqnarray}
Comparing the coefficients of $(\partial\cdot A^a )^2 $, $(\eta\cdot A^a )^2 $, $(\partial\cdot A^a 
)(\eta\cdot A^a )$, $ (c^a M^{ab} \bar{c}^b -gf^{abc}\bar{c}^
b c^c\partial\cdot A^
a)$ and $(c^a\tilde{M}^{ab}\bar{c}^b - g f^{abc}\bar{c}^b c^c\eta\cdot A^a)$
\ respectively, we get  
\begin{eqnarray}
&&\frac{d\xi_1}{d\kappa} -\frac{\gamma}{\lambda}+\gamma \left(2\xi_1 +\frac{\xi_4}{\lambda}\right) +\gamma
 \left(\xi_3 +\frac{\xi_5}{\lambda}\right) =0,\label{av} \\
 &&\frac{d\xi_2}{d\kappa} -\gamma \xi_3 -2\gamma \xi_2 =0, \\
 &&\frac{d\xi_3}{d\kappa} +\frac{\gamma}{\lambda}+\gamma \xi_3 -\gamma \left(2\xi_1 +\frac{\xi_4}{\lambda}
\right) +2\gamma \xi_2 -\gamma  \left(\xi_3 +\frac{\xi_5}{\lambda}\right) =0,\\
&&\frac{d\xi_4}{d\kappa}+\gamma =0,\nonumber\\
&&\frac{d\xi_5}{d\kappa}-\gamma =0. \label{az}
\end{eqnarray}
The solutions of the above equations (\ref{av}) to (\ref{az}) (for $\gamma =1$) are 
\begin{eqnarray}
\xi_1 &=&\frac{1}{2\lambda}\left[1-(\kappa -1)^2\right],\ \
\xi_2 =-\frac{\kappa^2}{2\lambda},\nonumber\\
\xi_3 &=&\frac{1}{\lambda}\kappa (\kappa -1),\ \
\xi_4 =-\kappa ,\ \
\xi_5 =\kappa .
\end{eqnarray}
Putting these in the expression for $S_1$, we have
\begin{eqnarray}
S_1(\kappa=1 )&=&\int d^4x\left[\frac{(\partial\cdot A^a )^2}{2\lambda}-\frac{(\eta\cdot A^a )^2}{2\lambda
}-c^
a M^{ab}\bar c^b +c^a\tilde M^{ab}\bar c^b \right.\nonumber\\
&+&\left. gf^{abc}\bar c^b c^
c \partial\cdot A^\alpha 
- gf^{abc}\bar c^b c^c \eta\cdot A^a\right].
\end{eqnarray}
The new effective action becomes $S^\prime_{eff}=S^L_{eff}+S_1 $. In this case, 
\begin{eqnarray}
S^L_{eff}+S_1&=&\int d^4x \left[-\frac {1}{4}F_{\mu\nu}^a F^{a\mu\nu} -\frac{1}{2\lambda}{(\eta\cdot A^a)}^2 
+c^a\tilde{M}^{ab}\bar{c}^b\right.\nonumber\\
 &-&\left. g f^{abc}\bar{c}^b c^c {(\eta\cdot 
A)}^a\right]
 =  S^A_{eff},
\end{eqnarray}
which is nothing but the FP effective action in axial gauge.

Thus, the generating functional corresponding to Lorentz gauge and axial gauge can also be related by FF-anti-BRST 
transformation. We observe FF-anti-BRST transformation plays exactly the same role as FFBRST transformation in this 
example.
\subsection{Lorentz gauge and Coulomb gauge in YM theory}
To show the connection between generating functional corresponding to the effective action in Lorentz gauge to that 
of the effective action in Coulomb gauge through FF-anti-BRST transformation, we choose the parameter, 
\begin{equation}
\Theta_{ab}^\prime =-i\gamma \int d^4x\ c^a(\partial\cdot A^a -\partial_j A^{ja}).
\end{equation}
Using equation (\ref{jaceva1}), we calculate the change in Jacobian as
\begin{eqnarray}
{\frac{1}{J}\frac{dJ}{d\kappa}}&=&i\gamma \int d^4x\left[\frac{(\partial\cdot A^a )^2}{
\lambda}-\frac{(\partial\cdot A^a )(\partial_j A^{ja})}{\lambda}- c^a M^{ab}\bar c^b + c^a(\tilde M^\prime \bar c)^a
\right.\nonumber\\
& +&\left. gf^{a bc}\bar c^b c^c
\partial\cdot A^a
 -  gf^{abc}\bar c^b c^c \partial_j A^{ja} \;\right],
\end{eqnarray}
where $\tilde M^\prime =\partial_jD^j$.

We make an ansatz for $S_1$ looking at the different terms in the effective action in Lorentz gauge and Coulomb 
gauge as
\begin{eqnarray}
S_1&=&\int d^4x\left[\xi_1 (\kappa)(\partial\cdot A^a )^2+\xi_2 (\kappa)(\partial_j A^{ja})^2+\xi_3 (
\kappa)(\partial\cdot A^
a )(\partial_j A^{ja})\right. \nonumber\\
&+&\left.\xi_4 (\kappa )\left (c^a M^{ab}\bar c^b
- gf^{abc}\bar c^b c^c \partial\cdot A^a\right )
+\xi_5 (\kappa)\left ( c^a (\tilde M^\prime\bar c)^a \right.\right.\nonumber\\  
&-&\left.\left. gf^{abc}\bar c^b c^c
\partial_j A^{ja}\right )\right].
\end{eqnarray}
$S_1$ will be the part of the new effective action if and only if the condition in Eq. (\ref{mcond}) is satisfied. 
The condition in this particular case reads as
\begin{eqnarray}
&&\int [{\cal{D}}\phi]  \; e^{i \left (S^L_{eff}+S_1 \right )} \left 
[  \left( M^{ab}\bar{c}^b - gf^{abc}\bar{c}^b \partial\cdot A^
c\right )\Theta_{ab}^\prime
 \left \{ (\partial\cdot A^a)\left( 2\xi_1 +\frac{\xi_4}{
\lambda}\right )  \right.\right. \nonumber \\ 
&+&\left.\left.(\partial_j A^{ja})\xi_3\right \}
+ \left ((\tilde{M}^\prime\bar{c})^a -gf^{abc}\bar{c}^b\partial_j {A^j}^
c\right ) \Theta_{ab}^\prime   \left\{(\partial\cdot A^a)\left( \xi_3 +\frac{\xi_5}{
\lambda}\right) \right.\right. \nonumber\\
&+&\left.\left. 2\xi_2(\partial_j A^{ja})\right \} + \left( \frac{d\xi_1}{d\kappa}-\frac{\gamma}{\lambda}\right 
){(\partial\cdot A^a)}^2 +
\frac{d\xi_2}{d\kappa}{(\partial_j A^{ja})}^2\right. \nonumber\\
&+&\left. (\partial\cdot A^a)(\partial_j A^{ja}) 
\left(\frac{d\xi_3}{d\kappa}+\frac{\gamma}{\lambda}\right) +\left(\frac{d\xi_4}{d\kappa}+\gamma\right) c^a M^{ab} \bar
{c}^b +\left( 
\frac{d\xi_5}{d\kappa}-\gamma \right)c^a (\tilde{M}^\prime\bar{c})^a  \right. \nonumber\\
&-&\left. gf^{abc}\bar{c}^
bc^c (\partial\cdot A^a )\left (\frac{d\xi_4}{d\kappa}+\gamma\right )-g f^{abc}\bar{c}^b c^c
(\partial_j A^{ja}) 
\left(\frac{d\xi_5}{d\kappa}-\gamma\right) \right ]=0. 
\end{eqnarray}
Following the similar procedure as in the previous subsection, we obtain $S_1$ at $\kappa =1$ as
\begin{eqnarray}
S_1&=&\int d^4x\left[\frac{(\partial\cdot A^a )^2}{2\lambda}-\frac{(\partial_j A^{ja})^2}{2\lambda
}-c^a M^{ab}\bar c^b +c^a(\tilde M'\bar c)^a +gf^{abc}\bar c^b
 c^c \partial\cdot A^a \right.\nonumber\\
&-&\left. gf^{abc}\bar c^b c^c \partial_j 
A^{ja}\right].
\end{eqnarray}
Adding this part to $S_{eff}^L$ we obtain 
\begin{eqnarray}
S^L_{eff}+S_1(\kappa=1) &=&\int d^4x \left[-\frac{1}{4}F^{a \mu \nu }F^{a}_{\mu 
\nu}-\frac{(\partial_j A^{ja})^2}{2\lambda}+c^a(\tilde M^\prime\bar 
c)^a\right.\nonumber\\
&-&\left. gf^{abc}\bar c^b c^c \partial_j A^{ja}\right]\nonumber\\
&=&\int d^4x \left[-\frac{1}{4}F^{a \mu \nu }F^{a}_{\mu 
\nu}-\frac{(\partial_j A^{ja})^2}{2\lambda}-\bar c^a(\tilde M^\prime 
c)^a\right]\nonumber\\
&=& S^C_{eff},
\end{eqnarray}
which is effective action in Coulomb gauge.
Thus, the generating functionals corresponding to Lorentz gauge and Coulomb gauge can also be related by 
FF-anti-BRST transformation.

\subsection{ FP theory to most general BRST/anti-BRST invariant theory }
In all previous examples we consider the effective theory in different gauges. However, similar to 
FFBRST transformation, FF-anti-BRST transformation can also relate the different effective theories. 
In order to connect   two different theories viz. YM effective action in Lorentz gauge and the most 
general BRST/anti-BRST invariant action in Lorentz gauge, we consider infinitesimal field dependent parameter as
\begin{equation}
\Theta_{ab}^\prime =-i\gamma\int d^4x\ c^a f^{abc}\bar c^b c^c.
\end{equation}
Then, corresponding to the above $\Theta_{ab}^\prime$ the change in Jacobian, using the Eq.(\ref{jaceva1}), is 
calculated as
\begin{equation}
\frac{1}{J}\frac{dJ}{d\kappa}=i\gamma\int d^4x\left[2\frac{\partial\cdot A^a}{\lambda}f^{
abc}\bar c^b c^c +gf^{abc}\bar c^b c^c f^{
alm}\bar c^l c^m \right].
\end{equation}
Looking at the kind of terms present in the FP effective action in Lorentz gauge 
and in the most general BRST/anti-BRST invariant effective action in Lorentz gauge, we try an ansatz for $S_1$ as
\begin{equation}
S_1=\int d^4x\left[\xi_1 (\kappa)f^{abc}\partial\cdot A^a \bar c^b c^c +\xi_2 (\kappa) 
f^{abc}\bar c^b \bar c^c f^{alm}c^l c^m\right].
\end{equation}
$S_1$ can be expressed as $e^{iS_1}$ iff it satisfies the condition mentioned in Eq. (\ref
{mcond}). The condition for this case is calculated as 
\begin{eqnarray}
&&\int  [{\cal{D}}\phi]  \; e^{ i \left (S^L_{eff}+S_1 \right )} \left
[f^{abc }\partial\cdot A^a\bar c^b c^c \left (\frac{d\xi_1}{d\kappa} -
\frac{2\gamma}{\lambda}\right )\right.\nonumber\\
&+& f^{abc}\bar c^b \bar c^c f^{alm}c^l c^m \left (
\frac{d\xi_2}{d\kappa} +\frac{\gamma g}{2} -\gamma\xi_1 \right)\nonumber\\
&+& \left. f^{abc}\bar c^b \bar c^c f^{alm} c^
l\partial\cdot A^m \Theta_{ab}^\prime\left(\frac{\xi_1^2}{2} -\frac{g \xi_1}{2}-\frac{2
\xi_2}{\lambda}\right )\right ]=0.
\end{eqnarray}
We look for a special solution  corresponding to the condition 
\begin{equation}
\frac{\xi_1^2}{2} -\frac{g \xi_1}{2}-\frac{2\xi_2}{\lambda}=0.
\end{equation}
The coefficient of $ f^{abc}\partial\cdot A^a \bar c^b c^c $ and
$ f^{abc}\bar c^b \bar c^c f^{alm}c^l c^m $ gives 
respectively
\begin{equation}
\frac{d\xi_1}{d\kappa} -\frac{2\gamma}{\lambda}=0,
\end{equation}
\begin{equation}
\frac{d\xi_2}{d\kappa} +\frac{g\gamma}{2}-\gamma\xi_1 =0.
\end{equation}
For a particular $\gamma =\frac{\alpha\lambda g}{4}$,
the solutions of above two equations are 
\begin{equation}
\xi_1 =\frac{\alpha}{2}g\kappa, 
\end{equation}
\begin{equation}
\xi_2 =-\frac{\alpha}{8}\lambda g^2\kappa +\frac{\alpha^2}{16}\lambda g^2\kappa^2.
\end{equation}
At $\kappa=1$
\begin{equation}
S_1=\int d^4x\left[\frac{\alpha}{2}gf^{abc}\partial\cdot A^a \bar c^b c^c -\frac
{\alpha}{8}\left(1-\frac{\alpha}{2}\right)\lambda  g^2f^{abc}\bar c^b \bar c^c
f^{alm}c^l c^m\right].
\end{equation}
Hence,
\begin{eqnarray}
S^L_{eff}+S_1&=& \int d^4x \left[-\frac{1}{4}F^{a \mu \nu }F^{a}_{\mu 
\nu} - \frac{(\partial\cdot A^a)^2}{2\lambda }  + \partial ^\mu\bar{c}^aD_\mu^{ab} c^b\right. \nonumber 
\\
&+&\left.\frac{\alpha}{2} gf^{abc }\partial\cdot A^a\bar{c}^b c^
c -\frac{1}{8}\alpha (1-\frac{1}{2}\alpha )\lambda g^2 f^{abc}\bar{c}^
b\bar{c}^c f^{alm}c^l c^m \right ]\nonumber\\
&=& S_{eff}^{AB}[A,c,\bar{c}].
\end{eqnarray}
which is most general BRST/anti-BRST invariant effective action.
 
Thus, the generating functional corresponding to most general effective actions in Lorentz gauge can also be related 
through FF-anti-BRST transformation. We observe that FF-anti-BRST transformation plays the same role
as of the FFBRST transformation but with different parameters. 
\section{ Off-shell nilpotent FF-anti-BRST transformation }
The FF-anti-BRST transformation, we have constructed in previous section, is on-shell nilpotent. In this section, we 
construct FF-anti-BRST transformation which is off-shell nilpotent. For this purpose, we consider the following 
effective action for YM theories in auxiliary field formulation in Lorentz gauge 
\begin{eqnarray}
S^L_{eff}&=&\int d^4x \left[-\frac{1}{4}F^{a \mu \nu }F^{a}_{\mu 
\nu}+\frac{\lambda}{2}{(B^a)}^2-B^a\partial\cdot A^a +c^a M^{ab}\bar 
c^b \right.\nonumber\\
&-&\left. gf^{abc}\bar c^b c^c {(\partial\cdot A)}^a \right]. \label{act}
\end{eqnarray}
This effective action is invariant under anti-BRST transformation mentioned in Eq. (\ref{antb}).
Following the procedure outlined in the section 3.2, we obtain the FF-anti-BRST transformation in auxiliary field 
formulation as,
\begin{eqnarray}
\delta_b A^a_\mu &=&{(D_\mu \bar c)}^a \ \Theta_{ab}  (A,c,\bar c,B),\ \ 
\delta_b\bar c^a  = -\frac{1}{2}gf^{abc }\bar c^b \bar c^c \ \Theta_{ab}
 (A,c,\bar c,B), \nonumber\\
\delta_b c^a &=&{(-B^a -gf^{abc}\bar c^b c^c )}\ \Theta_{ab}  (A,c,
\bar c,B), \nonumber\\
\delta_b B^a &=&-gf^{abc}B^b \bar c^c\ \Theta_{ab}  (A,c,\bar c,B),
\end{eqnarray}
which also
leaves the effective action in Eq. (\ref{act}) invariant.
Now, we consider the different choices of the parameter $\Theta_{ab}^\prime (A,c,\bar c,B)$ in auxiliary 
field formulation to connect different theories. We redo the same examples using off-shell 
FF-anti-BRST transformation. 

\subsection{ YM theory in Lorentz gauge to Coulomb gauge}
The effective action for YM theory in Coulomb gauge can be written after rearranging the ghost term as
\begin{eqnarray}
S^C_{eff}&=&\int d^4x \left[-\frac{1}{4}F^{a \mu \nu }F^{a}_{\mu 
\nu}+\frac{\lambda}{2}{(B^a)}^2-B^a\partial ^j A^a _j +c^a(\tilde 
M^\prime\bar c)^a \right.\nonumber\\
&-&\left. gf^{abc}\bar c^b c^c \partial _j A^{ja }\right].
\end{eqnarray}
To show the connection of this theory with the theory in Lorentz gauge, we choose the following infinitesimal field 
dependent 
parameter:
\begin{equation}
\Theta_{ab}^\prime =-i\int d^4x\ c^a[\gamma _1\lambda B^a +\gamma _2(\partial\cdot A^
a -\partial ^jA_j^a )].
\end{equation}
Using the above $\Theta_{ab}^\prime$, we find the change in Jacobian as
\begin{eqnarray}
\frac {1}{J}\frac{dJ}{d\kappa}&=&i\int d^4x\left[\lambda\gamma_1 (B^a )^2+\gamma_2 B^
a\partial\cdot A^a -\gamma_2 B^a \partial^j A_j^a +\gamma_2 gf^{abc}\bar c^b c^c\partial\cdot A^a \right.\nonumber\\
&-&\left .\gamma_2 gf^{abc}\bar c^b c^c\partial^j A^a _j -
\gamma_2 c^a M^{ab}\bar c^b +\gamma_2 c^a(\tilde M^\prime\bar c)^a\right ].
\end{eqnarray}
We make an ansatz for $S_1$ as
\begin{eqnarray}
S_1&=&\int d^4x\left[\xi _1 (\kappa)(B^a )^2+\xi _2 (\kappa)B^a\partial\cdot A^a+\xi_3 (\kappa) B^
a\partial ^jA_j^
a \right. \nonumber\\
&+&\left. \xi _4 (\kappa )\left (c^a M^{ab}\bar c^b
- gf^{abc}\bar c^b c^c
\partial\cdot A^a \right )
+\xi_5 (\kappa)\left (c^a(\tilde M^\prime\bar c)^a \right.\right.\nonumber\\
&-&\left.\left. gf^{abc}\bar c^b c^c \partial _j 
A^{ja}\right )\right].
\end{eqnarray}
The essential requirement for replacing the Jacobian as $e^{iS_1}$ mentioned in Eq. (\ref{mcond}) is satisfied iff
\begin{eqnarray}
&&\int [{\cal{D}}\phi]  \; e^{ i \left (S^L_{eff}+S_1 \right )}\left
[ (B^a )^2\left(\frac{d\xi_1}{d\kappa}-\lambda\gamma_1 \right) +B^a\partial\cdot A^a 
\left(\frac{d\xi_2}{d\kappa}-\gamma_2 \right)\right.\nonumber\\
&+&\left. B^a\partial^j A_j^a \left(\frac{d\xi_3}{d\kappa}+\gamma_2 \right)
-\bar c^a M^{ab}c^a \left(\frac{d\xi_4}{d\kappa}+\gamma_2 \right)-\bar c^a (\tilde M'c)^a \left(
\frac{d\xi_5}{d\kappa}-\gamma_2 \right) \right.\nonumber\\
&+&\left.  \{M^{ab}\bar c^b -gf^{abc}\bar c^b\partial\cdot 
A^c \}\Theta_{ab}^\prime\{B^a (\xi_2 +\xi_4 )\}
+\{(\tilde M^\prime\bar c)^a -gf^{abc}\bar c^b\partial^j A_j^c \}
\Theta_{ab}^\prime\right.\nonumber\\
&&\left.\{B^a (\xi_3 +\xi_5 )\}\right]=0. \label{ac}
\end{eqnarray}
The last two terms in the integrand of Eq. (\ref{ac}) are dependent on $ c$ in a local 
fashion. The contribution of these terms can possibly vanish by ghost equation of motion
given in Eq. (\ref{gh}),
This can only happen if the ratio of coefficients of the two terms is identical to the ratio 
of coefficients of $\bar c^{a}M^{ab}c^{b}$ and $\bar c^{a} (\tilde M'c)^a$ in $S^L_{eff}+S_1$.
The nonlocal terms become local, only if, it satisfy the following conditions:
\begin{eqnarray}
 \frac{{d\xi_4}/{d\kappa}+\gamma_2}{\xi_4 +1}=\frac{{d\xi_5}/{d\kappa}-\gamma _2}{\xi_5},
\end{eqnarray}
and
\begin{eqnarray} 
 \frac{\xi_2 +\xi_4}{\xi_4 +1}=\frac{\xi_3 +\xi_5}{\xi_5}. 
\end{eqnarray}
We further obtain equations for the parameter $\xi_i $ by vanishing the coefficient of different independent terms 
in the L.H.S. of the Eq. (\ref{ac}) as 
\begin{eqnarray}
&&\frac{d\xi_1}{d\kappa} -\gamma_1\lambda +\gamma_1\lambda(\xi_2 +\xi_4 ) + \gamma_1 \lambda(\xi_3 +\xi_5) =0, 
\nonumber\\
&&\frac{d\xi_2}{d\kappa} -\gamma_2 +\gamma_2 (\xi_2 +\xi_4 ) + \gamma_2 (\xi_3 +\xi_5) =0, \nonumber\\
&&\frac{d\xi_3}{d\kappa} +\gamma_2 -\gamma_2 (\xi_2 +\xi_4 ) - \gamma_2 (\xi_3 +\xi_5) =0, \nonumber\\
&&\frac{d\xi_4}{d\kappa}+\gamma_2 =0,\nonumber\\
&&\frac{d\xi_5}{d\kappa} -\gamma_2 =0. 
\end{eqnarray}
We determine the parameter $\xi_i$ subjected to the initial condition in Eq. (\ref{xcond}) as 
\begin{eqnarray}
\xi_1 &=& \gamma_1\lambda\kappa, \ \
\xi_2 = \kappa, \ \
\xi_3 = -\kappa, \nonumber\\
\xi_4 &=& -\kappa, \ \
\xi_5 = \kappa.
\end{eqnarray}
Using the above solutions for $\xi_i$, we write  $S_1$ at $\kappa =1$ as
\begin{eqnarray}
S_1&=&\int d^4x\left[\lambda\gamma_1 (B^a )^2 +B^a\partial\cdot A^a -B^
a\partial_j A^{j
a} -c^a M^{ab}\bar c^b +gf^{abc}\bar c^a c^b\partial\cdot A^
c\right.\nonumber\\
&+&\left. c^a(\tilde M^\prime\bar c)^a
- gf^{abc}\bar c^b c^c\partial_j A^{ja}\right].
\end{eqnarray}
Now, when this $S_1$ is added to the effective action $S^L_{eff}$, it provides effective action in Coulomb gauge as
\begin{eqnarray} 
S^L_{eff} +S_1&=&\int d^4x \left[-\frac{1}{4}F^{a\mu \nu }F^{a}_{\mu 
\nu}+\frac{\zeta}{2}{(B^a)}^2-B^a\partial ^j A^a _j 
+c^a(\tilde M^\prime\bar c)^a \right.\nonumber\\
&- &\left. gf^{abc}\bar c^b c^c
\partial _j A^{ja}\right], \nonumber\\
&=&\int d^4x \left[-\frac{1}{4}F^{a \mu \nu }F^{a}_{\mu 
\nu}+\frac{\zeta}{2}{(B^a)}^2-B^a\partial ^j A^a _j 
-\bar c^a(\tilde M^\prime c)^a \right],\nonumber\\
&=& S^{\prime C}_{eff},
\end{eqnarray}
which is the effective action in Coulomb gauge with gauge parameter $\zeta$. 
 
 Thus, FF-anti-BRST in auxiliary field formulation produces the same result as expected, even though the finite 
finite parameter is different. 
\subsection { Lorentz gauge to axial gauge theory}
We repeat the same steps as in the previous subsection again for this case, with a different finite field dependent 
parameter  obtainable from
\begin {equation}
\Theta_{ab}^\prime =-i\int d^4x c^a\left[\gamma_1\lambda B^a +\gamma_2 (\partial\cdot A^
a -\eta\cdot A^a )\right ],\label{the}
\end{equation}
using Eq. (\ref{81}) and consider the ansatz for $S_1$ as
\begin{eqnarray}
S_1&=&\int d^4x\left[\xi_1 (\kappa)(B^a )^2 +\xi_2 (\kappa)B^a\partial\cdot A^a +\xi_3 (
\kappa)B^a\eta\cdot A^
a \right.\nonumber\\ 
&+&\xi_4 (\kappa) (c^a M^{ab}\bar c^b - gf^{a
bc}\bar c^b c^c \partial\cdot A^a )
+\left.\xi_5 (\kappa )( c^a\tilde M^{ab}\bar c^b\right.\nonumber\\
&-&\left. gf^{abc}\bar c^b c^c \eta\cdot A^a )\right].
\end{eqnarray}
The condition for which the Jacobian of FF-anti-BRST transformation in auxiliary field formulation corresponding to 
the parameter given in Eq. (\ref{the}) can be replaced as $e^{iS_1}$ is,
\begin{eqnarray}
&&\int [{\cal{D}}\phi]  \; e^{i \left (S^L_{eff}+S_1 \right )} \left
[ \left \{ M^{ab}\bar{c}^b - gf^{abc}\bar{c}^b (\partial\cdot A^
c)\right \}\Theta_{ab}^\prime   \left\{ B^a(\xi_2 +\xi_4 )\right \}  \right. \nonumber\\
&+&\left.\left \{ \tilde M^{ab}\bar{c}^b - gf^{abc}\bar{c}^
b (\eta\cdot A^
c)\right \}\Theta_{ab}^\prime \left \{ B^a (\xi_3+\xi_5 )\right \} +(B^a )^2 \left(
\frac{d\xi_1}{d\kappa} -\lambda\gamma_1 \right)\right.\nonumber\\
&+&\left. B^a\partial\cdot A^a \left(\frac{d\xi_2}{d\kappa} -\gamma_2 \right) + B^a\eta\cdot A^
a\left(\frac{d\xi_3}{d\kappa} +\gamma_2 \right) +c^a M^{ab}\bar c^b \left(\frac{d\xi_4}{d\kappa} +\gamma_2 \right)
\right.\nonumber\\
&+&\left. c^a \tilde M^{ab}\bar c^b \left(\frac{d\xi_5}{d\kappa} 
-\gamma_2 \right)-gf^{abc}
\bar c^b c^c \partial\cdot A^a \left(\frac{d\xi_4}{d\kappa} +\gamma_2 \right)\right.\nonumber\\
&-&\left. gf^{abc}\bar c^b c^c \eta\cdot A^a \left(\frac{d\xi_5}{d\kappa} -
\gamma_2 \right)\right] =0.
\end{eqnarray}
Following the previous subsection, we solve the differential equations, which are obtained from the above condition,
to calculate the values of parameters $\xi_i$.

After calculating the exact values of $\xi_i$, the extra piece of the action $S_1$ becomes as 
\begin{eqnarray}
S_1&=&\int d^4x\left[\lambda\gamma_1 ( B^a )^2 + B^a\partial\cdot A^a - B^a\eta\cdot A^
a +\bar c^a M^{ab}\bar c^b- \bar c^a\tilde M^{ab}\bar c^b\right].
\end{eqnarray}
Now,
\begin{eqnarray}
 S^L_{eff}+S_1= \int d^4x \left[-\frac{1}{4}F^{a \mu \nu }F^{a}_{\mu 
\nu} +\frac{\zeta}{2} (B^a)^2 - B^a\eta\cdot A^a - 
\bar c^a\tilde M^{ab} c^b\right]
 = S^{\prime A}_{eff},
\end{eqnarray}
 where $S^{\prime A}_{eff}$ is the effective action in axial gauge
with gauge parameter $\zeta$.
 
 This implies off-shell nilpotent FF-anti-BRST also produces the same result even though calculations are different.
\subsection{  Lorentz gauge theory to  most general BRST/anti-BRST invariant theory}
We consider one more example in FF-anti-BRST formulation using auxiliary field. We show that the most general 
BRST/anti-BRST invariant theory can be obtained from FP theory in Lorentz gauge. 
The most general effective action which is invariant under BRST/anti-BRST transformation in auxiliary field is given 
in Eq. (\ref{seffab}). 
We choose
\begin{equation}
\Theta_{ab}^\prime =-i\int d^4x\ {c}^a\left [ \gamma_1 \lambda  B^a +\gamma_2 f^{abc}\bar{c}^b
 c^c \right ].
\end{equation} 
and make an ansatz for $S_1$ as
\begin{equation}
S_1=\int d^4x\left[\xi_1 (\kappa )(B^a)^2 +\xi_2 (\kappa)B^a f^{abc}\bar c^b c^c +
\xi_3 (\kappa)f^{abc}\bar
 c^b \bar c^c f^{alm}c^l c^m\right].
\end{equation}
The condition in Eq. (\ref{mcond}) leads to 
\begin{eqnarray}
&&\int [{\cal{D}}\phi ]\; e^{i \left (S^L_{eff}+S_1 \right )}  \left 
[ \left (\frac{d\xi_1}{d\kappa} -\gamma_1\lambda \right ){B^a}^2 
+ \left(\frac{d\xi_2}{d\kappa}-2\gamma_2\right) 
B^a f^{abc}\bar{c}^b c^
c  \right.  \nonumber \\
&+&\left.\left (\frac{d\xi_3}{d\kappa} +\frac{g}{2}\gamma_2 \right ) f^{abc}
\bar{c}^b \bar{c}^c f^{alm}c^l c^m - \left ( \frac{g}{2} \xi_2 
+2 \xi_3 \right )
  f^{abc} B^b{c}^c f^{alm}\bar c^l \bar c^m\Theta_{ab}^\prime \right ]\nonumber\\
&=&0.
\end{eqnarray}
Using the same procedure we obtain the solutions for the parameter $\xi_i$, exactly same as in Eq. (\ref{soln}).
Even if the finite parameter is different in FFBRST and FF-anti-BRST, we obtain the same contribution from Jacobian 
in this case as
\begin{equation}
S_1=\int d^4x\left[\gamma_1\lambda (B^a )^2+\frac{\alpha g\zeta}{2}B^a f^{abc}\bar{c}^b
c^c -\frac{1}{8}
\alpha g^2\zeta f^{abc}\bar{c}^b\bar{c}^c f^{alm}c^l c^
m\right].
\end{equation}
Now, $
S^L_{eff}+S_1 =S_{eff}^{\prime AB},$
which is nothing but most general BRST/anti-BRST invariant effective action with gauge 
parameter $\zeta$ mentioned in Eq. (\ref{mga}).
In all three cases we show that off-shell nilpotent FF-anti-BRST transformation plays exactly same role as 
on-shell nilpotent FF-anti-BRST transformation but with different finite parameters.
\section{ Conclusions}
In this chapter we have  reformulated   the FFBRST transformation in an 
auxiliary field formulation where the BRST transformation is off-shell nilpotent. We have 
considered several examples with different choices of finite parameter to connect the 
different effective theories.  Most of the 
results of the FFBRST transformation are also obtained in auxiliary field formulation.  
 We have further introduced and developed for first time the concept of the FF 
anti-BRST transformation analogous to the FFBRST transformation. FF-anti-BRST transformation 
is also used to connect the different generating functionals corresponding to different 
effective theories. 
Several examples have been worked out explicitly to show that the FF-anti-BRST transformation 
also plays the same role as of FFBRST transformation but with different finite parameters. Lastly, we consider the 
FF-anti-BRST transformation  in auxiliary field formulation to make it off-shell 
nilpotent. The overall multiplicative antighost fields 
in the finite parameters of the FFBRST transformation are replaced by ghost fields in case of the FF-anti-BRST 
transformation. Even though the finite parameters and hence the calculations are different, the same results are 
also produced in 
an auxiliary field formulation of the FF-anti-BRST transformation. The BRST and the anti-BRST 
transformations are not independent transformations in the YM theories. We observe that the FF-anti-BRST 
transformation plays exactly the same role in connecting theories in 1-form gauge 
theory as expected. In 2-form gauge theories the BRST and anti-BRST transformations play some sort of 
independent roles. Therefore, it will be interesting to study the finite field dependent BRST 
and anti-BRST transformations in 2-form gauge theories \cite{sm1}, which we will discuss in the next chapter.

\label{Chap:chapter3}

\chapter{FFBRST formulation in 2-form gauge theory}
In this chapter we extend FFBRST formulation to Abelian 2-form gauge theories and show that it relates  generating  
functional corresponding to different effective theories \cite{sm1}. We further construct FF-anti-BRST transformation for such 
theories to show that it plays same role as in the case of 1-form gauge theories. 
Field/antifield formulation of 2-form gauge theories are also studied in the context of FFBRST transformation.  

\section{ Preliminary: gauge theory of Abelian rank-2 antisymmetric tensor field}

We consider the Abelian gauge theory for rank-2 antisymmetric tensor field $B_{\mu\nu}$ 
defined by the action
\begin{equation}
S_0=\frac{1}{12}\int d^4x F_{\mu \nu \rho}F^{\mu \nu \rho},\label{kin}
\end{equation}
where $F_{\mu \nu \rho}\equiv \partial_\mu B_{\nu\rho}+\partial_\nu B_{\rho\mu}+\partial_\rho 
B_{\mu\nu}.$ This action is invariant under the gauge transformation $\delta 
B_{\mu\nu}=\partial_{\mu}\zeta_{\nu} -\partial_{\nu}\zeta_{\mu}$ with a vector gauge 
parameter $\zeta_{\mu}(x)$.

To quantize this theory using BRST transformation, it is necessary to introduce the following 
ghost and auxiliary fields: anticommuting vector fields $\rho_{\mu}$ and $\tilde\rho_{\mu}$, 
a commuting vector field $\beta_{\mu}$, anticommuting scalar fields $\chi$ and $\tilde\chi$, 
and commuting scalar fields $\sigma, \varphi,$ and $ \tilde\sigma $. The BRST transformation 
is then defined for $B_{\mu\nu}$ by replacing $\zeta_{\mu}$ in the gauge transformation by 
the ghost field $\rho_{\mu}$.

The complete effective action for this theory in covariant gauge, using  the BRST 
formulation, is given by
\begin{equation}
S^L_{eff}=S_0+S_{gf}+S_{gh}, \label{2act}
\end{equation}
with the gauge-fixing and ghost term
\begin{eqnarray}
S_{gf}+S_{gh}&=&\int d^4x\left[-i\partial_\mu\tilde\rho_\nu (\partial^\mu\rho^\nu -
\partial^\nu\rho^\mu )+\partial_\mu\tilde\sigma\partial^\mu\sigma +\beta_\nu(\partial_\mu B^{
\mu\nu} +\lambda_1\beta^\nu\right.\nonumber\\ 
&-&\left. \partial^\nu\varphi) -i\tilde\chi\partial_\mu\rho^\mu -i\chi (\partial_\mu\tilde\rho^\mu -
\lambda_2\tilde\chi)\right], \label{gfix}
\end{eqnarray}
where $\lambda_1$ and $\lambda_2$ are gauge parameters.
This effective action is invariant under following BRST and anti-BRST symmetries.

BRST:
\begin{eqnarray}
\delta_b B_{\mu\nu} &=& -(\partial_\mu\rho_\nu -\partial_\nu\rho_\mu)\Lambda, \ \
\delta_b\rho_\mu = -i\partial_\mu\sigma \Lambda,\nonumber\\
 \delta_b\sigma &=& 0, \ \
\delta_b\tilde\rho_\mu =i\beta_\mu \Lambda,   \ \ \ \delta_b\tilde\sigma =-\tilde\chi\Lambda,
\nonumber\\
 \delta_b\beta_\mu &=& 0, \ \ \ 
\delta_b\tilde\chi =0,\ \
\delta_b\varphi = -\chi\Lambda,  \ \ \delta_b\chi =0,\label{2sym}
\end{eqnarray}
anti-BRST:
\begin{eqnarray}
\delta_{ab} B_{\mu\nu}&=&-(\partial_\mu\tilde\rho_\nu -\partial_\nu\tilde\rho_\mu)\Lambda,\ \ \
\delta_{ab}\tilde\rho_\mu = -i\partial_\mu\tilde\sigma \Lambda,\nonumber\\ 
\delta_{ab}\tilde\sigma &=& 0, \ \ \ 
\delta_{ab}\rho_\mu =-i\beta_\mu \Lambda,\ \ \ \delta_{ab}\sigma =-\chi\Lambda,\nonumber\\
\delta_{ab}\beta_\mu &=& 0,\ \ \
 \delta_{ab}\chi =0, \ \
\delta_{ab}\varphi = \tilde\chi\Lambda,\ \
\delta_{ab}\tilde\chi =0,\label{2asym}
\end{eqnarray}
where the BRST parameter $\Lambda$ is global, infinitesimal and anticommuting in nature.
The anti-BRST transformation is similar to the BRST transformation, where the role of 
ghost and antighost field is interchanged with some modification in coefficients.
The generating functional for this Abelian rank 2 antisymmetric tensor field theory in a
covariant gauge is defined as,
\begin{equation}
Z^L=\int {\cal D}\phi\exp [iS_{eff}^L[\phi ]],\label{2zfun}
\end{equation}
where $\phi$ is the generic notation for all the fields ($B_{\mu\nu}, \rho_\mu, 
\tilde\rho_\mu, \beta_\mu, \varphi, \sigma, \tilde\sigma, \chi, \tilde\chi$).

The BRST and anti-BRST transformations in Eqs. (\ref{2sym}) and (\ref{2asym}) respectively 
leave the
above generating functional invariant as, the path integral measure ${\cal D}\phi \equiv {\cal D}B {\cal D}\rho 
{\cal D}\tilde\rho {\cal D}\beta {\cal D}\sigma {\cal D}\varphi {\cal D}\tilde\sigma {\cal D}\chi {\cal D}\tilde\chi$ 
is invariant under such 
transformations.
\section{FFBRST formulation of Abelian rank 2 anti-symmetric tensor field}
 To generalized the BRST transformation for
this theory we follow the method outlined in chapter 2. We start 
by making the  infinitesimal BRST parameter field dependent by introducing a parameter $\kappa\ 
(0\leq \kappa\leq 1)$. All 
 the fields  $\phi(x,\kappa)$ dependent on $\kappa$  in such a way that 
$\phi(x,\kappa =0)=\phi(x)$ and $\phi(x,\kappa 
=1)=\phi^\prime(x)$, the transformed field.
It can easily be shown that such off-shell nilpotent BRST transformation with finite field 
dependent parameter is symmetry  of the effective action in Eq. (\ref{2act}). However, the 
path integral measure in Eq. (\ref{2zfun}) is not invariant under such transformation as the 
BRST parameter is finite.

The Jacobian of the path integral measure for such transformations can be evaluated for some 
particular choices of the finite field dependent parameter, $\Theta_b [\phi(x)]$, as
\begin{eqnarray}
{\cal D}B^\prime {\cal D}\rho^\prime {\cal D}\tilde\rho^\prime {\cal D}\beta^\prime {\cal D}\sigma^\prime {\cal D}
\tilde\sigma^\prime {\cal D}
\chi^\prime {\cal D}\tilde\chi^\prime &=&J(
\kappa) {\cal D}B(\kappa) {\cal D}\rho(\kappa) {\cal D}\tilde\rho(\kappa) {\cal D}\beta(\kappa) {\cal D}
\sigma(\kappa)\nonumber\\
&& {\cal D}\tilde\sigma(\kappa ){\cal D}\chi(\kappa) {\cal D}\tilde\chi(\kappa).
\end{eqnarray}
The Jacobian, $J(\kappa )$ can be replaced (within the functional integral) as
\begin{equation}
J(\kappa )\rightarrow \exp[iS_1[\phi(x,\kappa) ]],
\end{equation}
 iff the  condition given in Eq. (\ref{mcond}) is satisfied. However, the
 infinitesimal change in the $J(\kappa)$ can be calculated using Eq. (\ref{jaceva}). 

By choosing appropriate $\Theta_b$, we can change $S_{eff}$ either to another effective action 
for same theory or to an effective action for another theory. The resulting effective action also 
be invariant under same BRST transformation.
\section{ FFBRST transformation in 2-form gauge theory: examples}
In this section, we would like to show explicitly that the FFBRST transformation 
interrelates the different effective 2-form gauge theories. In particular, we are interested to obtain the 
effective theories for Abelian rank-2 tensor field in noncovariant gauges by applying 
FFBRST transformation to the effective theories in covariant gauge.
\subsection{Effective theory in axial gauge}
We start with the generating functional corresponding to the effective theory in Lorentz gauge
given in Eq. (\ref{2zfun}), where the Lorentz gauge effective action
$S^L_{eff}$ is invariant under following FFBRST:
\begin{eqnarray} 
\delta_b B_{\mu\nu} &=& -(\partial_\mu\rho_\nu -\partial_\nu\rho_\mu)\Theta_b [\phi ] \nonumber\\
\delta_b\rho_\mu &=& -i\partial_\mu\sigma \Theta_b [\phi ],  \ \ \ \ \ \ \ \ \ \ \ \ \ \ \ 
\delta_b\sigma = 0 \nonumber\\
\delta_b\tilde\rho_\mu &=&i\beta_\mu \Theta_b [\phi ],   \ \ \ \ \ \ \ \ \ \ \ \ \ \ \ \ \ \ 
\delta_b\beta_\mu = 0\nonumber\\
\delta_b\tilde\sigma &=&-\tilde\chi\Theta_b [\phi ],  \ \ \ \ \ \ \ \ \ \ \ \ \ \ \ \ \ \ \ \ 
\delta_b\tilde\chi =0\nonumber\\
\delta_b\varphi &=& -\chi\Theta_b [\phi ], \ \ \ \ \ \ \ \ \ \ \ \ \ \ \ \ \ \ \ \ \delta_b\chi 
=0,\label{2fsym}
\end{eqnarray}
where $\Theta_b$ is a finite, anticommuting BRST parameter depends on the fields in global manner.

To obtain the effective theories in axial gauge we choose the finite parameter obtainable from  
\begin{eqnarray}
\Theta^\prime_b &=&\int d^4x\left[\gamma_1\tilde\rho_\nu (\partial_\mu B^{\mu\nu}-\eta_\mu B^{
\mu\nu}-\partial^\nu\varphi -\eta^\nu\varphi )+\gamma_2 \lambda_1\tilde\rho_\nu\beta^\nu 
\right. \nonumber\\
&+&\left.\gamma_1\tilde\sigma (\partial_\mu\rho^\mu -\eta_\mu\rho^\mu )+\gamma_2 
\lambda_2\tilde\sigma\chi\right],\label{2afin}
\end{eqnarray}
where $\gamma_1$ and $\gamma_2$ are arbitrary  parameters (depend on $\kappa$) and 
$\eta_\mu$ is arbitrary
constant four vector.
Now we apply this FFBRST transformation to the generating functional $Z^L$ given in 
Eq. (\ref{2zfun}). The path integral measure is not invariant  and  give rise a nontrivial 
functional $e^{iS_1^A}$ (explicitly shown in Appendix A), 
where 
\begin{eqnarray}
S_1^A&=&\int d^4x [-\beta_\nu\partial_\mu B^{\mu\nu} +\beta_\nu\eta_\mu B^{\mu\nu} +\gamma_2
 \lambda_1\beta_\nu\beta^\nu -i\tilde\rho_\nu\partial_{\mu}(\partial^\mu\rho^\nu -
\partial^\nu\rho^\mu)\nonumber\\
&+&i\tilde\rho_\nu\eta_{\mu}(\partial^\mu\rho^\nu -\partial^\nu\rho^\mu )+i\gamma_2 
\lambda_2\chi\tilde\chi +i\tilde\chi\partial_\mu\rho^\mu -i
\tilde\chi\eta_\mu\rho^\mu\nonumber\\
&+&\tilde\sigma\partial_\mu\partial^\mu\sigma -\tilde\sigma\eta_\mu\partial^\mu\sigma - 
\partial_\mu\beta^\mu\varphi + \eta_\mu\beta^\mu\varphi +i\chi\partial_\mu\tilde\rho^\mu 
-i\chi\eta_\mu\tilde\rho^\mu ].\label{2s1}
\end{eqnarray}
Now, adding this $S_1^A$ to $S^L_{eff}$, we get 
\begin{equation}
S^L_{eff} +S_1^A\equiv S^A_{eff},
\end{equation}
where
\begin{eqnarray}
S^A_{eff}&=&\int d^4x\left[\frac{1}{12}F_{\mu\nu\rho}F^{\mu\nu\rho}+i\tilde\rho_\nu\eta_\mu (
\partial^\mu\rho^\nu -\partial^\nu\rho^\mu )-\tilde\sigma\eta_\mu\partial^\mu\sigma +
\beta_\nu(\eta_\mu B^{\mu\nu}\right.\nonumber\\ 
&+&\left. \lambda_1^\prime\beta^\nu +\eta^\nu\varphi) 
- i\tilde\chi\eta_\mu\rho^\mu -i\chi (\eta_\mu\tilde\rho^\mu -
\lambda_2^\prime\tilde\chi)\right],
\end{eqnarray}
is the effective action in axial gauge with new gauge parameters $\lambda_1^\prime = 
(1+\gamma_2)\lambda_1 $ and $\lambda_2^\prime = (1+\gamma_2)\lambda_2 $.
Thus, the FFBRST transformation with the finite parameter given in Eq. (\ref{2fsym}) 
takes
\begin{equation}
Z^L\left(=\int {\cal D}\phi e^{iS_{eff}^L}\right)\stackrel{FFBRST}{-----\longrightarrow
} Z^A\left(=\int {\cal D}\phi e^{iS_{eff}^A}\right).
\end{equation}
The effective theory of Abelian rank-2 antisymmetric field in axial gauge is convenient in
 many different situations. The generating functional in axial gauge with a suitable 
axis is same as the generating functional obtained by using Zwanziger's formulation for 
electric and magnetic charges \cite{zwan,deg}. Using the FFBRST transformation with the 
parameter 
given in Eq. (\ref{2afin}) we have linked generating functionals corresponding to the effective 
theories in covariant and noncovariant 
gauges.
\subsection{Effective theory in Coulomb gauge}
The generating functional for the effective theories in Coulomb gauge is also obtained by using 
the FFBRST transformation with a different parameter
\begin{eqnarray}
\Theta^\prime_b &=&\int d^4x\left[\gamma_1\tilde\rho_\nu (\partial_\mu B^{\mu\nu}-\partial_i 
B^{i\nu}-\partial^\nu\varphi )+\gamma_1\tilde\rho_i\partial^i\varphi +\gamma_2 
\lambda_1\tilde\rho_\nu\beta^\nu \right. \nonumber\\
&+&\left.\gamma_1\tilde\sigma (\partial_\mu\rho^\mu -\partial_i\rho^i )+\gamma_2 
\lambda_2\tilde\sigma\chi\right].\label{2par}
\end{eqnarray} 
The effective action in Lorentz gauge, $S_{eff}^L$ as given in Eq. (\ref{2act}) is invariant 
under FFBRST transformation in Eq. (\ref{2fsym}) corresponding to above mentioned
finite parameter. Now we consider the effect of this FFBRST transformation on the 
generating functional in Lorentz gauge.

In Appendix A, it has been shown that the Jacobian for the path integral measure 
corresponding to this FFBRST transformation is replaced by $e^{iS_1^C}$, where 
$S_1^C$ is the local functional of fields calculated as
\begin{eqnarray}
S^C_1&=&\int d^4x \left [-\beta_\nu\partial_\mu B^{\mu\nu} +\beta_\nu\partial_i B^{i\nu} +\gamma_2 
\lambda_1\beta_\nu\beta^\nu -i\tilde\rho_\nu\partial_{\mu}(\partial^\mu\rho^\nu -
\partial^\nu\rho^\mu)\right.\nonumber\\
&+&\left. i\tilde\rho_\nu\partial_i(\partial^i\rho^\nu -\partial^\nu\rho^i )+i\gamma_2 
\lambda_2\chi\tilde\chi +i\tilde\chi\partial_\mu\rho^\mu -i
\tilde\chi\partial_i\rho^i\right.\nonumber\\
&+&\left. \tilde\sigma\partial_\mu\partial^\mu\sigma -\tilde\sigma\partial_i\partial^i\sigma - 
\partial_\mu\beta^\mu\varphi + \partial_i\beta^i\varphi +i\chi\partial_\mu\tilde\rho^\mu 
-i\chi\partial_i\tilde\rho^i\right],
\end{eqnarray}
 and this extra piece of the action can be added to the effective action in covariant gauge 
to lead a new effective action
\begin{eqnarray}
S^L_{eff} +S_1^C&=&\int d^4x\left[\frac{1}{12}F_{\mu\nu\rho}F^{\mu\nu\rho}+i
\tilde\rho_\nu\partial_i (\partial^i\rho^\nu -\partial^\nu\rho^i )-
\tilde\sigma\partial_i\partial^i\sigma \right.\nonumber\\ 
&+&\left.\beta_\nu(\partial_i B^{i\nu}+ \lambda_1^\prime\beta^\nu) -\beta_i\partial^i\varphi 
- i\tilde\chi\partial_i\rho^i -i\chi (\partial_i\tilde\rho^i -\lambda_2^\prime\tilde\chi )
\right]\nonumber\\
&\equiv &S_{eff}^C,
\end{eqnarray}
which is an effective action in Coulomb gauge for Abelian rank 2 tensor field. Thus, we  
study the Abelian 2-form gauge theory in Coulomb gauge more rigorously through its connection 
with Lorentz gauge via finite BRST transformation.

\section{FF-anti-BRST transformation in 2-form gauge theories}
In this section, we consider the generalization of anti-BRST transformation following the 
similar method as discussed in section 2 and show it plays exactly similar role in 
connecting the generating functionals in different effective theories of Abelian rank-2 
antisymmetric tensor field.
\subsection{ Lorentz to axial gauge theory using FF-anti-BRST formulation}
For sake of convenience we recast the effective action in covariant gauge given in Eq. (\ref{2act}) as
\begin{eqnarray}
S_{eff}^L&=&\int d^4x\left[\frac{1}{12}F_{\mu\nu\rho}F^{\mu\nu\rho}+i\partial_\mu\rho_\nu (
\partial^\mu\tilde\rho^\nu -\partial^\nu\tilde\rho^\mu )+
\partial_\mu\sigma\partial^\mu\tilde\sigma +\beta_\nu(\partial_\mu B^{\mu\nu} \right.\nonumber\\ 
&+&\lambda_1\beta^\nu -\partial^\nu\varphi) -\left. i\chi\partial_\mu\tilde\rho^\mu -i\tilde\chi (
\partial_\mu\rho^\mu +\lambda_2\chi )
\right],
\end{eqnarray}
which is invariant under following FF-anti-BRST transformation:
\begin{eqnarray}
\delta_{ab} B_{\mu\nu} &=& -(\partial_\mu\tilde\rho_\nu -\partial_\nu\tilde\rho_\mu)\Theta_{ab} [
\phi ] \nonumber\\
\delta_{ab}\tilde\rho_\mu &=& -i\partial_\mu\tilde\sigma \Theta_{ab} [\phi ],\ \ \ \ \ \ \ \ \ \ \ 
\ \ \ \ \delta_{ab}\tilde\sigma = 0 \nonumber\\
\delta_{ab}\rho_\mu &=&-i\beta_\mu \Theta_{ab} [\phi ],\ \ \ \ \ \ \ \ \ \ \ \ \ \ \ 
\delta_{ab}\beta_\mu = 0\nonumber\\
\delta_{ab}\sigma &=&-\chi\Theta_{ab} [\phi ],\ \ \ \ \ \ \ \ \ \ \ \ \ \ \ \ \ \ \ \delta_{ab}\chi 
=0\nonumber\\
\delta_{ab}\varphi &=& \tilde\chi\Theta_{ab} [\phi ],\ \ \ \ \ \ \ \ \ \ \ \ \ \ \ \ \ \ \ \ \ 
\delta_{ab}\tilde\chi =0, \label{2antisym}
\end{eqnarray}
where $\Theta_{ab}$ is finite field dependent anti-BRST parameter.

To obtain the generating functional in axial gauge using FF-anti-BRST
 transformation we choose,
\begin{eqnarray}
\Theta^\prime_{ab} &=&-\int d^4x\left[\gamma_1\rho_\nu (\partial_\mu B^{\mu\nu}-\eta_\mu B^{
\mu\nu}-\partial^\nu\varphi -\eta^\nu\varphi )+\gamma_2 \lambda_1\rho_\nu\beta^\nu \right. 
\nonumber\\
&-&\left.\gamma_1\sigma (\partial_\mu\tilde\rho^\mu -\eta_\mu\tilde\rho^\mu )+\gamma_2 
\lambda_2\sigma\tilde\chi\right]. \label{abfin}
\end{eqnarray}
This parameter is similar to $\Theta^\prime_b$ in Eq. (\ref{2afin}) except the (anti)ghost and 
ghost 
of (anti)ghost fields are replaced by their antighost fields respectively.

Similar to FFBRST case, this FF-anti-BRST transformation 
changes the Jacobian in the path integral measure by a factor $e^{iS_1^A}$, 
where $S_1^A$ is a local 
functional of fields and is given by
\begin{eqnarray}
S_1^A&=&\int d^4x [-\beta_\nu\partial_\mu B^{\mu\nu} +\beta_\nu\eta_\mu B^{\mu\nu} +\gamma_2
\lambda_1\beta_\nu\beta^\nu +i\rho_\nu\partial_{\mu}(\partial^\mu\tilde\rho^\nu -
\partial^\nu\tilde\rho^\mu)\nonumber\\
&-&i\rho_\nu\eta_{\mu}(\partial^\mu\tilde\rho^\nu -\partial^\nu\tilde\rho^\mu )+i\gamma_2 
\lambda_2\chi\tilde\chi +i\chi\partial_\mu\tilde\rho^\mu -i
\chi\eta_\mu\tilde\rho^\mu\nonumber\\
&+&\sigma\partial_\mu\partial^\mu\tilde\sigma -\sigma\eta_\mu\partial^\mu\tilde\sigma - 
\partial_\mu\beta^\mu\varphi + \eta_\mu\beta^\mu\varphi +i\tilde\chi\partial_\mu\rho^\mu 
-i\tilde\chi\eta_\mu\rho^\mu ].
\end{eqnarray}
It is easy to verify that
\begin{eqnarray}
S^L +S_1^A&=&\int d^4x\left[\frac{1}{12}F_{\mu\nu\rho}F^{\mu\nu\rho}-i\rho_\nu\eta_\mu (
\partial^\mu\tilde\rho^\nu -\partial^\nu\tilde\rho^\mu )-
\sigma\eta_\mu\partial^\mu\tilde\sigma \right.\nonumber\\ 
&+&\left.  \beta_\nu(\eta_\mu B^{\mu\nu}+\lambda_1^\prime\beta^\nu +\eta^\nu\varphi) 
- i\tilde\chi\eta_\mu\rho^\mu -i\chi (\eta_\mu\tilde\rho^\mu -
\lambda_2^\prime\tilde\chi)\right]\nonumber\\
&\equiv &S^A,
\end{eqnarray}
which is the action in axial gauge in 2-form gauge theory.
Thus, the FF-anti-BRST transformation  with the finite parameter given in Eq. (\ref{2antisym}) 
takes
\begin{equation}
Z^L\left(=\int {\cal D}\phi e^{iS_{eff}^L}\right)\stackrel{FF-anti-BRST}{-------
\longrightarrow
} Z^A\left(=\int {\cal D}\phi e^{iS_{eff}^A}\right),
\end{equation}
which shows that the FF-anti-BRST transformation plays the similar role as FFBRST transformation
in 2-form gauge theory but with different finite field dependent parameter.
\section{Field/Antifield formulation of Abelian rank-2 antisymmetric tensor field theory}
In this section we construct field/antifield formulation for Abelian rank-2 antisymmetric
tensor field theory to show that techniques of FFBRST formulation can also be applied in
this modern approach of quantum field theory. For this purpose we express the 
generating functional in Eq. (\ref{2zfun}) in field/antifield formulation by introducing 
antifield $\phi^\star $ corresponding to each field $\phi$ with opposite statistics as, 
\begin{eqnarray}
Z^L &=&\int\left[{\cal D}Bd\rho {\cal D}\tilde{\rho}{\cal D}\sigma {\cal D}\tilde{\sigma}{\cal D}\varphi {\cal D}\chi
 {\cal D}\tilde{\chi}{\cal D}
\beta\right]\exp\left[i\int d^4x\left\{\frac{1}{12}F_{\mu\nu\lambda}F^{\mu\nu\lambda}-B^{
\mu\nu\star}\right.\right.\nonumber\\
&&\left.\left.\left(\partial_\mu\rho_\nu - i \partial_\nu\rho_\mu\right) -\rho^{\mu\star}\partial_\mu\sigma +i{\tilde
{\rho}}^{\nu\star}
\beta_\nu-\tilde{
\sigma}^\star\tilde\chi-\varphi^\star\chi\right\}\right].
\end{eqnarray}
This can be written in compact form as
\begin{equation}
Z^L = \int [{\cal D}\phi] \exp{\left [iW_{\Psi^L}(\phi,\phi^\star)\right]},
\end{equation}
where $W_{\Psi^L}(\phi,\phi^\star)$ is an extended action for 2-form gauge theory in Lorentz 
gauge corresponding the gauge-fixed fermion $\Psi^L $  having Grassmann parity 1 and ghost 
number {-1}. The expression for
 $\Psi^L $ is
\begin{equation}
\Psi^L =-i\int d^4x \left[\tilde\rho_\nu \left(\partial_\mu B^{\mu\nu}+
\lambda_1\beta^\nu \right) +\tilde\sigma \partial_\mu\rho^\mu +\varphi 
\left(\partial_\mu\tilde{\rho}^
\mu-\lambda_2\tilde\chi \right)\right].
\end{equation}
 The generating functional $Z^L$ does not depend on the choice of gauge-fixed fermion.
This extended quantum action, $W_{\Psi^L}(\phi,\phi^\star)$, satisfies the quantum master equation 
 given  in Eq. (\ref{2mq}).
The antifields $\phi^\star$ corresponding to each field $\phi$ for this particular theory can 
be obtained from the gauge-fixed fermion as
\begin{eqnarray}
B^{\mu\nu\star }&=&\frac{\delta \psi^L}{\delta B_{\mu\nu}}=i\partial^\mu\tilde\rho^\nu, \ \ \ \ \ \ \ \ \ \ \ \ \ 
\tilde\rho^{\nu\star}=\frac{\delta \psi^L}{\delta \tilde \rho_\nu}=-i(\partial_\mu B^{\mu\nu}+
\lambda_1\beta^\nu -\partial^\nu\varphi )\nonumber\\
\rho^{\mu\star }&=&\frac{\delta \psi^L}{\delta \rho_{\mu}}=i\partial^\mu\tilde\sigma, \ \ \ \ \ \  \ \ \ \ \ \ \ \ \ \
\tilde\sigma^\star =\frac{\delta \psi^L}{\delta\tilde\sigma}=-i\partial_\mu\rho^\mu\nonumber\\
\sigma^\star &=&\frac{\delta \psi^L}{\delta \sigma}=0, \ \ \ \ \ \ \ \ \ \ \ \ \ \ \ \ \ \ \ \ \beta^{\nu\star}=\frac{
\delta \psi^L}{\delta \beta_\nu}=-i\lambda_1\tilde\rho^\nu\nonumber\\
\varphi^\star &=&\frac{\delta \psi^L}{\delta\varphi}=-i(
\partial_\mu\tilde\rho^\mu-\lambda_2\tilde\chi ), \ \ \tilde\chi^\star =\frac{\delta \psi^L}{
\delta\tilde\chi}=i\lambda_2\varphi\nonumber\\
\chi^\star &=&\frac{\delta \psi^L}{\delta\chi}=0.
\end{eqnarray}
Now we apply the FFBRST transformation with the finite parameter given in Eq. (\ref{2afin})
to this generating functional in Lorentz gauge. But the path integral measure is not invariant 
under such a finite transformation and give rise to a factor which can be written as 
$e^{iS_1}$, where the functional $S_1$ is calculated in Appendix A and also given in Eq. 
(\ref{2s1}).
The transformed generating functional
\begin{eqnarray}
Z^\prime&=&\int {\cal D}\phi \exp[i\{ W_{\Psi^L}+S_1\}],\nonumber\\
        &=&\int{\cal D}\phi \exp[i W_{\Psi^A}]\equiv Z^A.
\end{eqnarray}
The generating functional in axial gauge 
\begin{eqnarray}
Z^A &=&\int\left[{\cal D}Bd\rho {\cal D}\tilde{\rho}{\cal D}\sigma {\cal D}\tilde{\sigma}{\cal D}\varphi {\cal D}\chi 
{\cal D}\tilde{\chi}{\cal D}
\beta\right ]\exp\left[i\int d^4x\left\{\frac{1}{12}F_{\mu\nu\lambda}F^{\mu\nu\lambda}\right.\right.\nonumber\\
&-&i\left.\left. \bar B^{
\mu\nu\star}\left(\partial_\mu\rho_\nu-\partial_\nu\rho_\mu\right) -\bar{\rho}^{\mu\star}\partial_\mu\sigma +i\bar{
\tilde{\rho}}^{\nu\star}
\beta_\nu-\bar{\tilde{\sigma}}^\star\tilde\chi-\bar{\varphi}^\star\chi\right\}\right ].
\end{eqnarray}
The extended action $ W_{\Psi^A}$ for 2-form gauge theory in axial gauge also
satisfies  
the quantum master
equation given in (\ref{2mq}).
The gauge-fixed fermion for axial gauge   
\begin{equation}
\Psi^A =-i\int d^4x \left[\tilde\rho_\nu \left(\eta_\mu B^{\mu\nu}+
\lambda_1\beta^\nu \right) +\tilde\sigma \eta_\mu\rho^\mu +\varphi \left(\eta_\mu\tilde{\rho}^
\mu-\lambda_2\tilde\chi \right)\right].
\end{equation}
and corresponding antifields are
\begin{eqnarray}
 \bar B^{\mu\nu\star }&=&\frac{\delta \psi^A}{\delta B_{\mu\nu}}=-i\eta^\mu\tilde\rho^\nu, \ \ \ \ \ \ \ \ \ \ \ \ \  
\bar{\tilde\rho}^{\nu\star}=\frac{\delta \psi^A}{\delta \tilde \rho_\nu}=-i(\eta_\mu B^{\mu\nu
}+
\lambda_1^\prime\beta^\nu +\eta^\nu\varphi )\nonumber\\
\bar{\rho}^{\mu\star }&=&\frac{\delta \psi^A}{\delta \rho_{\mu}}=-i\eta^\mu\tilde\sigma, \ \ \ \ \ \  \ \ \ \ \ \ \ \ \ \ 
\bar{\tilde\sigma}^\star =\frac{\delta \psi^A}{\delta\tilde\sigma}=-i
\eta_\mu\rho^\mu\nonumber\\
\bar{\sigma}^\star &=&\frac{\delta \psi^A}{\delta \sigma}=0, \ \ \ \ \ \ \ \ \ \ \ \ \ \ \ \ \ \ \ \ \ \bar\beta^{
\nu\star}=\frac{\delta \psi^A}{\delta \beta_\nu}=-i\lambda_1^\prime\tilde\rho^\nu\nonumber\\
\bar\varphi^\star &=&\frac{\delta \psi^A}{\delta\varphi}=-i(
\eta_\mu\tilde\rho^\mu-\lambda_2^\prime\tilde\chi ), \ \ \ \bar{\tilde\chi}^\star =\frac{\delta 
\psi^A}{\delta\tilde\chi}=i\lambda_2^\prime\varphi\nonumber\\
\bar{\chi}^\star &=&\frac{\delta \psi^A}{\delta\chi}=0.
\end{eqnarray}
Thus, the FFBRST transformation with finite parameter given in Eq. (\ref{2afin}) relates the 
different  solutions of quantum master equation
in field/antifield formulation.

\section{Conclusions}
The usual BRST transformation has been generalized for the Abelian rank-2 antisymmetric 
tensor field theory by making the parameter finite and field dependent. Such FFBRST 
transformation is nilpotent and leaves the effective action invariant. However, being 
finite in nature such transformations do not leave the path integral measure invariant. We 
have shown that for certain choices of finite field dependent parameter, the Jacobian for the 
path integral of such FFBRST transformation always can be written as $e^{iS_1}$, 
where $S_1$ is some local functional of fields and depends on the choice of the finite BRST 
parameter. $S_1$ can be added with $S_{eff}^L$ to produce the new effective action. Thus,
the generating functional corresponding to one effective theory is then linked to the 
generating functional corresponding  to another effective theory through the FFBRST 
transformation. In this present work we have shown that the generating functional 
corresponding to covariant gauge viz. Lorentz gauge is connected to the generating functional 
in noncovariant gauges viz. axial gauge and Coulomb gauge. Thus, the generalized BRST 
transformation is helpful in the study of Abelian rank-2 antisymmetric tensor field theory 
in noncovariant gauges, which is very useful in certain situation \cite{deg}.
We further have considered the generalization of anti-BRST transformation and show that even 
FF-anti-BRST transformation can connect generating functionals for different 
effective theories. The FFBRST transformation is also very useful 
in modern approach of quantum field theory, namely field/antifield 
formulation. With the help of an explicit example we have shown
 that the different solutions of the master 
equation are related through FFBRST transformation in the field/antifield formulation
of Abelian 2-form antisymmetric tensor field theory.

\label{Chap:Chapter4}
\chapter{FFBRST formulation for Gribov-Zwanziger (GZ) theory}
 In this chapter, we develop the FFBRST formulation for the multiplicatively renormalizable  GZ 
 theory which is free from Gribov ambiguity. 
The mapping between GZ theory and Yang-Mills (YM) theory, which contains the Gribov copies, is 
established in Euclidean space \cite{sudbm}.
 We further extend this formulation using BV techniques \cite{sudbpm}.
\section{GZ theory: brief introduction}
In YM theories even after gauge-fixing the redundancy of gauge
fields is not completely removed in certain gauges for large gauge fields (Gribov ambiguity)
\cite{gri}.  In
order to resolve such problem,  Gribov and Zwanziger proposed a theory,
which restricts the domain of integration in the functional integral within the first Gribov
horizon \cite{zwan}.  
It has been shown in Ref. \cite{zwan2} that the restriction to the Gribov region $\Omega$ (defined in such a way 
that
the Faddeev-Popov (FP) operator is strictly positive. i.e.
 \begin{equation}
\Omega \equiv \{ A_\mu^a, \partial_\mu A^{\mu a}=0, {\cal{M}}^{ab} >0\} ),
 \end{equation} 
can be imposed by adding a nonlocal term $S_h$ 
 to the standard YM 
action
\begin{equation}
S_{YM}=S_0+S_{GF+FP},
\end{equation}
where $S_0$ is the kinetic part and $S_{GF+FP}$ is the ghost and gauge(Landau gauge)-fixing 
part of the YM action respectively,
\begin{eqnarray}
S_0&=&\int d^4x\left[\frac{1}{4}F^a_{\mu \nu }F^{\mu \nu a}\right],\nonumber\\
S_{GF+FP}&=&\int d^4x \left[B^a\partial_\mu A^{\mu  a} +\bar c^a\partial^\mu D_{\mu}^{ 
ab}c^b\right],
\end{eqnarray}
and the nonlocal horizon term in $4$-dimensional Euclidean space is written as
\begin{equation}
S_h=\int d^4x h(x), \label{gact}
\end{equation}
where the integrand $h(x)$ is called the horizon function. There exist many different choices
for the horizon function  in literature \cite{sor}. One such horizon term is
\begin{equation}
h_1(x)=\gamma^4\int d^4y\ g^2 f^{abc}A_\mu^b(x) ({\cal{M}}^{-1})^{ce}(x,y) f^{ade}A^{\mu d}
(y).
\label{ht1}
\end{equation}
$({\cal{M}}^{-1})^{ce}$ is the inverse of the Faddeev-Popov operator 
${\cal{M}}^{ab}\equiv -\partial^\mu D^{ab}_\mu  =-\partial^\mu 
(\partial_\mu\delta^{ab}+gf^{acb}A^c_\mu)$. The Gribov parameter $\gamma$ can be obtained in 
a consistent way by solving a gap equation (also known as horizon condition) \cite{zwan,zwan2}
\begin{equation}
\left<{h(x)}\right> = 4(N^2-1), \label{hc}
\end{equation}
where $N$ is the number of colors.
Another horizon term which gives the correct multiplicative
renormalizability of the GZ theory is given as \cite{ sor}
\begin{eqnarray}
h_2(x) = \lim_{\gamma(x)\rightarrow \gamma}\int d^4y \left[\left(D_\mu^{ac}(x)
\gamma^2(x)\right) 
({\cal{M}}^{-1})^{ce} (x,y) 
 \left(D^{\mu  ae}(y)\gamma^2(x)\right)\right].\label{ht2}
\end{eqnarray}
The nonlocal term (\ref{gact}) corresponding to the horizon function (\ref{ht2}) can be 
localized as 
\cite{zwan, zwan2}
\begin{equation}
e^{-S_{h_2}}=\int {\cal D}\varphi{\cal D}\bar\varphi{\cal D}\omega{\cal D}\bar\omega
e^{S_{loc}},
\end{equation}
with
\begin{eqnarray}
S_{loc} = \int d^4x\left[\bar\varphi^{ai}\partial^\mu D_\mu^{ab}
\varphi^{b}_i-\bar\omega^{ai}\partial^\mu D_\mu^{ab}
\omega^{b}_i -\gamma^2 D^{\mu{ca}} (\varphi_\mu^{ac}(x)+\bar\varphi_\mu^{ac}(x))\right],
\end{eqnarray}
where a pair of complex conjugate bosonic field  $ 
(\bar\varphi_i^{a}, \varphi_i^{a})=(\bar\varphi_\nu^{ac}, \varphi_\nu^{ac})$ 
and anticommuting auxiliary fields
$( \omega_i^{a}, \bar\omega_i^{a})=( \omega_\nu^{ac}, \bar\omega_\nu^{ac})$, with composite 
index $i=(\nu, c)$, have been introduced. As at the level of the 
action, total derivatives are always neglected, $S_{loc}$ becomes
\begin{eqnarray}
S_{loc} = \int d^4x\left[\bar\varphi^{ai}\partial^\mu D_\mu^{ab}
\varphi^{b}_i-\bar\omega^{ai}\partial^\mu D_\mu^{ab}
\omega^{b}_i
-\gamma^2gf^{abc}A^{\mu a} (\varphi_\mu^{bc}(x) +\bar\varphi_\mu^{bc}(x))\right].
\end{eqnarray}
Here it is concluded that at the local level horizon functions (\ref{ht1}) and (\ref{ht2}) 
are same.
So that the localized GZ action becomes
\begin{eqnarray}
S_{GZ}&=& S_{YM} +S_{loc} =S_{YM} +
  \int d^4x\left[\bar\varphi^{ai}\partial^\mu D_\mu^{ab}
\varphi^{b}_i \right.\nonumber\\
&-&\left. \bar\omega^{ai}\partial^\mu D_\mu^{ab}
\omega^{b}_i - \gamma^2gf^{abc}A^{\mu a} (\varphi_\mu^{bc} +\bar\varphi_\mu^{bc})\right].
\end{eqnarray}
 Thus, the local action $S_{GZ}$ and the nonlocal action $S_{YM}+S_h$ are related as
 the following:
 \begin{equation}
\int [{\cal D}\phi_1]e^{-\{S_{YM}+S_{h_2}\}}=\int [{\cal D}\phi]e^{-S_{GZ}},\label{nlact}
\end{equation}
with $\int[{\cal D}\phi_1]\equiv \int[{\cal D}A {\cal D}B {\cal D}c {\cal D}\bar c]$ and 
$\int[{\cal D}\phi]\equiv \int[{\cal D}A {\cal D}B {\cal D}c {\cal D}\bar c {\cal D}\varphi {\cal D}\bar\varphi
{\cal D}\omega {\cal D}\bar\omega]$.
By differentiating Eq. (\ref{nlact}) with respect to $\gamma^2$ and noting $\left < 
\partial^\mu \varphi^{aa}_\mu \right > = \left < \partial^\mu \bar\varphi^{aa}
_\mu \right >=0$, 
 the horizon condition in Eq. (\ref{hc}) is recast as 
\begin{equation}
\left<gf^{abc}A^{\mu a} (\varphi_\mu^{bc}+\bar\varphi_\mu^{ bc})\right> +8\gamma^2(N^2-1)
  =0.
\end{equation}
The horizon condition can further be written as \cite{sor,sor1}
 \begin{equation}
\frac{\partial \Gamma }{\partial \gamma^2}=0,\label{hori}
\end{equation}
with $\Gamma $, the quantum action defined as 
\begin{equation}
e^{-\Gamma }=\int [{\cal D}\phi] e^{-S_{GZ}}.
\end{equation}

We see that the horizon condition (\ref{hori}) is equivalent to
\begin{eqnarray}
\left < 0\mid gf^{abc}A^{\mu a}\varphi^{bc}_\mu \mid 0\right >  + \left < 0\mid gf^{abc}
A^{\mu a}\bar\varphi^{ bc}_\mu\mid 0\right > 
 = -8\gamma^2(N^2-1),
\end{eqnarray}
which, owing to the discrete symmetry of the action $S_{GZ}$ 
\begin{eqnarray}
\bar\varphi^{ac}_\mu \rightarrow  \varphi^{ac}_\mu,\ 
\varphi^{ac}_\mu\rightarrow \bar\varphi^{ac}_\mu,\
B^a\rightarrow (B^a -gf^{amn}\bar\varphi^{mc}_\mu \varphi^{\mu nc}),
\end{eqnarray}
becomes
\begin{eqnarray}
\left < 0\mid gf^{abc}A^{\mu a}\varphi^{bc}_\mu \mid 0\right >  = \left < 0\mid gf^{abc}
A^{\mu a}\bar\varphi^{ bc}_\mu\mid 0\right > 
  = -4\gamma^2 (N^2-1).\label{exp}
\end{eqnarray}
Further, the constant term $4\gamma^4(N^2-1)$ is introduced in $ S_{GZ}$, to incorporate 
the effect of horizon condition in the action as
\begin{eqnarray}
S_{GZ}&=& S_{YM} +\int d^4x \left[\bar\varphi^{ai}\partial^\mu D_\mu^{ab}
\varphi^{b}_i-\bar\omega^{ai}\partial^\mu D_\mu^{ab}
\omega^{b}_i \right.\nonumber\\
&-&\left.\gamma^2 g f^{abc} A^{\mu a}(\varphi^{bc}_\mu
+\bar\varphi^{bc}_\mu) -4(N^2 -1)\gamma^4 
\right].
\end{eqnarray}
For the GZ action to be renormalizable, it is crucial to shift the field 
$\omega^a_i$, \cite{zwan2}
\begin{equation}
\omega^a_i(x)\rightarrow \omega^a_i +\int d^4y ({\cal M}^{-1})^{ab} (x, y) gf^{bkl}\partial^\mu
[D_\mu^{ke} c^e (y)\varphi^{\mu al} (y)],
\end{equation}
so that the complete GZ action becomes 
\begin{eqnarray}
S_{GZ}&=& S_{YM} +\int d^4x \left[\bar\varphi^{ai}\partial^\mu D_\mu^{ab}
\varphi^{b}_i-\bar\omega^{ai}\partial^\mu D_\mu^{ab}
\omega^{b}_i-gf^{abc}\partial^\mu\bar\omega^{ai} D_\mu^{bd}c^d\varphi^c_i \right.\nonumber\\
&-&\left. 
\gamma^2 g\left( f^{abc} A^{\mu a}\varphi^{bc}_\mu + f^{abc} A^{\mu a}\bar\varphi^{bc}_\mu 
+\frac{4}{g}(N^2 -1)\gamma^2 
\right)\right],\label{cgz}
\end{eqnarray}
which has been shown to be multiplicative renormalizable \cite{sor}.
\section{The nilpotent BRST transformation of GZ action}
The complete GZ action after localizing the nonlocal horizon term in D dimensional Euclidean 
space can be recast as
\begin{equation}
S_{GZ}=S_{exact}+S_\gamma \label{gzact}
\end{equation}
with $S_{exact}$, the BRST exact action and $S_\gamma$, the action for horizon term, defined
  as \cite{sor}
\begin{eqnarray}
S_{exact}  &= & S_{YM} +\int d^4x \left[\bar\varphi^{ai}\partial^\mu D_\mu^{ab}
\varphi^{b}_i-\bar\omega^{ai}\partial^\mu D_\mu^{ab}
\omega^{b}_i - gf^{abc}\partial^\mu\bar\omega^{ai} D_\mu^{bd}c^d\varphi^c_i
\right],\nonumber\\ 
S_\gamma & =& -\gamma^2 g\int d^4x \left[f^{abc} A^{\mu a}\varphi^{bc}_\mu +
f^{abc} A^{\mu a}\bar\varphi^{bc}_\mu  
 + \frac{4}{g}(N^2 -1)\gamma^2\right].\label{sgm}
\end{eqnarray}

The conventional BRST transformation for all the fields is given by
\begin{eqnarray}
\delta_b A_\mu^a &=&  D_\mu^{ab}c^b\ \Lambda,\ \ \delta_b c^a =\frac
{1}{2}gf^{abc}c^bc^c \ \Lambda, \ \ \
 \delta_b \bar c^a = B^a\ \Lambda,\ \ \delta_b B^a =0, \nonumber\\
 \delta_b\varphi_i^a &=& -\omega_i^a \ \Lambda,\ \ \delta_b\omega_i^a =0,
 \ \
\delta_b\bar\omega_i^a =\bar \varphi_i^a \ \Lambda,\ \ \delta_b\varphi_i^a
 =0,\label{gzsym}
\end{eqnarray}
where $\Lambda$ is usual infinitesimal BRST parameter. 
But one can check that the BRST symmetry is broken softly for the GZ action \cite{zwan},
\begin{eqnarray}
\delta_b S_{GZ} &=& \delta_b (S_{exact} +S_\gamma )=\delta_b S_\gamma\nonumber\\
&=& \gamma^2 g\int d^4x f^{abc}\left ( A^{\mu a}\omega_\mu^{bc} 
 -  (D^{\mu am}c^m)
(\bar\varphi^{bc}_\mu +\varphi_\mu^{bc})\right ),
\end{eqnarray}
 the breaking is due to the presence of $\gamma$ dependent term, $S_\gamma$. 

To discuss the renormalizability of $S_{GZ}$, 
  $S_\gamma$ is embedded into a larger action with 3 doublets of sources 
$(U_\mu^{ai}, M_\mu^{ai}), (V_\mu^{ai}, N_\mu^{ai}) $  and $(T_\mu^{ai}, R_\mu^{ai})$  as  \cite{sor}
\begin{eqnarray}
\Sigma_\gamma &=& \delta_b \int d^4 x \left (-U_\mu^{ai}D^{\mu ab}\varphi_i^b -
V_\mu^{ai}D^{\mu ab}\bar\omega_i^b -U_\mu^{ai}V^{\mu a}_i 
 +  gf^{abc}
T_\mu^{ai}D^{\mu bd} c^d\bar\omega_i^c \right )\nonumber\\
&=&\int d^4 x \left ( -M^{ai}_\mu D^{\mu ab}\varphi^b_i 
-gf^{abc}U^{ai}_\mu D^{\mu bd}c^d\varphi^c_i +U^{ai}_\mu D^{\mu ab}
\omega^b_i \right. \nonumber\\
&-&\left.  N^{ai}_\mu D^{\mu ab}\bar\omega^b_i -V^{ai}_\mu
D^{\mu ab}\bar\varphi^b_i +gf^{abc}V^{ai}_\mu D^{\mu bd}c^d
\bar\omega^c_i\right.\nonumber\\
&-&\left.  M^{ai}_\mu V^{\mu a}_i +U^{ai}_\mu N^{\mu a }_i
 -  gf^{abc}R^{ai}_\mu D^{\mu bd}c^d\bar\omega^c_i
+gf^{abc}T^{ai}_\mu D^{\mu bd}c^d\bar\varphi^c_i \right ),
\end{eqnarray}
whereas the sources involved $M_\mu^{ai}, V_\mu^{ai}, R_\mu^{ai}$ are commuting and 
$U_\mu^{ai}, N_\mu^{ai}, T_\mu^{ai}$ are fermionic in nature.
The above action is invariant under following BRST transformation:
\begin{eqnarray}
\delta_b U^{ai}_\mu &=&M_\mu^{ai} \ \Lambda,\ \ \delta_b M^{ai}_\mu =0,\ \
\delta_b V^{ai}_\mu =-N_\mu^{ai}\ \Lambda,\nonumber\\
 \delta_b N^{ai}_\mu &=&0,\ \
\delta_b T^{ai}_\mu =-R_\mu^{ai}\ \Lambda,\ \ \delta_b R^{ai}_\mu =0.\label{syma}
\end{eqnarray}
Therefore, the   BRST symmetry has been restored at the cost of introducing new sources.
The different quantum numbers (to study the system properly) of fields and sources, 
involved in this theory, are discussed in Ref. \cite{sor}.
However, we do not want to change our original theory (\ref{sgm}) and  therefore 
we choose the sources to have the following values at the end:
\begin{eqnarray}
U^{ai}_\mu |_{phys} &=& N^{ai}_\mu |_{phys}=T^{ai}_\mu |_{phys}=0\nonumber\\
M^{ab}_{\mu\nu} |_{phys} &=& V^{ab}_{\mu\nu} |_{phys}=R^{ab}_{\mu\nu} |_{phys}=\gamma^2
\delta^{ab}\delta_{\mu\nu}.
\end{eqnarray}
It follows that $\Sigma_\gamma|_{phys}=S_\gamma $.  

The generating functional for the effective GZ action in Euclidean space is 
defined as
\begin{equation}
Z_{GZ}=\int [{\cal D}\phi ] e^{-S_{GZ}},\label{gzfun} 
\end{equation}
where $\phi$ is generic notation for all fields used in GZ action.

\section{FFBRST transformation in Euclidean space}

The Jacobian, $J(\kappa )$, of the path integral measure (as given in Eq. (\ref{jac}) in the Euclidean space can be 
replaced (within the functional 
integral) as
\begin{equation}
J(\kappa )\rightarrow \exp[-S_1[\phi(x,\kappa) ]],
\end{equation}
 iff the following condition is satisfied \cite{sdj}:
\begin{equation}
\int [{\cal D}\phi (x)] \;  \left [ \frac{1}{J}\frac{dJ}{d\kappa}+\frac
{d S_1[\phi (x,\kappa )]}{d\kappa}\right ]\exp{[-(S_{GZ}+S_1)]}=0 \label{mconde}
\end{equation}
where $ S_1[\phi ]$ is local functional of fields.
The infinitesimal change in the $J(\kappa)$ can be calculated using Eq. (\ref{jaceva}).

Now, we generalize the BRST transformation given in Eqs. (\ref{gzsym}) and
(\ref{syma}) by making usual BRST parameter finite and field dependent as  
 \begin{eqnarray}
\delta_b A_\mu^a &=&  D_\mu^{ab}c^b\ \Theta_b,\ \ \delta_b c^a =\frac
{1}{2}gf^{abc}c^bc^c
\ \Theta_b,\ \ \
 \delta_b \bar c^a =B^a\ \Theta_b, \nonumber\\
 \delta_b\varphi_i^a &=& -\omega_i^a \ \Theta_b, 
\ \delta_b\bar\omega_i^a =\bar \varphi_i^a \ \Theta_b,\ 
 \delta_b U^{ai}_\mu =M_\mu^{ai}\ \Theta_b,\  
\delta_b V^{ai}_\mu =-N_\mu^{ai}\ \Theta_b,\nonumber\\
\delta_b T^{ai}_\mu &=&-R_\mu^{ai}\ \Theta_b,\ \delta_b [ B^a, \omega_i^a, \varphi_i^a,  M^{ai}_\mu, 
N^{ai}_\mu, R^{ai}_\mu ]=0,
\end{eqnarray}
where $\Theta_b$ is finite, field dependent, anticommuting and space-time independent parameter.
One can easily check that the above FFBRST transformation is also symmetry of the effective GZ action ($S_{GZ}$).
\section{A mapping between GZ theory and YM theory} 
In this section we establish the connection between the theories with GZ action and YM action 
by using finite field 
dependent BRST transformation. In particular, we show that the generating functional for 
GZ theory in path integral formulation is directly related to that of YM theory 
with proper choice of finite field dependent BRST transformation. The nontrivial Jacobian
of the path integral measure is responsible for such a connection. For this purpose
 we choose a
 finite field dependent parameter $\Theta_b$ obtainable from 
\begin{eqnarray}
\Theta_b^\prime &=&\int d^4x \left[\bar \omega^{ai}\partial^\mu D_\mu^{ab} 
\varphi_i^b
-U^{ai}_\mu D^{\mu ab} \varphi_i^b -V^{ai}_\mu D^{\mu ab}\bar \omega_i^b 
\right.\nonumber\\
&-&\left. U^{ai}_\mu V^{\mu a}_i  +T^{\mu ai} gf^{abc}D_{\mu}^{ bd}c^d\bar\omega^c_i\right],
\end{eqnarray}
using Eq. (\ref{80}).
The infinitesimal change in Jacobian for above $\Theta_b^\prime$ using Eq. (\ref{jaceva}) is 
calculated as
\begin{eqnarray}
\frac{1}{J}\frac{dJ}{d\kappa}&=&-\int d^4x \left[-\bar \varphi^{ai}\partial^\mu D_\mu^
{ab}\varphi^b_i +\bar \omega^{ai}\partial^\mu D_\mu^{ab} 
\omega_i^b 
+ gf^{abc}\partial^\mu\bar\omega^{ia} D_\mu^{bd}c^d\varphi^c_i \right.\nonumber\\
&+&\left. M^{ai}_\mu D^{\mu ab}\varphi^b_i -U^{ai}_\mu D^{\mu ab}\omega^b_i 
+ gf^{abc}U^{ai}_\mu D^{\mu bd}c^d\varphi^c_i +
N^{ai}_\mu D^{\mu ab}\bar\omega^b_i \right.\nonumber\\
&+&\left. V^{ai}_\mu 
D^{\mu ab}\bar\varphi^b_i -gf^{abc}V^{ai}_\mu D^{\mu bd}c^d\bar\omega^c_i +M^{ai}_\mu V^{\mu a}_i 
-U^{ai}_\mu N^{\mu a}_i
 \right.\nonumber\\
& +&\left. gf^{abc}R^{ai}_\mu D^{\mu bd}c^d\bar\omega^c_i -gf^{abc}T^{ai}_\mu D^{\mu bd}c^d\bar\varphi^c_i
\right].
\end{eqnarray}
Now, the Jacobian for path integral measure 
in the generating functional (\ref{gzfun}) can be replaced by $e^{-S_1}$ iff condition (\ref
{mconde}) is satisfied.
We consider an ansatz for $S_1$ as 
\begin{eqnarray}
S_1&=&\int d^4x \left[\chi_1(\kappa)\bar \varphi^{ai}\partial^\mu D_\mu^
{ab}\varphi^b_i +\chi_2(\kappa)\bar \omega^{ai}\partial^\mu D_\mu^{ab} 
\omega_i^b 
+ \chi_3(\kappa) gf^{abc}\partial^\mu\bar\omega^{ia} D_\mu^{bd}c^d\varphi^c_i \right.\nonumber\\
&+&\left.\chi_4(\kappa) M^{ai}_\mu D^{\mu ab}\varphi^b_i + \chi_5(\kappa) 
U^{ai}_\mu D^{\mu ab}\omega^b_i +\chi_6(\kappa) gf^{abc}U^{ai}_\mu D^{\mu bd}c^d\varphi^c_i
\right.\nonumber\\
&+&\left.\chi_7(\kappa) N^{ai}_\mu D^{\mu ab}\bar\omega^b_i + 
\chi_8(\kappa) V^{ai}_\mu 
D^{\mu ab}\bar\varphi^b_i 
+ \chi_9(\kappa) gf^{abc}V^{ai}_\mu D^{\mu bd}c^d
\bar\omega^c_i \right.\nonumber\\
&+&\left. \chi_{10}(\kappa) M^{ai}_\mu V^{\mu a}_i  
 +\chi_{11}(\kappa) U^{ai}_\mu N^{\mu a}_i  +
\chi_{12}(\kappa) gf^{abc}R^{ai}_\mu D^{\mu bd}c^d\bar\omega^c_i
\right.\nonumber\\
& +&\left. \chi_{13}(\kappa) gf^{abc}T^{ai}_\mu D^{\mu bd}c^d\bar\varphi^c_i
\right].  
\end{eqnarray}
where $\chi_j(\kappa) (j=1,2,.....,13)$ are arbitrary constants which depend on the parameter $\kappa$ and satisfy 
following 
initial conditions:
\begin{equation}
\chi_j(\kappa =0)=0.\label{inc}
\end{equation}   
The condition (\ref{mconde}) with the above $S_1$ leads to 
\begin{eqnarray}
&&\int [{\cal D}\phi ]\ e^{-(S_{eff}+S_1)}\left[\bar \varphi^{ai}\partial^\mu D_\mu^
{ab}\varphi^b_i (\chi_1' +1)
 +\bar \omega^{ai}\partial^\mu D_\mu^{ab} 
\omega_i^b (\chi_2' -1)\right.\nonumber\\
 &+ &\left. gf^{abc}\partial^\mu\bar\omega_i^a D_\mu^{bd}c^d\varphi^c_i
(\chi_3' -1)
+  M^{ai}_\mu D^{\mu ab}\varphi^b_i (\chi_4' -1)+ 
U^{ai}_\mu D^{\mu ab}\omega^b_i(\chi_5' -1) 
\right.\nonumber\\
&+&\left. gf^{abc}U^{ai}_\mu D^{\mu bd}c^d\varphi^c_i (\chi_6' +1)
+N^{ai}_\mu D^{\mu ab}\bar\omega^b_i (\chi_7' -1)
+ V^{ai}_\mu 
D^{\mu ab}\bar\varphi^b_i (\chi_8' -1)\right.\nonumber\\
 &+&\left. gf^{abc}V^{ai}_\mu D^{\mu bd}c^d \bar\omega^c_i (\chi_9' +1)
 +  M^{ai}_\mu V^{\mu ai}  (\chi_{10}' -1)+ U^{ai}_\mu N^{\mu ai} (\chi_{11}' +1)
 \right.\nonumber\\
& +&\left.
 gf^{abc}R^{ai}_\mu D^{\mu bd}c^d\bar\omega^c_i (\chi_{12}' -1)
 +  gf^{abc}T^{ai}_\mu D^{\mu bd}c^d\bar\varphi^c_i (\chi_{13}' +1)
\right.\nonumber\\
& +&\left. gf^{abc}\bar\varphi^a_i\partial_\mu D^{\mu bd}c^d\varphi^c_i\Theta_b' (\chi_1 +
\chi_3) - \bar \varphi^a_i\partial_\mu D^{\mu ab}\omega^b_i\Theta_b' (\chi_1 +\chi_2 )\right.\nonumber\\
 & -&\left. gf^{abc}\bar\omega^a_i\partial_\mu D^{\mu bd}c^d\omega^c_i\Theta' (\chi_2-\chi_3) 
 + gf^{abc}M^{ai}_\mu D^{\mu bd}c^d\varphi^c_i\Theta_b' 
(\chi_4-\chi_5)\right.\nonumber\\ 
& -&\left.  M_\mu^{ai} D^{\mu ab}\omega_i^b\Theta_b'(\chi_4 +\chi_6) 
+ gf^{abc}U^{ai}_\mu D^{\mu bd}c^d\omega^c_i\Theta_b' 
(\chi_5 +\chi_6)\right.\nonumber\\ 
& -&\left.  N_\mu^{ai} D^{\mu ab}\bar\varphi_i^b\Theta_b'(\chi_7 -\chi_8)
- gf^{abc}N^{ai}_\mu D^{\mu bd}c^d\bar\omega^c_i\Theta_b' 
(\chi_7 +\chi_9)\right.\nonumber\\ 
& +&\left.  gf^{abc}V^{ai}_\mu D^{\mu bd}c^d\bar\varphi^c_i\Theta_b' (\chi_8 +
\chi_9)
+ M^{ai}_\mu N^{\mu a}_i\Theta_b' (\chi_{10}+\chi_{11})\right.\nonumber\\
& +&\left. gf^{abc}R^{ai}_\mu D^{\mu bd}c^d\bar\varphi^c_i\Theta_b' (\chi_{12} +\chi_{13})
\right] =0
\end{eqnarray}
where prime denotes the differentiation with respect to the parameter $\kappa$.
Equating the coefficient of terms $ \bar \varphi^{ai}\partial^\mu D_\mu^
{ab}\varphi^b_i,\ \bar \omega^{ai}\partial^\mu D_\mu^{ab} 
\omega_i^b,\  gf^{abc}\partial^\mu\bar\omega{ai} D_\mu^{bd}c^d\varphi^c_i,\ 
M^{ai}_\mu D^{\mu ab}\varphi^b_i,$
$\ U^{ai}_\mu D^{\mu ab}\omega^b_i, \ gf^{abc}U^{ai}_\mu D^{\mu bd}c^d\varphi^c_i, 
N^{ai}_\mu
D^{\mu ab}\bar\omega^b_i,\
 V^{ai}_\mu D^{\mu ab}\bar\varphi^b_i $,
$ gf^{abc}V^{ai}_\mu D^{\mu bd}c^d
\bar\omega^c_i,\ M^{ai}_\mu V^{\mu a}_i,$ \\ $ U^{ai}_\mu N^{\mu a}_i,\ gf^{abc}R^{ai}_\mu  
D^{\mu bd}c^d\bar\omega^c_i $ and $gf^{abc}T^{ai}_\mu D^{\mu bd}c^d\bar\varphi^c_i$
 from both sides
of above condition, 
we get following differential equations:
\begin{eqnarray}
\chi_1^\prime +1&=&0,\ \ \chi_2^\prime -1=0,\ \ 
\chi_3^\prime -1=0,\ \ \chi_4^\prime -1=0,\nonumber\\
\chi_5^\prime -1&=&0,\ \ \chi_6^\prime +1=0,\ \
\chi_7^\prime -1=0,\ \ \chi_8^\prime -1=0,\nonumber\\
\chi_9^\prime +1&=&0,\ \ \chi_{10}^\prime -1=0,\ \
\chi_{11}^\prime +1=0,\ \ \chi_{12}^\prime -1=0,\nonumber\\
\chi_{13}^\prime +1&=&0.\label{diff1}
\end{eqnarray}
The $\Theta_b^\prime$ dependent terms will be cancelled separately and comparing the coefficients 
 of $\Theta_b^\prime$ dependent terms, we obtain 
\begin{eqnarray}
\chi_1 +\chi_2 &=&\chi_1 +\chi_3 =\chi_2-\chi_3 =\chi_4-\chi_5 = 0\nonumber\\
\chi_4 +\chi_6 &=&\chi_5 +\chi_6 =\chi_7-\chi_8 =\chi_7 +\chi_9 = 0\nonumber\\
\chi_8 +\chi_9 &=&\chi_{10} +\chi_{11} =\chi_{12}+\chi_{13}=0.\label{scond}
\end{eqnarray}
The particular solution of Eq. (\ref{diff1}) subjected to the condition (\ref{inc})
and Eq. (\ref{scond}) is
\begin{eqnarray}
\chi_1 &=& -\kappa,\ \ \chi_2 =\kappa, \ \ \chi_3 =\kappa, \ \ \chi_4 =\kappa \nonumber\\
\chi_5 &=& \kappa,\ \ \chi_6 =-\kappa, \ \ \chi_7 =\kappa, \ \ \chi_8 =\kappa \nonumber\\
\chi_9 &=& -\kappa,\ \ \chi_{10} =\kappa, \ \ \chi_{11} =-\kappa, \ \ \chi_{12} =\kappa \nonumber\\
\chi_{13} &=& -\kappa.
\end{eqnarray}
Therefore, the expression for $S_1$ in term of $\kappa$ is
\begin{eqnarray}
S_1&=&\int d^4x \left[-\kappa\ \bar \varphi^{ai}\partial^\mu D_\mu^
{ab}\varphi^b_i +\kappa\ \bar \omega^{ai}\partial^\mu D_\mu^{ab} 
\omega_i^b +\kappa\ gf^{abc}\partial^\mu\bar\omega^{ai} D_\mu^{bd}c^d\varphi^c_i 
\right.\nonumber\\
&+&\left. 
\kappa\ M^{ai}_\mu D^{\mu ab}\varphi^b_i 
+ \kappa\ U^{ai}_\mu D^{\mu ab}\omega^b_i -\kappa\ gf^{abc}U^{ai}_\mu D^{\mu bd}c^d\varphi^c_i 
\right.\nonumber\\
& +&\left. \kappa\ N^{ai}_\mu D^{\mu ab}\bar\omega^b_i +\kappa\ V^{ai}_\mu
D^{\mu ab}\bar\varphi^b_i 
 -\kappa\ gf^{abc}V^{ai}_\mu D^{\mu bd}c^d
\bar\omega^c_i +\kappa\ M^{ai}_\mu V^{\mu a}_i \right.\nonumber\\
& -&\left.\kappa\ U^{ai}_\mu N^{\mu a}_i  +
\kappa\ gf^{abc}R^{ai}_\mu D^{\mu bd}c^d\bar\omega^c_i
-\kappa\ gf^{abc}T^{ai}_\mu D^{\mu bd}c^d\bar\varphi^c_i
\right].
\end{eqnarray}
The transformed action is obtained by adding $S_1(\kappa=1)$ to $S_{GZ}$ as,
\begin{equation}
S_{GZ}+S_1=\int d^4x [\frac{1}{4}F_{\mu \nu }^aF^{\mu \nu a}+B^a\partial^\mu A_\mu^a+\bar 
c^a\partial^\mu D_\mu^{ ab}c^b ].
\end{equation}
We left with the YM effective action in Landau gauge.
 \begin{equation}
 S_{GZ}+S_1=S_{YM}.
 \end{equation}
Note the new action is independent of horizon parameter $\gamma$, and hence horizon 
condition 
$\left(\frac{\partial \Gamma }{\partial \gamma^2}=0\right) $ 
 leads trivial relation for $S_{YM}$.
Thus, using  FFBRST transformation we have mapped the generating functionals in 
Euclidean space as
\begin{equation}
Z_{GZ}\left( =\int[ {\cal D}\phi] e^{-S_{GZ}}\right)\stackrel{ FFBRST}{--\longrightarrow }Z_{YM}
\left( =\int [{\cal D}\phi] e^{-S_{YM}}\right),
\end{equation}
where $Z_{YM}$ is the generating functional for Yang-Mils action $S_{YM}$.

\section{Connecting GZ theory and YM theory in BV formalism}
The generating functional of YM theory in the BV formulation can be written by  introducing
antifields $\phi^\star $ corresponding to the all fields $\phi$
 with opposite statistics as,
{\begin{eqnarray}
Z_{YM} = \int [{\cal D}\phi] e^{-\int d^4x\left\{\frac{1}{4}F^a_{\mu \nu }F^{\mu \nu a}
+A_\mu^{a\star}D^{\mu ab}c^b 
+ \bar c^{a \star}B^a \right\}}.
\end{eqnarray}}
This can further be written in a compact form as
 \begin{equation}
Z_{YM} = \int [{\cal D}\phi] e^{-\left  [W_{\Psi_1 }(\phi,\phi^\star)\right]},
\end{equation} 
where the expression for gauge-fixing fermion  is  given as 
$\Psi_1 =  \int d^4x\ \left[\bar c^a\partial^\mu A_\mu^a \right] $.
The generating functional does not depend on the choice of gauge-fixing fermion \cite{ht}.
The extended quantum action, $W_{\Psi_1}(\phi,\phi^\star)$, satisfies the quantum master equation given in Eq. 
(\ref{2mq}).

The antifields  can be evaluated from { $\Psi_1$} as
{ \begin{eqnarray}
A_\mu^{a\star }=\frac{\delta\Psi_1}{\delta A^{\mu a}}= -\partial_\mu\bar c^a,\ \
 \bar c^{a\star}=\frac{\delta\Psi_1}{\delta \bar c^a}= 
 \partial^\mu A_\mu^a,\ \ B^{a\star}=\frac{\delta\Psi_1}{\delta B^a}= 
 0.
\end{eqnarray}}
Similarly, the generating functional of GZ theory in BV formulation can be written as, 
{\begin{eqnarray}
Z_{GZ} & =&  \int [{\cal D}\phi] \exp\left[ -\int d^4x\left\{\frac{1}{4}F^a_{\mu \nu }F^{\mu \nu a}
+A_\mu^{a\star}D^{\mu ab}c^b  
 +  \bar c^{a \star}B^a -\varphi_i^{b\star}\omega^{bi}
 \right.\right.\nonumber\\
&+&\left.\left. +\bar\varphi^{ai}\bar\omega_i^{a\star}+U_\mu^{ai\star}M^{\mu a}_i 
-V_\mu^{ai\star}N^{\mu a}_i 
 -   T_\mu^{ai\star}R^{\mu a}_i \right\}\right].
\end{eqnarray}}
This can further be written in compact form using gauge-fixed fermion ($\Psi_2$) as 
\begin{eqnarray}
Z_{GZ} &= &\int [{\cal D}\phi] e^{-\left  [W_{\Psi_2 }(\phi,\phi^\star)\right]}, 
\Psi_2  = \int d^4x  \left[\bar c^a\partial^\mu A_\mu^a +
\bar \omega^{ai}\partial^\mu D_\mu^{ab}  
\varphi_i^b   \right.\nonumber\\ 
  &-&\left. U_\mu^{ai}D^{\mu ab}\varphi^{  b}_i -V_\mu^{ai}D^{\mu ab}\bar\omega^{  b}_i
- U_\mu^{ai}V^{\mu a}_i 
+   gf^{abc}
T_\mu^{ai}D^{\mu bd} c^d\bar\omega^{ c}_i \right].\label{1gzfun} 
\end{eqnarray}
The antifields are
obtained from  $\Psi_2$  as 
{ \begin{eqnarray}
A_\mu^{a\star } =  -\partial_\mu\bar c^a -gf^{abc}\partial_\mu\omega^{c i}\varphi^{ b}_i 
-gf^{abc}U_\mu^{ci}\varphi^{  b}_i   
-gf^{abc}V_\mu^{ci}\bar\omega^{  b}_i, \ \
 \bar c^{a\star} =  
 \partial_\mu A_\mu^a,\nonumber \\ 
U_\mu^{ai
\star} =  -D_\mu^{ab}\varphi^{  bi}  -V_\mu^{ai},\ \
\bar\omega_i^{a\star}= \partial^\mu D_
\mu^{ab}\varphi^b_i -V^{\mu b}_iD_\mu^{ba}
+ gf^{abc} 
T_{ i}^{\mu b }D^{cd}_\mu c^d,\nonumber\\ 
V_\mu^{ai\star}  =  -D_\mu^{ab}\bar\omega^{  bi}
-U_\mu^{ai}, T_\mu^{ai\star} =gf^{abc}
D^{bd}_\mu c^d\bar\omega^{ ci},
\varphi_i^{b\star } =\bar \omega^a_i\partial^\mu D_\mu^{ab} 
 -U_{i}^{\mu a}D_\mu^{ab}.
\end{eqnarray} 
To connect these two theories we construct the following finite field dependent parameter $\Theta_b [\phi]$: 
\begin{eqnarray}
\Theta_b [\phi,\phi^\star] =\int_0^1 d\kappa \int d^4x \left[\varphi_i^{b\star}
\varphi^{bi} +\bar\omega_i^{b\star}\bar \omega^{bi} 
  + V^{ai\star}_\mu V^{\mu a}_i \right].
\end{eqnarray} 
The Jacobian of path integral measure  in the generating functional (\ref{1gzfun}) for the FFBRST 
with this parameter can be replaced by $e^{-S_1}$ iff condition (\ref
{mconde}) is satisfied.
To find $S_1$ we start with an ansatz for $S_1$ as 
\begin{eqnarray}
S_1 = \int d^4x \left[\chi_1 \varphi_i^{b\star}\omega^{bi}
+\chi_2 \bar\varphi^{ai}\bar\omega_i^{a\star} + \chi_3  U_\mu^{ai\star}M^{\mu a}_i 
+\chi_4  V_\mu^{ai\star}N^{\mu a}_i
+\chi_5  T_\mu^{ai\star}R^{\mu a}_i 
\right].
\end{eqnarray}
where $\chi_j(\kappa) (j=1,2,..,5)$ are arbitrary but $\kappa$-dependent constants and satisfy the
following 
initial conditions: $\chi_j(\kappa =0)=0.$
These constants are calculated using Eq. (\ref{mconde}) subjected to the initial condition to find the $S_1$ as
\begin{eqnarray}
S_1=\int d^4x \left[\kappa\ \varphi_i^{b\star}\omega^{bi}
-\kappa\ \bar\varphi^{ai}\bar\omega_i^{a\star}  -\kappa\ U_\mu^{ai\star}M^{\mu a}_i 
+ \kappa\  V_\mu^{ai\star}N^{\mu a}_i
+ \kappa\  T_\mu^{ai\star}R^{\mu a}_i 
\right].
\end{eqnarray}
By adding $S_1(\kappa=1)$ to $S_{GZ}$, we get
$
S_{GZ}+S_1(\kappa=1)= S_{YM}.
$
Hence, 
\begin{eqnarray}
Z_{GZ}\left(= \int [{\cal D}\phi]\ e^{-W_{\Psi_2 }} \right)  \stackrel{FFBRST}{----\longrightarrow }
Z_{YM}\left(= \int [{\cal D}\phi]\ e^{- W_{\Psi_1 }(\phi,\phi^\star) }\right) 
\end{eqnarray}
Thus, using  FFBRST transformation we connect the GZ theory to YM theory in BV formulation. 

\section{Conclusions}
The GZ theory which is free from Gribov copies as the domain of integration is restricted 
to the first Gribov horizon, is not  invariant under usual BRST 
transformation due 
to the presence of the nonlocal horizon term. Hence the KO criterion for color 
confinement in a manifestly covariant gauge fails for GZ theory. A nilpotent BRST   
transformation which leaves GZ action invariant was developed recently and can be applied to 
KO analysis for color confinement. This nilpotent BRST symmetry is generalized by
allowing the transformation parameter finite and field dependent. This generalized BRST 
transformation is nilpotent and symmetry of the GZ effective action.
We have shown that this nilpotent
BRST with an appropriate choice of finite field dependent parameter relates the GZ theory 
with a correct horizon term 
to the YM theory in Euclidean space where horizon condition becomes a trivial one. 
We have shown the same connection  in BV formulation also by considering the appropriate 
finite parameter. Thus, we 
have shown that 
the theory, free from Gribov copies (i.e. GZ 
theory with proper horizon term), is related through a  nilpotent BRST 
transformation with a finite parameter to a theory with Gribov copies (i.e. YM 
theory in Euclidean space). 
The nontrivial Jacobian of such finite transformation is responsible for this important 
connection. This implies our formulation is very useful for the better understanding of Gribov
ambiguity.

\label{Chap:chapter5}

\chapter{Finite nilpotent symmetry for gauge theories}
Earlier it was shown that FFBRST and FF-anti-BRST transformations  
are symmetry of the effective action but not of the generating functional. In this chapter we construct  a nilpotent 
finite BRST transformation which leaves both the effective action and generating functional invariant \cite{epjc}. 
To construct such a finite transformation we combine 
usual BRST and anti-BRST transformations with finite parameters.
The field/antifield (or BV) formulation in the context such transformation is also studied in this chapter. 

\section{The infinitesimal  mixed BRST (MBRST) transformation}
The generating functional for the Green's function in an effective theory described by the effective action $S_{eff} 
[\phi]$  is defined as
\begin{equation}
Z=\int  [{\cal D}\phi]\ e^{iS_{eff}[\phi]},
\end{equation}
 \begin{equation}
S_{eff} [\phi]= S_0[\phi] +S_{gf}[\phi] +S_{gh}[\phi],
\end{equation} 
where  $\phi$ is the generic notation for all fields 
 involved in the effective theory.
The infinitesimal  
 BRST ($\delta_b$) and anti-BRST 
($\delta_{ab}$) transformations are defined as
\begin{equation}
\delta_b \phi =s_b\phi\ \delta\Lambda_1,\ \ s_b^2=0
\end{equation}
\begin{equation}
\delta_{ab} \phi =s_{ab}\phi\ \delta\Lambda_2,\ \ s_{ab}^2=0,
\end{equation}
where $\delta\Lambda_1$ and $\delta\Lambda_2$ are infinitesimal, anticommuting but global parameters.  
 Such transformations leave the generating functional as well as effective action 
invariant, separately, as
\begin{equation}
\delta_b Z=0=\delta_b S_{eff},
\end{equation}
\begin{equation}
\delta_{ab} Z =0=\delta_{ab} S_{eff}.
\end{equation}
This implies that the effective action $S_{eff} [\phi]$ and the generating functional are also  invariant 
under the MBRST  ($\delta_m =\delta_b +\delta_{ab}$) transformation 
\begin{equation}
\delta_m Z =0= \delta_m S_{eff}. \label{mix}
\end{equation}
Further such MBRST transformation is nilpotent because,
\begin{equation}
\{  s_b, s_{ab}\} =0.
\end{equation}
Now, in the next section we  construct the finite version of the following infinitesimal MBRST symmetry 
transformation: 
  \begin{equation}
\delta_m \phi = s_b\phi\ \delta\Lambda_1 +s_{ab}\phi\ \delta\Lambda_2.\label{infi}
\end{equation}

\section{Construction of finite field dependent MBRST (FFMBRST) transformation}
To construct the FFMBRST transformation, we follow the similar method of 
constructing FFBRST transformation \cite{jm}. However, in this case unlike FFBRST transformation
 we have to deal with two 
parameters, one for 
the BRST transformation and the other for anti-BRST transformation. We introduce a numerical parameter $\kappa$ $ 
(0\leq \kappa\leq 1)$ and make all the fields ($\phi (x,\kappa)$) $\kappa$-dependent in such a way that
 $\phi(x,\kappa =0)\equiv \phi(x)$ and $\phi(x,\kappa 
=1)\equiv \phi^\prime(x)$, the transformed field. Further, we make the infinitesimal parameters $\delta\Lambda_1$ and
$\delta\Lambda_2$  field dependent as
\begin{eqnarray}
\delta\Lambda_1&=&\Theta_b^\prime[\phi(x,\kappa)]d\kappa\\
\delta\Lambda_2&=&\Theta_{ab}^\prime[\phi(x,\kappa)]d\kappa,
\end{eqnarray}
 where the prime denotes the derivative with respect to $\kappa$ and 
 $\Theta^\prime_i [\phi(x,\kappa)] (i=b, ab)$ are  infinitesimal field dependent parameters.
The infinitesimal but field dependent MBRST transformation, thus can be written generically as 
\begin{equation}
\frac{d\phi(x,\kappa)}{d\kappa}= s_b \phi (x,\kappa ) \Theta_b^\prime [\phi (x,\kappa )] +
s_{ab} \phi (x,\kappa ) \Theta_{ab}^\prime [\phi (x,\kappa )]. 
\label{mdiff}
\end{equation}
Following the work in Ref.\cite{jm} it can be shown that the parameters 
 $\Theta_i^\prime[\phi(x, \kappa)]$ $(i= b, ab),$ contain  the factors
$\Theta_i^\prime[\phi(x, 0)]$ $(i=b,ab),$ which are considered to be nilpotent. Thus $\kappa$ dependency from 
$\delta_b\phi(x,\kappa)$ and $\delta_{ab}\phi(x,\kappa)$ can be dropped. 
Then   Eq. (\ref{mdiff}) can be written as
\begin{equation}
\frac{d\phi(x,\kappa)}{d\kappa}= s_b \phi (x,0 ) \Theta_b^\prime [\phi (x,\kappa )] +
s_{ab} \phi (x, 0 ) \Theta_{ab}^\prime [\phi (x,\kappa )].
\end{equation}
The FFMBRST transformation with the finite field dependent parameters then can be 
constructed by integrating such infinitesimal transformations from $\kappa =0$ 
to $\kappa= 1$, such that
\begin{equation}
\phi^\prime\equiv \phi (x,\kappa =1)=\phi(x,\kappa=0)+s_b \phi (x) \Theta_b [\phi (x
 )] + s_{ab} \phi (x) \Theta_{ab} [\phi (x)],
\label{mkdep} 
\end{equation}
where 
\begin{equation}
\Theta_i[\phi(x)]=\int_0^1 d\kappa^\prime\Theta_i^\prime [\phi(x,\kappa^\prime)],
\end{equation}
 are the finite field dependent parameters with $i=b, ab$.  

Therefore, the FFMBRST transformation corresponding to MBRST transformation mentioned in Eq. (\ref{infi})
is given by
\begin{equation}
\delta_m \phi = s_b\phi\ \Theta_b +s_{ab}\phi\ \Theta_{ab} 
\end{equation}
It can be shown that above FFMBRST transformation with some specific choices of the finite 
parameters $\Theta_b$ and $\Theta_{ab}$ is the symmetry transformation of the both effective action
and the generating functional as the 
path integral measure is   invariant under such transformation.
\section{Method for evaluating the  Jacobian} 
To show the invariance of the generating functional we need to calculate the Jacobian of the 
path integral measure in the expression of generating functional.
The Jacobian  of the path integral measure for FFMBRST transformation $J  $ can be evaluated for some 
particular choices of the finite field dependent parameters $\Theta_b[\phi(x)]$ and 
$\Theta_{ab}[\phi(x)]$. We start with the definition,
\begin{eqnarray}
{\cal D}\phi  &=&J( \kappa)\ {\cal D}\phi(\kappa) \nonumber\\ 
&=& J(\kappa+d\kappa)\ {\cal D}\phi(\kappa+d\kappa),
\end{eqnarray} 
 Now the transformation from $\phi(\kappa)$ to $\phi(\kappa +d\kappa)$ is infinitesimal 
in nature, thus the infinitesimal change in Jacobian can be calculated as 
\begin{equation}
\frac{J (\kappa)}{J (\kappa +d\kappa)}=\int d^4 x\sum_\phi\pm 
\frac{\delta\phi(x, \kappa)}{
\delta\phi(x, \kappa+d\kappa)} 
\end{equation}
where $\Sigma_\phi $ sums over all fields involved in the path integral measure  
and  $\pm$ 
sign refers to whether $\phi$ is a bosonic or a fermionic field.
Using the Taylor expansion we calculate the above expression as 
\begin{eqnarray}
1-\frac{1}{J (\kappa)}\frac{dJ (\kappa)}{d\kappa}d\kappa 
&=&1+\int d^4x \sum_\phi\left [ \pm s_b  
\phi (x,\kappa )\frac{
\delta\Theta_b^\prime [\phi (x,\kappa )]}{\delta\phi (x,\kappa )} \right.\nonumber\\
&\pm&   \left. s_{ab}  \phi (x,\kappa )\frac{
\delta\Theta_{ab}^\prime [\phi (x,\kappa )]}{\delta\phi (x,\kappa )}\right ]d\kappa.\label{mjac}
\end{eqnarray}

The Jacobian, $J(\kappa )$, can be replaced (within the functional integral) as
\begin{equation}
J(\kappa )\rightarrow e^{i (S_1[\phi ] + S_2[\phi ] )}
\end{equation}
iff the following condition is satisfied: 
\begin{equation}
\int [{\cal D}\phi (x)] \;  \left [ \frac{1}{J }\frac{dJ }{d\kappa} -i\frac
{d  S_1[\phi (x,\kappa )]}{d\kappa} -i\frac{d S_2[\phi (x,\kappa )] }{d\kappa}\right ] e^{ i(S_{eff}+S_1+S_2) }=0, 
\label{mcondf}
\end{equation}
where $ S_1[\phi ]$ and  $ S_2[\phi ]$ are some local functionals of fields and satisfy the initial
condition
 \begin{equation}
S_i[\phi(\kappa =0) ]=0,\ i =1, 2.\label{inco}
\end{equation}
The finite parameters $\Theta_b$ and $\Theta_{ab}$ are arbitrary and we can construct them in such a way that
the infinitesimal change in Jacobian $J$ (Eq. (\ref{mjac})) with respect to $\kappa$ vanishes  
\begin{equation}
\frac{1}{J}\frac{dJ}{d\kappa}=0.\label{mj}
\end{equation}
Therefore, with the help of Eqs. (\ref{mcondf}) and (\ref{mj}), we see that
\begin{equation}
\frac
{d S_1[\phi (x,\kappa )]}{d\kappa} +\frac{S_2[\phi (x,\kappa )] }{d\kappa}=0.
\end{equation}
It means the $S_1 +S_2$ is independent of $\kappa $ (fields) and must vanish to satisfy the initial condition
 given in Eq. (\ref{inco}) satisfied.
Hence the generating functional is not effected by the Jacobian $J$ as $J =e^{i(S_1+S_2)} = 1$. 
The nontrivial Jacobian arising from finite BRST parameter  $\Theta_b$ compensates the same arising due to finite 
anti-BRST parameter  $\Theta_{ab}$. It is straightforward to see that the  effective action $S_{eff}$ is invariant 
under such FFMBSRT transformation.
\section{Examples} 
To demonstrate the results obtained in the previous section we would like to consider several explicit examples in 
1D as well as in 4D. In particular we consider the 
bosonized self-dual chiral model in 2D, 
Maxwell's theory in 4D, and non-Abelian YM theory in 4D. In all these cases we construct 
explicit finite parameters $\Theta_b$ and $\Theta_{ab}$ of FFMBRST transformation such that
the generating functional  remains invariant.
\subsection{Bosonized chiral model}
We start with the generating functional for the bosonized self-dual chiral model as \cite{pp, pp1}
\begin{equation}
Z_{CB}=\int [{\cal D}\phi ]\ e^{iS_{CB}},\label{mz}
\end{equation}
where   $[{\cal D}\phi]$ is the path integral measure in generic notation. The  effective action $S_{CB}$  in 2D 
 is given as
 \begin{eqnarray}
{S}_{CB}&=& \int d^2x [\pi_\varphi\dot\varphi +\pi_\vartheta\dot\vartheta 
+p_u\dot u-\frac{1}{2}
\pi_\varphi^2
+\frac{1}{2}\pi_\vartheta^2 
+\pi_\vartheta (\varphi'-\vartheta' +\lambda)
+ \pi_\varphi\lambda \nonumber\\
&+&\frac{1}{2}B^2 +B(\dot \lambda -\varphi -\vartheta ) 
 + \dot{\bar c}\dot c
-2\bar c c],
\end{eqnarray}
where the fields $\varphi, \vartheta, u, B, c$ and $\bar c$  are  the self-dual field, Wess-Zumino field, multiplier 
field, auxiliary 
field, ghost field and antighost field, respectively.
The   nilpotent 
BRST and anti-BRST transformations for this theory are

BRST:
\begin{eqnarray}
\delta_{b}\varphi &=&  c\ \delta \Lambda_1,\ \ \ \delta_{b} \lambda =-\dot { c
}\ \delta \Lambda_1, \ \ \ \delta_{b} \vartheta = c \ \delta \Lambda_1,\nonumber\\
\delta_{b}  \pi_\varphi &=& 0,\ \ \ \delta_{b}  u =0, \ \ \ \delta_{b}  \pi_\vartheta =0,\ \ \ 
\delta_{b} \bar c =  B \ \delta \Lambda_1,\nonumber\\
 \delta_{b}  B &=&0, \ \ \ \delta_{b}   c =0,\ \ \delta_{b}  p_u 
=0,\label{msym}
\end{eqnarray}
anti-BRST: 
\begin{eqnarray}
\delta_{ab}\varphi &=& -\bar c\ \delta \Lambda_2,\ \ \ \delta_{ab}\lambda =\dot {\bar c
}\ \delta \Lambda_2, \ \ \ \delta_{ab}\vartheta = -\bar c\ \delta \Lambda_2,\nonumber\\
\delta_{ab} \pi_\varphi &=& 0,\ \ \ \delta_{ab} u =0, \ \ \ \delta_{ab} \pi_\vartheta =0,\ 
\delta_{ab}c  =  B\ \delta \Lambda_2,\nonumber\\
 \delta_{ab} B &=&0, \ \ \ \delta_{ab} \bar c =0,\ \ \delta_{ab} p_u 
=0,\label{mab}
\end{eqnarray}
where $\delta\Lambda_1$ and $\delta\Lambda_2$  are infinitesimal, anticommuting and global parameters.
Note that $s_{b}$ and $s_{ab}$ are absolutely anticommuting i.e.
$(s_{b}s_{ab}+s_{ab}s_{b})\phi=0$. In this case the   MBRST symmetry transformation
  ($\delta_m\equiv \delta_{b}+\delta_{ab} $), as constructed in section II, reads 
\begin{eqnarray}
\delta_m\varphi &=&  c\ \delta \Lambda_1 -\bar c\ \delta \Lambda_2,\ \ \ \delta_m\lambda =-\dot { c
}\ \delta \Lambda_1 +\dot {\bar c}\ \delta \Lambda_2,\nonumber\\
 \delta_m\vartheta &=& c \ \delta \Lambda_1 -\bar c\ \delta \Lambda_2,\ \delta_m \pi_\varphi = 0,\ \ \ \delta_m u =0, 
\nonumber\\
 \delta_m \pi_\vartheta &=&0,\ \delta_m\bar c = B \ \delta \Lambda_1,\ \ \ \delta_m B =0,\
\delta_m  c =  B\ \delta \Lambda_2,\nonumber\\ 
 \delta_m p_u &=&0.\label{mb}
\end{eqnarray}
The FFMBRST transformation corresponding to the above MBRST transformation
is constructed as
\begin{eqnarray}
\delta_m\varphi &=&  c\ \Theta_b -\bar c \ \Theta_{ab},\   \delta_m\lambda =-\dot { c
}\ \Theta_b +\dot {\bar c}\ \Theta_{ab},  \ \delta_m\vartheta = c\ \Theta_b -\bar c\ \Theta_{ab}\nonumber\\
\delta_m \pi_\varphi &=& 0,\ \ \ \delta_m u =0, \ \ \ \delta_m \pi_\vartheta =0,\ 
\delta_m\bar c  =  B \ \Theta_b,\ \ \ \delta_m B =0, \nonumber\\
 \delta_m  c &=& B\ \Theta_{ab},\ \ \delta_m p_u 
=0,\label{mc}
\end{eqnarray}
where $\Theta_b$ and $\Theta_{ab}$ are finite field dependent parameters and are still 
anticommuting 
in nature. 
 We construct  the finite  parameters $\Theta_b$ and $\Theta_{ab}$ as 
\begin{equation}
\Theta_b= \int\Theta^\prime_b d\kappa= \gamma\int d\kappa\int d^2x [\bar c(\dot\lambda-\varphi -\vartheta )],\label{mfi}
\end{equation}
\begin{equation}
\Theta_{ab}=\int\Theta^\prime_{ab} d\kappa= -\gamma\int d\kappa \int d^2x [ c(\dot\lambda-\varphi -\vartheta )],\label{mfia}
\end{equation}
where $\gamma$ is an arbitrary parameter.

Using Eq. (\ref{mjac}), the infinitesimal change in Jacobian for the FFMBRST transformation given in Eq. (\ref{mc})  
is calculated 
 as 
\begin{equation}
\frac{1}{J}\frac{dJ}{d\kappa} = 0.
\end{equation}
The contributions from second and third terms in the R.H.S. of Eq. (\ref{mjac}) cancel each other.
This implies that the Jacobian for path integral measure is unit under FFMBRST transformation. Hence, the generating 
functional as well as the effective action are invariant under FFMBRST 
transformation
\begin{eqnarray}
Z_{CB} \left(= \int [{\cal D}\phi ]\ e^{iS_{CB}} \right) \stackrel{FFMBRST}{----\longrightarrow }
Z_{CB}.
\end{eqnarray} 

Now, we would like to consider of the effect of FFBRST transformation with finite parameter $\Theta_b$ and 
FF-anti-BRST transformation with finite parameter $\Theta_{ab}$ independently.
The infinitesimal change in Jacobian $J_1$ for the FFBRST  transformation with the parameter $\Theta_b$  
is calculated as
\begin{equation}
\frac{1}{J_1}\frac{dJ_1}{d\kappa} =  \gamma\int d^4x \left[B (\dot\lambda -\varphi -\vartheta) +\dot{\bar c}\dot c -
2\bar c c\right].\label{mjac1}
\end{equation}
To write the Jacobian $J_1$ as $e^{iS_1}$ in case of BRST transformation, we make the following ansatz for $S_1$:
\begin{eqnarray}
S_1 =i\int d^4 x \left[\xi_1(\kappa)\ B (\dot\lambda -\varphi -\vartheta)  +\xi_2(\kappa)\ \dot{\bar c}\dot c
+\xi_3 (\kappa)\ \bar c c\right],\label{msn1}
\end{eqnarray}
where $\xi_i (i=1, 2, 3)$ are arbitrary $\kappa$-dependent constants and satisfy the
initial conditions $\xi_i (\kappa=0)=0$.

The essential condition in Eq. (\ref{mcondf})  satisfies with Eqs. (\ref{mjac1}) and (\ref{msn1}) iff
\begin{eqnarray}
 &&\int d^4x
\left[- B (\dot\lambda -\varphi -\vartheta)(\xi_1' +\gamma) - \dot{\bar c}\dot c(\xi_2' +\gamma) 
-\bar c c(2\gamma -\xi_3') \right.\nonumber\\
 &+&\left. B\ddot c \Theta_b^\prime (\xi_1 -\xi_2) + Bc\Theta_b^\prime (2\xi_1 +\xi_3) \right] =0,
\end{eqnarray}
where prime denotes the derivative with respect to $\kappa$.
Equating the both sides of the above equation, we get the following  equations: 
\begin{eqnarray}
\xi_1' +\gamma =0, \ \ \xi_2' +\gamma =0,\ \ \xi_3' -2\gamma =0,\ \ \xi_1 -\xi_2 =0=2\xi_1 +\xi_3. 
\end{eqnarray}
The solution of above equations satisfying the initial conditions is
\begin{eqnarray}
\xi_1 =-\gamma\kappa, \ \ \xi_2 =-\gamma\kappa, \ \ \xi_3 =2\gamma\kappa.
\end{eqnarray}
Then, the expression for $S_1$ in terms of $\kappa$ becomes 
\begin{eqnarray}
S_1 =i\int d^4 x \left[-\gamma\kappa B (\dot\lambda -\varphi -\vartheta)  -\gamma\kappa \dot{\bar c}\dot c
+2\gamma\kappa \bar c c\right]. 
\end{eqnarray}
On the other hand the infinitesimal change in Jacobian $J_2$ for the FF-anti-BRST parameter $\Theta_{ab}$  
is calculated as
\begin{equation}
\frac{1}{J_2}\frac{dJ_2}{d\kappa} = - \gamma\int d^4x \left[B (\dot\lambda -\varphi -\vartheta) +\dot{\bar c}\dot c -
2\bar c c\right].
\end{equation}
Similarly, to write the Jacobian $J_2$ as $e^{iS_2}$ in the anti-BRST case, we make an ansatz for $S_2$ as
\begin{eqnarray}
S_2 =i\int d^4 x \left[\xi_4(\kappa)\ B (\dot\lambda -\varphi -\vartheta)  +\xi_5(\kappa)\ \dot{\bar c}\dot c
+\xi_6(\kappa)\ \bar c c\right], 
\end{eqnarray}
where arbitrary $\kappa$-dependent constants $\xi_i  (i=4, 5, 6)$ have to be calculated.

The essential condition in Eq. (\ref{mcondf}) for the above Jacobian $J_2$ and functional $S_2$  provides 
\begin{eqnarray}
 &&\int d^4x
\left[  B (\dot\lambda -\varphi -\vartheta)(\xi_4' -\gamma)+ \dot{\bar c}\dot c(\xi_5' -\gamma) 
-\bar c c(2\gamma +\xi_6') \right.\nonumber\\
 &+&\left. B\ddot {\bar c} \Theta_{ab}^\prime (\xi_4 -\xi_5) +B\bar c\Theta_{ab}^\prime (2\xi_4 +\xi_6) \right] =0.
\end{eqnarray}
Comparing the L.H.S. and R.H.S. of the above equation, we get the following equations: 
\begin{eqnarray}
\xi_4' -\gamma=0,\ \xi_5' -\gamma=0,\ \xi_6'+ 2\gamma =0,\ \xi_4 -\xi_5 =0 =2\xi_4 +\xi_6.
\end{eqnarray}
Solving the above equations, we get the following values for the $\xi_i$'s:
\begin{eqnarray}
\xi_4 = \gamma\kappa, \ \ \xi_5 = \gamma\kappa, \ \ \xi_6 = -2\gamma\kappa.
\end{eqnarray}
Putting these values in expression of $S_2$, we get
\begin{eqnarray}
S_2 =i\int d^4 x \left[ \gamma\kappa B (\dot\lambda -\varphi -\vartheta)  +\gamma\kappa \dot{\bar c}\dot c
-2\gamma\kappa \bar c c\right]. 
\end{eqnarray}
Thus, under successive FFBRST and FF-anti-BRST transformations the generating functional transformed
as 
\begin{eqnarray}
Z_{CB} \left( = \int [{\cal D}\phi]\ e^{iS_{CB}} \right) \stackrel{(FFBRST)(FF-anti-BRST)}{--------\longrightarrow }
Z_{CB}\left( = \int [{\cal D}\phi]\ e^{iS_{CB}+S_1+S_2} \right).
\end{eqnarray}
Note, for the particular choices of $\Theta_b$ and $\Theta_{ab}$, the $S_1$ and $S_2$ cancel each other.
Hence $Z_{CB}$ remains invariant under successive FFBRST and FF-anti-BRST transformations.
It is interesting to note that the effect of FFMBRST transformation is equivalent to successive operation of
FFBRST and FF-anti-BRST transformations.
\subsection{Maxwell's theory}
We consider FFMBRST transformation here for a more basic model.  
The generating functional for Maxwell theory, using Nakanishi-Lautrup type auxiliary field ($B$),
can be given as
\begin{equation}
Z_{M}=\int [{\cal D}\phi ]\ e^{iS_{eff}^M},\label{mzy}
\end{equation}
where the effective action in covariant (Lorentz) gauge with the ghost term is
\begin{equation}
S_{eff}^M=\int d^4 x\left[-\frac{1}{4}F_{\mu\nu}F^{\mu\nu}+\lambda B^2-B\partial_\mu A^\mu
-\bar c\partial_\mu\partial^\mu c\right].
\end{equation}
The infinitesimal off-shell nilpotent BRST and anti-BRST transformations under which the  effective action
$S_{eff}^M$ as well as generating functional $Z_{M}$ remain invariant, are given as

BRST:
\begin{eqnarray}
\delta_b A_\mu &=& \partial_\mu c\ \delta \Lambda_1,\ \ 
\delta_b c =  0\nonumber\\ 
\delta_b \bar c &=& B\ \delta \Lambda_1,\ \ 
\delta_b B = 0.
\end{eqnarray}
Anti-BRST:
\begin{eqnarray}
\delta_{ab} A_\mu &=& \partial_\mu \bar c\ \delta \Lambda_2,\ \ 
\delta_{ab} \bar c = 0,\nonumber\\ 
\delta_{ab} c& = & - B\ \delta \Lambda_2,\ \ 
\delta_{ab} B =  0.
\end{eqnarray}
The nilpotent BRST transformation ($s_b$) and anti-BRST transformation 
($s_{ab}$) mentioned above are absolutely anticommuting in nature i.e.
$\{s_b, s_{ab}\}\equiv s_bs_{ab}+s_{ab}s_b=0$. Therefore,
the sum of these two transformations
($s_b$ and $s_{ab}$) is also a nilpotent symmetry transformation. Let us define 
MBRST transformation ($\delta_m \equiv \delta_b+\delta_{ab} $) in this case, which is characterized by 
two infinitesimal parameters $\delta \Lambda_1$ and $\delta \Lambda_2$, as
\begin{eqnarray}
\delta_m A_\mu &=& \partial_\mu c\ \delta \Lambda_1 +\partial_\mu \bar c\ \delta 
\Lambda_2,\nonumber\\
\delta_m c&=& - B\ \delta \Lambda_2,\nonumber\\
\delta_m \bar c &=& B\ \delta \Lambda_1,\nonumber\\
\delta_m B&=& 0.
\end{eqnarray}
The FFMBRST symmetry transformation for this theory is then
constructed as
\begin{eqnarray}
\delta_m A_\mu &=& \partial_\mu c\ \Theta_b +\partial_\mu \bar c\ \Theta_{ab},\nonumber\\
\delta_m c&=& - B\ \Theta_{ab},\nonumber\\
\delta_m \bar c &=& B\ \Theta_b,\nonumber\\
\delta_m B&=& 0,\label{mfin}
\end{eqnarray}
where $\Theta_b$ and $\Theta_{ab}$ are finite, field dependent  and anticommuting parameters.
We choose particular    $\Theta_b$ and $\Theta_{ab}$ in this case as
\begin{equation}
\Theta_b=\int\Theta_b^\prime d\kappa =\gamma\int d\kappa\int d^4x [ \bar c \partial_\mu A^\mu], \label{mfi1}
\end{equation}
and \begin{equation}
\Theta_{ab}=\int\Theta_{ab}^\prime d\kappa =\gamma\int d\kappa\int d^4x[  c\partial_\mu A^\mu ],\label{mfib}
\end{equation}
where $\gamma$ is an arbitrary parameter. The infinitesimal change in Jacobian 
using Eq. (\ref{mjac}) for the FFMBRST transformation with the above finite parameters vanishes.   
It means that the path integral measure and hence the generating functional is invariant under
FFMBRST transformation.

Now, the infinitesimal change in Jacobian  for the FFBRST transformation with 
the  parameter $\Theta_b$  is calculated as 
\begin{equation}
\frac{1}{J_1}\frac{dJ_1}{d\kappa} =  \gamma\int d^4x \left[B\partial_\mu A^\mu +\bar c \partial_\mu\partial^\mu c
\right].
\end{equation}
To write the Jacobian $J_1$ as $e^{iS_1}$ in the BRST case, we make the following ansatz for $S_1$:
\begin{eqnarray}
S_1 =i\int d^4 x \left[\xi_1 B\partial_\mu A^\mu +\xi_2 \bar c \partial_\mu\partial^\mu c\right].\label{msn}
\end{eqnarray}
The essential condition in Eq. (\ref{mcondf}) is satisfied subjected to
\begin{eqnarray}
 \int d^4x
\left[ B\partial_\mu A^\mu (\xi_1' +\gamma) + \bar c \partial_\mu\partial^\mu c (\xi_2' +\gamma) 
-B\partial_\mu\partial^\mu c \Theta_b^\prime (\xi_1 -\xi_2) \right] =0,
\end{eqnarray}
where the prime denotes the derivative with respect to $\kappa$.
Equating the both sides of the above equation, we get the following: 
\begin{eqnarray}
\xi_1' +\gamma =0, \ \ \xi_2' +\gamma =0,\ \ \xi_1 -\xi_2 =0.
\end{eqnarray}
The solution of above equations satisfying the initial conditions $\xi_i =0, (i=1,2)$ is
\begin{eqnarray}
\xi_1 =-\gamma\kappa, \ \ \xi_2 =-\gamma\kappa.
\end{eqnarray}
Putting these value  in Eq. (\ref{msn}), the expression of $S_1$ becomes 
\begin{eqnarray}
S_1 =-i\gamma\kappa\int d^4 x \left[  B\partial_\mu A^\mu + \bar c \partial_\mu\partial^\mu c\right].\label{ms1}
\end{eqnarray}
However, the infinitesimal change in Jacobian $J_2$ for the FF-anti-BRST transformation with 
the parameter $\Theta_{ab}$  
is calculated as
\begin{equation}
\frac{1}{J_2}\frac{dJ_2}{d\kappa} = -\gamma\int d^4x \left[B\partial_\mu A^\mu +\bar c \partial_\mu\partial^\mu c
\right].\label{mj22}
\end{equation}
Similarly, to write the Jacobian $J_2$ as $e^{iS_2}$ in the anti-BRST case, we make the following ansatz for $S_2$:
\begin{eqnarray}
S_2 =i\int d^4 x \left[\xi_3 B\partial_\mu A^\mu +\xi_4 \bar c \partial_\mu\partial^\mu c\right].\label{ms22}  
\end{eqnarray}
The essential condition in Eq. (\ref{mcondf}) for Eqs. (\ref{mj22}) and (\ref{ms22})  provides
\begin{eqnarray}
 \int d^4x
\left[ B\partial_\mu A^\mu (\xi_3' -\gamma) + \bar c \partial_\mu\partial^\mu c (\xi_4' -\gamma) 
-B\partial_\mu\partial^\mu \bar c \Theta_{ab}^\prime (\xi_3 -\xi_4) \right] =0.
\end{eqnarray}
Comparing the L.H.S. and R.H.S. of the above equation we get the following equations:
\begin{equation}
\xi_3' -\gamma =0,\ \xi_4' -\gamma =0,\ \xi_3 -\xi_4 =0.
\end{equation}
Solving the above equations, we get the following values for the $\xi$'s: 
\begin{eqnarray}
\xi_3 = \gamma\kappa, \ \ \xi_4 = \gamma\kappa.
\end{eqnarray}
Plugging back these value of $\xi_i (i=3, 4)$ in Eq. (\ref{ms22}), we obtain   
\begin{eqnarray}
S_2 = i\gamma\kappa\int d^4 x \left[  B\partial_\mu A^\mu + \bar c \partial_\mu\partial^\mu c\right].\label{ms2}
\end{eqnarray}
From Eqs. (\ref{ms1}) and (\ref{ms2}), one can easily see that 
$S_1+S_2=0$.
Therefore, under successive FFBRST and FF-anti-BRST transformations with these particular finite parameters $\Theta_b$
and  $\Theta_{ab}$, respectively, the generating functional 
transformed
as 
\begin{eqnarray}
Z_M \left( = \int [{\cal D}\phi ]\ e^{iS_{eff}^M} \right)\stackrel{(FFBRST)(FF-anti-BRST)}{--------\longrightarrow }
Z_M\left(  =\int [{\cal D}\phi ]\ e^{iS_{eff}^M+S_1+S_2} \right).
\end{eqnarray}
Hence, the successive operations of FFBRST and FF-anti-BRST transformations also  leave the generating functional 
$Z_M$.  This reconfirms that FFMBRST transformation has same effect on both  the effective action and  the generating 
functional as the successive FFBRST and FF-anti-BRST transformations.  
\subsection{Non-Abelian YM theory in Curci-Ferrari-Delbourgo-Jarvis (CFDJ) gauge}
The examples studied so far were   Abelian gauge theories. We now consider an example with non-Abelian
gauge theory.  CFDJ gauge is a standard gauge-fixing which has been studied extensively in non-Abelian gauge theories
\cite{cfdj}.
The generating functional for non-Abelian YM theory in CFDJ gauge is written as
\begin{equation}
Z^{CF}_{YM } =\int [{\cal D}\phi ]\ e^{iS^{CF}_{YM }[\phi] },\label{mzy1}
\end{equation}
where $\phi$ is generic notation for all the fields in the effective action 
$S^{CF}_{YM }$  
\begin{eqnarray}
S^{CF}_{YM }&=&\int d^4x \left[ -\frac{1}{4}F_{\mu\nu}^aF^{\mu\nu a}+\frac{\xi}{2}(h^a)^2 
+ih^a
\partial_\mu A^{\mu a}+\frac{1}{2}\partial_\mu{\bar c}^a(D^\mu c)^a  \right.\nonumber\\
&+&\left. \frac{1}{2}(D_\mu\bar 
c)^a\partial^\mu c^a -\xi\frac{g^2}{8}
 (f^{abc}{\bar c}^b c^c)^2\right],
\end{eqnarray}
with the  field strength tensor $F_{\mu\nu}^a=\partial_\mu A_\nu^a -\partial_\nu A_\mu^a +gf^{abc}A_\mu^b A_\nu^c$ 
and 
$h^a$ is the Nakanishi-Lautrup type auxiliary field. 
 The effective action as well as the generating functional are invariant under the following infinitesimal BRST 
and anti-BRST transformations 

 \begin{eqnarray}
\mbox{BRST:}\ \ \ \delta_b A_\mu^a&=& -(D_\mu c)^a \ \delta\Lambda_1, \ \ \ \delta_b c^a=-\frac{g}{2}f^{abc}
{ c}^b c^c \ \delta\Lambda_1
 \nonumber\\
\delta_b {\bar c}^a &=& \left(ih^a -\frac{g}{2}f^{abc}{\bar c}^b c^c\right)\ \delta
\Lambda_1\nonumber\\
\delta_b (ih^a)&=& -\frac{g}{2}f^{abc}\left( ih^b c^c  +\frac{g}{4} f^{cde}{\bar c}^b c^d c^e\right)
\ \delta\Lambda_1,\nonumber\\
\mbox{anti-BRST:}\ \ \ \delta_{ab} A_\mu^a&=& -(D_\mu \bar c)^a \ \delta\Lambda_2, \ \ \ \delta_{ab} {\bar c}^a=
-\frac{g}{2}f^{abc}{\bar c}^b {\bar c}^c \ \delta\Lambda_2
 \nonumber\\
\delta_{ab} { c}^a &=& \left(-ih^a -\frac{g}{2}f^{abc}{\bar c}^b c^c\right)\ 
\delta\Lambda_2\nonumber\\
\delta_{ab} (ih^a)&=& -\frac{g}{2}f^{abc}\left( ih^b {\bar c}^c  +\frac{g}{4} f^{cde}{ c}^b {\bar 
c}^d {\bar c}^e\right)
\ \delta\Lambda_2,
\end{eqnarray}
where $\delta\Lambda_1$ and $\delta\Lambda_2$ are infinitesimal, anticommuting and global parameters.
The infinitesimal MBRST symmetry transformation ($\delta_m= \delta_b +\delta_{ab}$) in this case is written as:
\begin{eqnarray}
\delta_m A_\mu^a&=& -D_\mu c^a\ \delta\Lambda_1 - D_\mu \bar c^a \ \delta\Lambda_2,\nonumber\\
  \delta_m
c^a&=&  -\frac{g}{2}f^{abc}{ c}^b c^c \ 
\delta\Lambda_1 -  \left(ih^a +\frac{g}{2}f^{abc}{\bar c}^b c^c\right)\ 
\delta\Lambda_2
 \nonumber\\
\delta_m {\bar c}^a &=&  \left(ih^a -\frac{g}{2}f^{abc}{\bar c}^b c^c\right)\ \delta
\Lambda_1  
-\frac{g}{2}f^{abc}{\bar c}^b {\bar c}^c \ \delta\Lambda_2\nonumber\\
\delta_m (ih^a)&=& -\frac{g}{2}f^{abc}\left( ih^b c^c  +\frac{g}{4} f^{cde}{\bar c}^b c^d c^e
\right)
\ \delta\Lambda_1 \nonumber\\
&-&\frac{g}{2}f^{abc}\left( ih^b {\bar c}^c  +\frac{g}{4} f^{cde}{ c}^b {
\bar 
c}^d {\bar c}^e\right)
\ \delta\Lambda_2. 
\end{eqnarray} 
Corresponding FFMBRST symmetry transformation 
is constructed as 
\begin{eqnarray}
\delta_m A_\mu^a &=& -D_\mu c^a\ \Theta_b - D_\mu \bar c^a \ \Theta_{ab}, \nonumber\\
 \delta_m c^a&= & -\frac
{g}{2}f^{abc}{ c}^b c^c \ 
\Theta_b -  \left(ih^a +\frac{g}{2}f^{abc}{\bar c}^b c^c\right)\ 
\Theta_{ab}
 \nonumber\\
\delta_m {\bar c}^a &=&  \left(ih^a -\frac{g}{2}f^{abc}{\bar c}^b c^c\right)\
\Theta_b  
-\frac{g}{2}f^{abc}{\bar c}^b {\bar c}^c \ \Theta_{ab}\nonumber\\
\delta_m (ih^a)&=& -\frac{g}{2}f^{abc}\left( ih^b c^c  +\frac{g}{4} f^{cde}{\bar c}^b c^d c^e
\right)
\ \Theta_b \nonumber\\
&-&\frac{g}{2}f^{abc}\left( ih^b {\bar c}^c  +\frac{g}{4} f^{cde}{ c}^b {\bar 
c}^d {\bar c}^e\right)
\ \Theta_{ab},\label{mfini}
\end{eqnarray}
with two arbitrary finite field dependent parameters  $\Theta_b$ and $\Theta_{ab}$.
The generating functional $Z^{CF}_{YM }$ is made invariant under the above FFMBRST
transformation  by constructing appropriate finite parameters  $\Theta_b$ and $\Theta_{ab}$.
We construct  the  finite nilpotent  parameters $\Theta_b$ and $\Theta_{ab}$ 
 as
\begin{equation}
\Theta_b =\int \Theta_b^\prime d\kappa =\gamma \int d\kappa \int d^4x[\bar c^a\partial_\mu A^{\mu a}], \label{mfi2}
\end{equation}
 \begin{equation}
\Theta_{ab} =\int \Theta_{ab}^\prime d\kappa =\gamma \int d\kappa \int d^4 x[  c^a\partial_\mu A^{\mu a}], \label{mfic}
\end{equation} 
where $\gamma$ is an arbitrary parameter.
Following the same method elaborated in the previous two examples, we show that the Jacobian for path  integral 
measure due to FFMBRST transformation given in Eq. (\ref{mfini}) 
with finite parameters  $\Theta_b$ and $\Theta_{ab}$ becomes unit.
It means that under such FFMBRST transformation the generating functional as well as effective action remain
invariant. The Jacobian contribution for path integral measure due to FFBRST 
transformation with parameter $\Theta_b$ compensates 
the same due to FF-anti-BRST transformation with parameter $\Theta_{ab}$.
Therefore, under the successive FFBRST and FF-anti-BRST transformations the 
generating functional remains invariant as
\begin{eqnarray}
Z_{YM}^{CF}\left(= \int [{\cal D}\phi ]\ e^{iS_{YM}^{CF}} \right) \stackrel{(FFBRST)(FF-anti-BRST)}{---------\longrightarrow }
Z_{YM}^{CF}.
\end{eqnarray}
Again we see the equivalence between FFMBRST transformation  and successive operations of FFBRST and FF-anti-BRST transformations.
  
We end up the section with conclusion that in all the three cases the FFMBRST transformation with appropriate
finite parameters is the finite nilpotent symmetry of the effective action as well as the 
generating functional of the effective theories. Here, we also note that the successive operations of FFBRST and
FF-anti-BRST  also leave the generating functional  as well as effective action invariant and hence equivalent 
to FFMBRST transformation. 
\section{FFMBRST symmetry in BV formulation} 
In this section, we consider the BV formulation using MBRST transformation.
Unlike BV formulation using either BRST or anti-BRST transformation, we need two sets of
antifields in BV formulation for MBRST transformation. We 
construct FFMBRST transformation in this context.
The change in Jacobian under FFBRST transformation in the path integral measure in the definition of  generating 
functional 
is used to adjust with the change in the gauge-fixing fermion $\Psi_1$ \cite{sudbpm}.  Hence,
 the FFBRST transformation
is used to connect the 
generating functionals of different solutions of quantum master equation \cite{ sm1}. However, in case of
 BV formulation for FFMBRST transformation we need to introduce two gauge-fixing fermions $\Psi_1$ and $\Psi_2$. We
construct the finite parameters in FFMBRST transformation in such a way that contributions from
 $\Psi_1$ and $\Psi_2$ adjust each other  to leave the extended action invariant. This implies that we can construct 
appropriate parameters in FFMBRST transformation such that generating functionals corresponding to different 
solutions of quantum master equations remain invariant under such transformation. 
These results can be demonstrated with the help of 
explicit examples. We would like to consider the same examples of previous section for this purpose. 
\subsection{Bosonized chiral model in BV formulation}
We recast the generating functional for (1+1) dimensional bosonized chiral model given in Eq. (\ref{mz}) 
using both BRST and anti-BRST exact terms as
\begin{eqnarray}  
Z_{CB}&=&\int D\phi\ e^{iS_{CB}} 
  =  \int D\phi\ \exp \left[i\int d^2 x\left\{\pi_\varphi\dot\varphi +
\pi_\vartheta\dot\vartheta
+p_u\dot u -\frac{1}{2}\pi_\varphi^2 \right.\right.\nonumber\\
&+& \left.\left.\frac{1}{2}\pi_\vartheta^2 
+\pi_\vartheta (\varphi'-\vartheta' +\lambda ) +\frac{1}{2}\pi_\varphi \lambda  +
\frac{1}{2}s_b \Psi_1 +\frac{1}{2}s_{ab}\Psi_2\right\}\right].\label{mac}
\end{eqnarray}
Here, the Lagrange multiplier field $u$ is considered as dynamical variable and 
expression for gauge-fixing fermions for BRST symmetry ($\Psi_1 $) and   anti-BRST symmetry ($\Psi_2$), 
respectively, are
\begin{eqnarray}
\Psi_1 &=&\int d^2x\ \bar c(\dot\lambda-\varphi-\vartheta +\frac{1}{2} B).\\
\Psi_2 &=&\int d^2x\  c (\dot\lambda-\varphi-\vartheta +\frac{1}{2} B).
\end{eqnarray}
The effective  action $S_{CB}$ is invariant under combined BRST and anti-BRST 
transformations given in Eq. (\ref{mb}).  
The   generating functional $Z_{CB}$ can be written in terms of antifields  $\phi_1^\star$ and $\phi_2^\star$ 
corresponding to
all fields $\phi$ as
\begin{eqnarray} 
Z_{CB}&=& \int D\phi\ \exp \left[i\int d^2 x\left\{\pi_\varphi\dot\varphi +\pi_\vartheta\dot\vartheta
+p_u\dot u -\frac{1}{2}\pi_\varphi^2 \right.\right.\nonumber\\
&+& \left.\left.\frac{1}{2}\pi_\vartheta^2 +\pi_\vartheta (\varphi'-\vartheta' +\lambda ) 
+\pi_\varphi \lambda +\frac{1}{2}
\varphi_1^\star
c  -\frac{1}{2}
\varphi_2^\star
 \bar c  \right.\right.\nonumber\\
&+&\left.\left. \frac{1}{2}
\vartheta_1^\star   c  -\frac{1}{2} \vartheta_2^\star
\bar c  +\frac{1}{2}\bar c_1^\star  B +\frac{1}{2}c_2^\star B 
- \frac{1}{2}\lambda_1^\star 
\dot c+\frac{1}{2} \lambda_2^\star 
\dot {\bar c}\right\}\right],
\end{eqnarray}
where $\phi_i^\star (i=1,2)$ is a generic notation for antifields arising from gauge-fixing fermions 
$\Psi_i$.
The above relation can further be written in compact form as 
\begin{equation}
Z_{CB} = \int [{\cal D}\phi ]\  e^{ iW_{\Psi_1 +\Psi_2}[\phi,\phi_i^\star ] },\label{mcq}
\end{equation}
where $W_{\Psi_1 +\Psi_2} [\phi,\phi_i^\star ]$ is an extended action for the theory of self-dual chiral 
boson corresponding the gauge-fixing fermions $\Psi_1$ and $\Psi_2$. 

This extended quantum action, $W_{\Psi_1 +\Psi_2} [\phi,\phi_i^\star ]$ satisfies certain rich 
mathematical
 relations commonly known as quantum master equation \cite{wei}, given by
\begin{equation}
\Delta e^{iW_{\Psi_1 +\Psi_2}[\phi,\phi_i^\star ]} =0  \ \mbox{ with }\ 
 \Delta\equiv \frac{\partial_r}{
\partial\phi}\frac{\partial_r}{\partial\phi_i^\star } (-1)^{\epsilon
+1}. 
\label{mq}
\end{equation}
The generating functional does not depend on the choice of gauge-fixing fermions \cite{ht} and therefore  
extended quantum action $W_{\Psi_i}$  with all possible $\Psi_i$  are the different 
solutions of quantum master equation.
The antifields $\phi^\star_1$ corresponding to each field $\phi$ for this particular theory 
can 
be obtained from the gauge-fixed fermion $\Psi_1$ as
\begin{eqnarray}
\varphi_1^\star &=& \frac{\delta \Psi_1}{\delta \varphi}= - \bar c,
\ \ \vartheta_1^\star = \frac{\delta \Psi_1}{\delta \vartheta}= - \bar c,
{ c}_1^\star =\frac{\delta \Psi_1}{\delta c}=0,\nonumber\\
 { \bar c}_1^\star &=&  \frac{\delta 
\Psi_1}{\delta  \bar c}=-\frac{1}{2}(
\dot\lambda -\varphi- \vartheta +\frac{1}{2}B),\nonumber\\
B_1^\star &=& \frac{\delta \Psi_1}{\delta B}= \frac{1}{2}\bar c,\ \ \lambda_1^\star =-\frac{
\delta \Psi_1}{\delta \lambda}= - \dot{\bar c}.
\end{eqnarray}
Similarly, the  antifields $\phi^\star_2$ can be
  calculated from the gauge-fixing fermion $\Psi_2$ as
\begin{eqnarray}
\varphi_2^\star &=& \frac{\delta \Psi_2}{\delta \varphi}= -  c,
\ \ \vartheta_2^\star = \frac{\delta \Psi_2}{\delta \vartheta}= -  c,\nonumber\\
{ c}_2^\star &=& \frac{\delta \Psi_2}{\delta  c}= (
\dot\lambda -\varphi- \vartheta +\frac{1}{2}B),\ \ {\bar c}_2^\star = \frac{\delta \Psi_2}{
\delta \bar c}=0,\nonumber\\
B_2^\star &=& \frac{\delta \Psi_2}{\delta B}=\frac{1}{2}   c,\ \ \lambda_2^\star =-\frac{
\delta \Psi_2}{\delta \lambda}= -  \dot {  c}.
\end{eqnarray}
Now, we apply the FFMBRST transformation given in Eq. (\ref{mc}) with the finite parameters written in Eqs. (\ref
{mfi}) 
and (\ref{mfia}) to this generating functional.
We see that the path integral measure in Eq. (\ref{mcq}) remains invariant under this FFMBRST transformation
as the Jacobian for path integral measure is $1$.
Therefore,
\begin{eqnarray}
Z_{CB}\left( = \int [{\cal D}\phi ]\ e^{i W_{\Psi_1 +\Psi_2}}\right)  \stackrel{(FFBRST)(FF-anti-BRST)}{ ---------
\longrightarrow}  
 Z_{CB}. 
    \end{eqnarray}
Thus, the solutions of quantum master equation in this model remain invariant under FFMBRST
 transformation  as well as under consecutive operation of FFBRST and FF-anti-BRST transformations. 
However,  the FFBRST (FF-anti-BRST) transformation connects the generating functionals corresponding to
the different solutions of the quantum master equation \cite{subm}.
\subsection{Maxwell's theory in BV formulation}
The generating functional for Maxwell's theory given in Eq. (\ref{mzy}) can be  recast 
using BRST and anti-BRST exact terms as  
\begin{eqnarray}
Z_{M}= \int [{\cal D}\phi ]\  e^{ i\int d^4 x\left[ -\frac{1}{4}F_{\mu\nu}F^{\mu\nu}+ \frac{1}{2}s_b \Psi_1 +
\frac{1}{2}s_{ab} \Psi_2 \right]},
\end{eqnarray}
where the  expressions for gauge-fixing fermions $\Psi_1 $ and $\Psi_2 $ are
\begin{eqnarray}
\Psi_1 &=& \int d^4x\ \bar c(\lambda B-\partial\cdot A),\nonumber\\ 
 \Psi_2 &=&-\int d^4x\ c (\lambda B-\partial\cdot A).
\end{eqnarray}
The generating functional for such theory can  further be expressed 
in fields/antifields formulation as 
\begin{eqnarray}
Z_{M}= \int [{\cal D}\phi ]\  e^{ i\int d^4 x\left[  -\frac{1}{4}F_{\mu\nu}F^{\mu\nu}+\frac{1}{2}A_{ \mu 1}^\star 
 \partial^\mu c +\frac{1}{2} A_{\mu 2}^\star\partial^\mu \bar c +\frac{1}{2}{\bar c}_1^\star B-\frac{1}{2}
{ c}_2^\star B  \right]}.
\end{eqnarray} 
In the compact form above generating functional is written as
\begin{equation}
Z_{M} = \int [{\cal D}\phi ] e^{ iW_{\Psi_1+\Psi_2} [\phi,\phi_i^\star ]},
\end{equation}
where $W_{\Psi_1 +\Psi_2}[\phi,\phi_i^\star ]$ is an extended action for  Maxwell's theory corresponding to
 the gauge-fixing fermions $\Psi_1 $ and $\Psi_2$.

The antifields for gauge-fixed fermion $\Psi_1$ are calculated as
\begin{eqnarray}
A_{ \mu 1}^{\star} &=&\frac{\delta \Psi_1}{\delta A^\mu}= 
\partial_\mu {\bar c}, \ \ \bar c_1^{\star} =\frac{\delta \Psi_1}{\delta \bar c}=
(\lambda B-\partial\cdot A),\nonumber\\
{ c}_1^{\star} &=&\frac{\delta \Psi_1}{\delta {c}}=0,
\ \ B_1^\star =\frac{\delta \Psi_1}{\delta B}=\lambda  \bar c.
\end{eqnarray}
The antifields $\phi_2^\star$ can 
be calculated from the gauge-fixed fermion $\Psi_2$ as 
\begin{eqnarray}
A_{ \mu 2}^{\star} &=&\frac{\delta \Psi_2}{\delta A^\mu}= - 
\partial_\mu { c}, \ \ \bar c_2^{\star} =\frac{\delta \Psi_2}{\delta \bar c}=0,\nonumber\\
{ c}_2^{\star} &=&\frac{\delta \Psi_2}{\delta {c}}=-(\lambda B-\partial\cdot A),
\  B_2^\star =\frac{\delta \Psi_2}{\delta B}= -\lambda c.
\end{eqnarray}
 Now, implementing the FFMBRST transformation mentioned in Eq. (\ref{mfin}) with parameters given in Eqs. (\ref{mfi1})
and (\ref{mfib}) to this generating functional  we see that the Jacobian  for the path integral
measure for such transformation becomes unit. Hence, the FFMBRST transformation given in Eq. (\ref{mfin}) is a finite 
symmetry of the solutions of quantum master equation for Maxwell's theory.

Now, we focus on the contributions arising from the Jacobian due to independent applications of FFBRST and 
FF-anti-BRST transformations. We construct the finite parameters of FFBRST and FF-anti-BRST transformations
in such a way that Jacobian remains invariant.
Therefore, the generating functional remains invariant under consecutive operations of FFBRST
and FF-anti-BRST transformations with appropriate parameters as 
\begin{eqnarray}
Z_{M}\left( = \int [{\cal D}\phi ]\ e^{i W_{\Psi_1 +\Psi_2}}\right)\stackrel{(FFBRST)(FF-anti-BRST)}{---------\longrightarrow} 
 Z_{M}.
    \end{eqnarray}
It also implies that the effect of consecutive FFBRST and FF-anti-BRST transformations is same as the effect 
of FFMBRST transformation on $Z_M$. 
\subsection{Non-Abelian YM theory in BV formulation}
  The generating functional for this theory can be written 
in both BRST and anti-BRST exact terms as  
\begin{eqnarray}
Z^{CF}_{YM } =  \int [{\cal D}\phi ]\ e^{i\int d^4 x\left[ -\frac{1}{4}F_{\mu\nu}^a
F^{a\mu\nu}+ \frac{1}{2}s_b \Psi_1 +\frac{1}{2}s_{ab} \Psi_2 \right ]},
\end{eqnarray}
with the expressions of gauge-fixing fermions  $\Psi_1$ and $\Psi_2$ as
\begin{eqnarray}
\Psi_1 &=&-\int d^4x\ \bar c^a(i\frac{\xi}{2}h^a-\partial\cdot A^a),\nonumber\\
\Psi_2 &=&\int d^4x\ c^a (i\frac{\xi}{2}h^a-\partial\cdot A^a).
\end{eqnarray}
We re-write the generating functional given in Eq. (\ref{mzy1}) using field/antifield formulation   as,
\begin{eqnarray}   
 Z^{CF}_{YM } &=&  \int [D\phi ]\ \exp \left[i\int d^4 x\left\{- \frac{1}{4}F_{\mu\nu}^a
F^{a\mu\nu}-\frac{1}{2}A_{ \mu 2}^{a\star} 
D^\mu c^a
-\frac{1}{2} A_{ \mu 2}^{a\star}D^\mu \bar c^a \right.\right.\nonumber\\
&+&\left.\left.\frac{1}{2}{\bar c}_1^{a\star} \left(
ih^a -\frac{g}{2}f^{abc}\bar c^b c^c\right) - \frac{1}{2}{ c}_2^{a\star} \left(
ih^a +\frac{g}{2}f^{abc}\bar c^b c^c\right)
-\frac{1}{2} 
h_1^{a\star}\left( \frac{g}{2}f^{abc}h^b c^c \right.\right.\right.\nonumber\\
&- &\left.\left.\left. i \frac{g^2}{8}f^{abc}f^{cde}\bar c^b{  c}^d{c}^e \right)
-
\frac{1}{2} 
h_2^{a\star}\left( \frac{g}{2}f^{abc}h^b \bar c^c - i \frac{g^2}{8}f^{abc}f^{cde} c^b{\bar c}^d{\bar c}^e 
\right)\right\}\right].
\end{eqnarray}
This generating functional $Z^{CF}_{YM }$ can be written compactly as
\begin{equation}
Z^{CF}_{YM } = \int [D\phi ]\ e^{  iW_{\Psi_1+\Psi_2}[\phi,\phi_i^\star ]},
\end{equation}
where $W_{\Psi_1+\Psi_2}[\phi,\phi_i^\star  ]$ is an extended quantum action for the non-Abelian YM theory 
in CFDJ gauge.

The antifields   are calculated with the help of gauge-fixed fermion $\Psi_1$ as
\begin{eqnarray} 
A_{ \mu 1}^{a\star} &=&\frac{\delta \Psi_1}{\delta A^{a\mu}}=-\partial_\mu {\bar c}^a,
\ \ {  c}_1^{a\star}  = \frac{\delta \Psi_1}{\delta {  c^a}}=0,\nonumber\\
 \bar c_1^{a\star} &=&\frac{\delta \Psi_1}{\delta \bar c^a}=- 
(i\frac{\xi}{2}h^a-\partial\cdot A^a),\nonumber\\
  h_1^{a\star} &=&\frac{\delta \Psi_1}{\delta h^a}=
-\frac{i}{2}\xi \bar c^a.
\end{eqnarray}
The explicit value of antifields  can be calculated with   $\Psi_2$ as  
\begin{eqnarray}
A_{ \mu 2}^{a\star} &=&\frac{\delta \Psi_2}{\delta A^{a\mu}} =\partial_\mu { 
c^a}, \ \ \bar c_2^{a\star} =\frac{\delta \Psi_2}{\delta \bar c^a}=0,\nonumber\\
{  c}_2^{a\star} &=&\frac{\delta \Psi_2}{\delta { c^a}}= 
(i\frac{\xi}{2}h^a-\partial\cdot A^a),\nonumber\\
 h_2^{a\star} &=&\frac{\delta \Psi_2}{\delta h^a}=
\frac{i}{2}\xi {  c}^a.
\end{eqnarray}
We observe here again that the Jacobian for path integral measure in the expression of generating functional 
$Z_{YM}^{CF}$ arising due to FFMBRST transformation and 
 due to successive operations of FFBRST and FF-anti-BRST transformations remains unit
for appropriate choice of finite parameters. 
Thus, the consequence of FFMBRST transformation given in Eq. (\ref{mfini}) with the finite parameters given in Eqs. 
(\ref{mfi2}) and (\ref{mfic}) is equivalent to the subsequent operations of FFBRST and FF-anti-BRST transformations
with same finite parameters.
\section{Conclusions}
FFBRST and FF-anti-BRST transformations are nilpotent symmetries of the effective action. However, these 
transformations do not leave the generating functional invariant as the path integral measure 
changes in a nontrivial way under these transformations. 
We have constructed infinitesimal MBRST transformation which is the combination of 
infinitesimal BRST and anti-BRST transformations.
Even though infinitesimal MBRST transformation does not play much significant role, its finite version
has very important  consequences. We have shown that it is possible to construct the finite field dependent
 MBRST (FFMBRST) transformation  which leaves the effective action as well as
the generating functional invariant. The finite parameters in the FFMBRST transformation have been chosen in such a 
way that the Jacobian contribution from the FFBRST part compensates the same arising from FF-anti-BRST part.
We have considered several explicit examples with diverse character in both gauge theories
as well as in field/antifield formulation to show these results.
It is interesting to point out that the effect of FFMBRST transformation is equivalent to successive operations of
FFBRST and FF-anti-BRST transformations. We have further shown that the generating functionals 
corresponding to different solutions of quantum
master equation remain invariant under such FFMBRST transformation whereas the independent FFBRST and FF-anti-BRST
 transformations  connect the generating functionals corresponding to
the different solutions of the quantum master equation. It will be interesting to see whether this 
FFMBRST transformation puts further restrictions on the relation of different Green's functions of the
theory to simplify the renormalization program. In particular, such FFMBRST transformation may be helpful 
for the theories where BRST and anti-BRST transformations play independent role.

\label{Chap:Chapter6}

\chapter{FFBRST transformation and constrained systems}
This chapter is devoted to study the different class of constraints theories in the framework of FFBRST
transformation.  
Here we develop the FFBRST and FF-anti-BRST transformations for  first-class theories \cite{aop}. 
Then we show that such generalization helps to connect first-class theories to second-class theories.
The results are established with the help of two explicit examples. 
\section{The theories with constraints: examples }
In the various field theories, all the dynamical phase space variables are not independent rather
some of the variables satisfy the constraints emerging from the structure of the theories. In other words,
the  relations between various dynamical variables are known as constraints of the theories. The usual
Poisson brackets may not represent the true brackets as they need not to satisfy the constraints
of the theories in such constrained systems.  A system is said to be first-class constrained system if
all the Poisson brackets among the constraints vanish weakly. On the other hand, if there exists at least
one non-zero  Poisson bracket among the constraints then the theory is called as second-class. 
In this section, we briefly outline the essential features of second-class and
first-class theories. In particular, we discuss the Proca theory for 
massive spin 1 vector field theory and gauge 
variant theory for the self-dual chiral boson,  which 
are second-class theories. Corresponding first-class theories i.e. the Stueckelberg 
theory for massive spin 1 vector fields  and gauge 
invariant theory for the self-dual chiral boson are also outlined in this section.
\subsection{ Theory for massive spin 1 vector field }
\subsubsection{Proca model}
We start with the action for a massive charge neutral spin 1
vector field $A_\mu$ in 4D  
\begin{equation}
S_P=\int d^4x\ {\cal L}_P,
\end{equation}
where the Lagrangian density is given as
\begin{equation}
{\cal {L}}_{P}=-\frac{1}{4} {F_{\mu\nu}}F^{\mu\nu} +\frac{M^2}{2}{A_\mu}A^\mu.\label{lag}
\end{equation}
The field strength tensor is defined as 
$
F_{\mu\nu}=\partial_\mu A_\nu -\partial_\nu A_\mu.
$
We recall the convention for $g^{\mu\nu}$,  i.e., $g^{\mu\nu}=$ diagonal $(1,-1,-1,-1)$, where $\mu, \nu =0, 1, 
2, 3$.
The canonically conjugate momenta for $A_\mu$ field is 
\begin{equation}
\Pi^\mu=\frac{\partial {\cal L}}{\partial \dot{A_\mu}}=F^{\mu 0}.
\end{equation}
This implies 
that the primary constraint of the theory is
\begin{equation}
\Omega_1\equiv \Pi^0 \approx 0.
\end{equation}
The Hamiltonian density of the theory is given by
\begin{eqnarray}
{\cal H}=\Pi_\mu\dot{A^\mu} -{\cal L} =\Pi_i\partial^i A^0-\frac{1}{2}\Pi_i^2+\frac{1}{2}F_{ij}F^{ij}-\frac
{1}{2}M^2A_\mu A^\mu.
\end{eqnarray}
The time evolution for the dynamical variable $\Pi^0$ can be written as
\begin{equation}
\dot{\Pi^0}=[\Pi^0, { H}], \label{pc}
\end{equation}
where the Hamiltonian $H=\int d^3 x\ {\cal H}$.
The constraints of the theory should be invariant under time evolution and using
(\ref{pc}) we obtain the secondary constraint
\begin{equation}
\Omega_2\equiv [\Pi^0, { H}]=\partial_i\Pi^i +M^2 A^0\approx 0.
\end{equation}
The constraint $\Omega_2$ contains $A^0$ which implies that
$
[\Omega_1, \Omega_2 ]\not=0.
$
Hence, the Proca theory for massive spin 1 vector field is endowed with the second-class 
constraint.

The propagator for this
theory can be written  in a simple
manner as
\begin{equation}
iG_{\mu\nu}(p)=-\frac{i}{p^2-M^2}\left(\eta_{\mu\nu}-\frac{p_\mu p_\nu}{M^2}\right).
\end{equation}
Note that the propagator in this theory does not fall  rapidly    
for large values of the momenta. This leads to difficulties in establishing
renormalizability of the (interacting) Proca theory for massive
photons.
Hence the limit $M \rightarrow  0$ of the Proca theory is clearly difficult to perceive.

The generating functional for the Proca theory is defined as 
\begin{equation}
Z_{P}\equiv \int [{\cal D}A_\mu]\ e^{iS_{P}}.
\end{equation}
\subsubsection{Stueckelberg theory}
To remove the difficulties in the Proca model, Stueckelberg considered the following
generalized action
 \begin{equation} 
S_{ST}=\int d^4x \left[-\frac{1}{4} {F_{\mu\nu}}F^{\mu\nu} +\frac{M^2}{2}\left(A_\mu -
\frac{1}{M}
\partial_\mu B\right)^2
\right],
\end{equation}
by introducing a real scalar field $B$.

This action is invariant under the following gauge transformation:
\begin{eqnarray}
A_\mu (x)\rightarrow A_\mu^\prime (x)&=& A_\mu (x)+\partial_\mu \lambda(x),\\
B(x)\rightarrow B^\prime (x)&=&B(x)+M\lambda (x),
\end{eqnarray}
where $\lambda$ is the gauge parameter. For the quantization of such theory one has to choose 
a gauge condition. By choosing the 't Hooft gauge condition,
${\cal L}_{gf}=-\frac{1}{2\chi}(\partial^\mu A_\mu +\chi MB)^2 $  where $\chi$ is any 
arbitrary gauge parameter, it is easy to see that 
 the propagators 
are well behaved at high momentum. As a result, there is no difficulty
in establishing renormalizability for such theory.
Now we turn to the BRST symmetry for the Stueckelberg theory. Introducing a 
ghost $(c)$ and an antighost field $(\bar c )$ the effective Stueckelberg action 
can be written as
\begin{eqnarray}
S_{ST }&=&\int d^4x \left[-\frac{1}{4} {F_{\mu\nu}}F^{\mu\nu} +\frac{M^2}{2}\left(A_\mu -
\frac{1}{M} \partial_\mu B\right)^2\right.\nonumber\\
&-&\left.\frac{1}{2\chi}(\partial^\mu A_\mu +\chi MB)^2 -\bar c
 (\partial^2  +
\chi M^2)c\right].\label{stl}
\end{eqnarray}
 This
action is invariant under the following on-shell BRST transformation:
\begin{eqnarray}
\delta_b  A_\mu &=&\partial_\mu c\ \ \Lambda,\ \
\delta_b  B  =  Mc\ \ \Lambda,\nonumber\\
\delta_b  c &=& 0,\ \
\delta_b \bar c = -\frac{1}{\chi}(\partial_\mu A^\mu +\chi MB)\ \ \Lambda,
\label{sym}
\end{eqnarray}
where $\Lambda$ is an infinitesimal, anticommuting and global 
parameter.
The generating functional for the Stueckelberg theory is defined as
\begin{equation}
Z_{ST }\equiv \int [{\cal D}\phi ]\ e^{iS_{ST }[\phi]},\label{zfun}
\end{equation} 
where $\phi$ is the generic notation for all fields involved in the 
theory. All the Green functions in this theory can be obtain from $Z_{ST }$.
\subsection{Theory for self-dual chiral boson}
A self-dual chiral boson can be described by the gauge variant as well as the gauge  invariant 
model. The purpose of this section is to introduce such models for a self-dual chiral boson. 
\subsubsection{Gauge  variant theory for self-dual chiral boson}
We start with the gauge  variant model \cite{pp} in 2D for
a single self-dual chiral boson.
The effective action for such a theory is given as 
\begin{equation} 
S_{CB}=\int d^2x\ {\cal L}_{CB} =\int d^2x\left[\frac{1}{2}\dot\varphi^2 -\frac{1}{2}{
\varphi' }^2 +\lambda ( \dot\varphi -\varphi' )\right],
\label{chintu}
\end{equation} 
where over dot and prime 
denote time and space derivatives respectively and $\lambda$ is the Lagrange multiplier.
The field $\varphi$ satisfies the self-duality condition $\dot\varphi =\varphi'$ 
in this case.  
We choose the Lorentz metric $g^{\mu\nu}=(1,-1)$ with  $\mu, \nu =0, 1$. The associated 
momenta for the field  $\varphi$ and the Lagrange multiplier are calculated as
\begin{eqnarray}
\pi_\varphi  = \frac{\partial {\cal L}_{CB}}{\partial \dot \varphi}=\dot\varphi +
\lambda,\ \
\pi_\lambda  = \frac{\partial {\cal L}_{CB}}{\partial \dot \lambda}=0,
\end{eqnarray}
which show that the model has the following primary constraint:
$
\Omega_1\equiv \pi_\lambda \approx 0.
$
The   Hamiltonian density corresponding to the above Lagrangian density
 ${\cal L}_{CB}$ 
in Eq. (\ref{chintu}) is
\begin{eqnarray}
{\cal H}_{CB}  = \pi_\varphi\dot\phi +\pi_\lambda\dot\lambda -{\cal L}_{CB}
 =  \frac{1}{2}(\pi_\varphi -\lambda )^2 +\frac{1}{2}\varphi'^2 +\lambda \varphi'.
\end{eqnarray}
Further, we can write the total Hamiltonian density corresponding to ${\cal L}_{CB}$ by 
introducing the Lagrange multiplier field $\eta$ for the primary constraint $\Omega_1$
 as
\begin{eqnarray}
{\cal H}_{CB}^T &=& \frac{1}{2}(\pi_\varphi -\lambda )^2 +\frac{1}{2}\varphi'^2 +\lambda 
\varphi' +
\eta\Omega_1,\nonumber\\
&=&\frac{1}{2}(\pi_\varphi -\lambda )^2 +\frac{1}{2}\varphi'^2 +\lambda 
\varphi' +
\eta\pi_\lambda.
\end{eqnarray}
Following the Dirac prescription \cite{dir}, we obtain the secondary constraint in this 
case as
\begin{eqnarray}
\Omega_2 & \equiv & \dot \pi_\lambda =[ \pi_\lambda, {\cal H}_{CB}] 
 = \pi_\varphi -
\lambda-\varphi'\approx 0.
\end{eqnarray}
The constraints $\Omega_1$ and $\Omega_2$ are of second-class as 
$[\Omega_1, \Omega_2] \not=0 $.
 This is an essential feature of a gauge variant theory.

 This model is quantized by establishing the following commutation relations \cite{pp}
\begin{eqnarray}
[\varphi(x), \pi_\varphi (y)]& =&[\varphi(x), \lambda(y) ]=+i\delta (x-y),\\
2[\lambda(x), \pi_\varphi(y)] &=&[\lambda(x), \lambda(y) ]=-2i\delta' (x-y),
\end{eqnarray}
where prime denotes the space derivative.
The rest of the commutator vanishes. 

The generating functional for the gauge  variant theory for a self-dual chiral boson is 
defined as
\begin{equation}
Z_{CB} =\int [{\cal D}\phi ]\ e^{iS_{CB}},
\end{equation}
where $D\phi$ is the path integral measure and $S_{CB}$ is the effective action
for a self-dual chiral boson. 
\subsubsection{Gauge invariant theory for self-dual chiral boson}
To construct a gauge invariant theory corresponding to the gauge non-invariant model
for chiral bosons, one generally introduces the WZ  term  in the 
Lagrangian density
${\cal L}_{CB}$.
For this purpose we need to enlarge the Hilbert space of the theory by introducing a new 
quantum field 
$\vartheta$, called the WZ field, through the redefinition of
 fields $\varphi$ and $\lambda$  as follows \cite{wz}:
$
\varphi\rightarrow \varphi -\vartheta, \ \ \lambda\rightarrow \lambda +\dot\vartheta
$.

 With this  redefinition of fields the modified Lagrangian density becomes
\begin{eqnarray} 
{\cal L}_{CB}^I={\cal L}_{CB} +{\cal L}_{CB}^{WZ},\label{wz}
\end{eqnarray}
where the WZ term
\begin{equation}
{\cal L}_{CB}^{WZ}=-\frac{1}{2}\dot\vartheta^2 -\frac{1}{2}{\vartheta'}^2  +
\varphi'\vartheta' +\dot\vartheta\vartheta' -\dot\vartheta\varphi' 
-\lambda (\dot\vartheta -\vartheta').
\end{equation}
The above Lagrangian density in Eq. (\ref{wz})  is invariant under 
time-dependent chiral gauge transformation:
\begin{eqnarray}
\delta\varphi &=&\mu (x, t),\ \ \delta\vartheta =\mu (x, t),\ \ \delta\lambda =-\dot \mu (x, 
t),
\nonumber\\
\delta\pi_\varphi &=&0,\ \ \delta\pi_\vartheta =0,\ \ \delta p_\lambda =0,
\end{eqnarray} 
where $\mu(x, t)$ is an arbitrary function of the space and time.

The  BRST invariant effective theory for the self-dual chiral boson \cite{sud}
can be written as
\begin{eqnarray}
S_{CB}^{II}&=&\int d^2x\ {\cal L}_{CB}^{II}, \label{lagc}\\{\mbox{where}}\ \
{\cal L}_{CB}^{II}&=&\frac{1}{2}\dot\varphi^2 -\frac{1}{2}{\varphi'}^2 +\lambda (\dot\varphi 
-\varphi' ) -\frac{1}{2}\dot\vartheta^2 -\frac{1}{2}{\vartheta'}^2 +
\varphi'\vartheta' \nonumber\\
&+&\dot\vartheta\vartheta' -\dot\vartheta\varphi' 
-\lambda (\dot\vartheta -\vartheta')
-\frac{1}{2}(\dot \lambda -\varphi -\vartheta )^2
+\dot{\bar c}\dot c 
-2\bar c c. \label{actcb}
\end{eqnarray} 
 $c$ and  $\bar c$ are ghost and antighost fields respectively.
The corresponding generating functional for the gauge invariant theory for the self-dual chiral boson is 
given as
\begin{equation}
Z_{CB}^{II} =\int [{\cal D}\phi ]\ e^{iS_{CB}^{II}},
\end{equation}
where $\phi$ is generic notation for all fields involved in the effective action.
The effective action $ S_{CB}^{II}$ and the generating functional $ Z_{CB}^{II}$ are 
invariant under the
 following nilpotent BRST transformation:  
\begin{eqnarray}
\delta_{b}\varphi &=&  c\ \Lambda,\ \ \ \delta_{b}\lambda =-\dot { c
}\ \Lambda, \ \ \ \delta_{b}\vartheta = c\ \Lambda,\nonumber\\
\delta_{b}\bar c &=& -(\dot \lambda -\varphi -\vartheta )\ \Lambda,\ \ \ \delta_{b}  c =0,
\end{eqnarray}
where $\Lambda$  is the infinitesimal and anticommuting  BRST parameter.          
\section{Relating the first-class and second-class theories through FFBRST formulation: examples}
In this section, we consider two examples to show the connection between the generating
 functionals for theories with first-class  and second-class constraints. First 
 we show the connection between the Stueckelberg theory and the Proca theory for massive
vector fields. In  the second example
 we link  the gauge invariant and the gauge variant theory for the self-dual chiral 
 boson.
\subsection{Connecting Stueckelberg and Proca theories } 
We start with the linearized form of the 
 Stueckelberg effective action (\ref{stl}) by introducing a Nakanishi-Lautrup type auxiliary
field ${\cal B}$ as  
\begin{eqnarray}
S_{ST}&=&\int d^4x \left[-\frac{1}{4}F_{\mu\nu}F^{\mu\nu} +\frac{1}{2}M^2\left(A_\mu -
\frac{1}{M}\partial_\mu B\right)^2 \right.\nonumber\\
 &+&\left.\frac{\chi}{2}{\cal B}^2-{\cal B}(\partial_\mu A^\mu 
+\chi MB) 
-\bar c (\partial^2 +\chi M^2)c\right],\label{s}
\end{eqnarray}
which is invariant under the following off-shell nilpotent BRST transformation:
\begin{eqnarray}
\delta_b A_\mu =\partial_\mu c\ \ \Lambda,\ \
\delta_b B  =  Mc\ \ \Lambda,\ \
\delta_b c  =  0,\ \
\delta_b \bar c  =  {\cal B}\ \ \Lambda,\ \
\delta_b {\cal B} =0.
\end{eqnarray}
The FFBRST transformation corresponding to the above BRST transformation is constructed as,
\begin{eqnarray}
\delta_b A_\mu =\partial_\mu c\ \Theta_b[\phi ], \ \
\delta_b B = Mc\ \Theta_b[\phi ],\ \
\delta_b c =0,\ \
\delta_b \bar c ={\cal B}\  \Theta_b[\phi ],\ \
\delta_b {\cal B}=0,\label{fin}
\end{eqnarray}
where $\Theta_b[\phi]$ is an arbitrary finite field dependent parameter but still anticommuting 
in nature.
To establish the connection we construct a finite field dependent parameter $\Theta_b$ 
obtainable from   
\begin{equation}
\Theta_b^\prime =i\gamma\int d^4x\left[\bar c \left(\chi MB-\frac{\chi}{2}{\cal B}+
\partial_\mu A^\mu\right)\right],
\end{equation}
via Eq. (\ref{80}), where $\gamma$ is an arbitrary parameter.

Using Eq. (\ref{jac}) the infinitesimal change in nontrivial Jacobian is calculated 
for this finite field dependent parameter as
\begin{eqnarray}
\frac{1}{J}\frac{dJ}{d\kappa}&=& i\gamma\int d^4x \left[ {\cal B}
\left(\chi MB-\frac{\chi}{2}{\cal B}+
\partial_\mu A^\mu\right) 
\right],
\end{eqnarray}
where the equation of motion for the antighost field,
$(\partial^2 +\chi M^2)c =0$, has been used.
 
We now make the following ansatz for  $S_1$:
\begin{eqnarray}
S_1=\int d^4x [\xi_1 (\kappa )\ {\cal B}^2+\xi_2 (\kappa )\ {\cal B} \partial_\mu A^
\mu+\xi_3
 (\kappa )\ \chi M{\cal B}B  ],
\end{eqnarray}
where $\xi_i,\ (i=1, 2, 3)$ are arbitrary $\kappa$ dependent parameter and
satisfy the following initial conditions: $\xi_i (\kappa =0)=0 $. 
Now, using the relation in Eq. (\ref{ibr}) we calculate $\frac{dS_1}{d\kappa}$ as 
\begin{eqnarray}
\frac{dS_1}{d\kappa}=\int d^4x \left [ {\cal B}^2 \xi_1^\prime 
 + {\cal B}\partial_\mu A^\mu 
\xi_2^\prime  + \chi M {\cal B}B \xi_3^\prime  
 \right],
\end{eqnarray}
where prime denotes the differentiation  with respect to $\kappa$.
The Jacobian contribution can be written  as $e^{S_1}$ if the  essential condition in 
Eq. (\ref{mcond}) is satisfied. This leads to
\begin{eqnarray}
\int [{\cal D}\phi]\ e^{i(S_{ST}+S_1)}\left [i{\cal B}^2(\xi_1^\prime +\gamma \frac{\chi}{2}
 )
+i{\cal B}\partial_\mu A^\mu(\xi_2^\prime -\gamma )
+i\chi M {\cal B}B(\xi_3^\prime 
-\gamma )
\right]
=0.\label{mcond1}
\end{eqnarray}
Equating the coefficients of terms $ i{\cal B}^2, i{\cal B}\partial_\mu A^\mu, $ and $
i\chi M {\cal B}B$
from both sides
of the above condition, 
we get the following  differential equations: 
\begin{eqnarray}
\xi_1^\prime +\gamma \frac{\chi}{2}&=&0, \ \
\xi_2^\prime -\gamma  = 0,\ \
\xi_3^\prime -\gamma  = 0.
\end{eqnarray}
To obtain the solution of the above equations we put $\gamma =1$ without any loss 
of generality. The solutions satisfying initial conditions are given as
\begin{equation}
\xi_1=-\frac{\chi}{2}\kappa,\ \xi_2=\kappa,\ \xi_3=\kappa.
\end{equation}
The transformed action can be obtained by adding $S_1(\kappa=1)$ to $S_{ST}$ as 
\begin{eqnarray} 
S_{ST}+S_1=\int d^4x \left[-\frac{1}{4}F_{\mu \nu } F^{\mu \nu }+\frac{1}{2}M^2
\left(A_\mu  
-\frac{1}{M}\partial_\mu B\right)^2 -\bar c (\partial^2 +\chi M^2)c\right].
\end{eqnarray}
Now the   generating functional under FFBRST transforms as
 \begin{eqnarray}
 Z'= \int  [{\cal D}A_\mu  {\cal D}B{\cal D}c {\cal D}\bar c] e^{i(S_{ST }+S_1)}.
 \end{eqnarray}
Integrating over the  $B, c,$ and $ \bar c$ fields, the above expression reduces to
the generating functional for the Proca model upto some renormalization constants as follows: 
 \begin{eqnarray}
 Z'= \int [{\cal D}A_\mu ] \ e^{iS_P}= Z_P.
 \end{eqnarray}
Therefore,
\begin{equation}
Z_{ST}\left( =\int [{\cal D}\phi ]\ e^{iS_{ST }}\right)\stackrel{FFBRST}{---\longrightarrow} Z_P
\left( =\int [{\cal D}A_\mu ]\ e^{iS_P}\right).
\end{equation}
Thus, by constructing appropriate finite field dependent 
parameter (given in Eq. (\ref{fin})) we have shown that the generating functional for Stueckelberg 
theory is connected to the 
generating functional  for the Proca theory through FFBRST transformation. This
indicates that the Green functions in these two theories are related through FFBRST 
formulation.
\subsection{Relating gauge invariant and the gauge variant theory for chiral boson  }
To see the connection between the gauge invariant and variant  theories for chiral
 boson, we start with the effective action  
for the  gauge invariant self-dual
 chiral boson theory as  
\begin{eqnarray} 
S_{CB}^{II}&=&\int d^2x
\left[\frac{1}{2}\dot\varphi^2 -\frac{1}{2}{\varphi'}^2 +\lambda (\dot\varphi 
-\varphi' ) -\frac{1}{2}\dot\vartheta^2 
-\frac{1}{2}{\vartheta'}^2 
+\varphi'\vartheta' \right.\nonumber\\
&+&\left.\dot\vartheta\vartheta' -\dot\vartheta\varphi' 
-\lambda (\dot\vartheta -\vartheta')
+\frac{1}{2}{\cal B}^2  
 + {\cal B}(\dot \lambda -\varphi -\vartheta )
+\dot{\bar c}\dot c 
-2\bar c c\right],\label{action}
\end{eqnarray}  
where we have linearized the gauge-fixing part of the effective action by introducing the
extra auxiliary field ${\cal B}$.
 This effective action is invariant under the following infinitesimal BRST transformation:
 \begin{eqnarray}
\delta_{b}\varphi &=&  c\ \Lambda,\ \ \ \delta_{b}\lambda =-\dot { c
}\ \Lambda, \ \ \ \delta_{b}\vartheta = c\ \Lambda,\nonumber\\
\delta_{b}\bar c &=& {\cal B}\ \Lambda,\ \ \ \delta_{b} {\cal B} =0, \ \ \ \delta_{b}  c =0.
\end{eqnarray} 
Corresponding  FFBRST transformation is written as
\begin{eqnarray}
\delta_{b}\varphi &=&  c\ \Theta_b [\phi],\ \ \ \delta_{b}\lambda =-\dot { c
}\ \Theta_b [\phi ], \ \ \ \delta_{b}\vartheta = c\ \Theta_b [\phi],\nonumber\\
\delta_{b}\bar c &=& {\cal B}\ \Theta_b [\phi],\ \ \ \delta_{b} {\cal B} =0, 
\ \ \ \delta_{b}  c =0,\label{fanbrs} 
\end{eqnarray} 
where $\Theta_b [\phi ]$ is arbitrary finite field dependent parameter,
which we have to constructed.
In this case we construct the  finite field dependent BRST parameter $\Theta_b [\phi]$
obtainable from
\begin{equation}
\Theta_b'= i\gamma\int d^2x \left[\bar c (\dot \lambda -\varphi-\vartheta+\frac{1}{2}{\cal B})
\right],
\end{equation} 
using Eq. (\ref{80})
and demand that the corresponding BRST transformation will lead to the gauge variant theory 
 for self-dual chiral boson.
  
To justify our claim we calculate the change in Jacobian, using equation of motion 
for antighost field, as
\begin{eqnarray}
\frac{1}{J}\frac{dJ}{d\kappa}&=& i\gamma\int d^2x \left[ {\cal B}(
\dot\lambda-\varphi-\vartheta 
+\frac{1}{2}{\cal B})\right].
\end{eqnarray}
We make an ansatz for local functional $S_1$ as, 
\begin{eqnarray}
S_1&=&\int d^2x \left[\xi_1(\kappa)\ {\cal B}^2 +\xi_2 (\kappa)\ {\cal B}
(\dot\lambda-\varphi-\vartheta )
 \right].
\end{eqnarray}
The  change in $S_1$ with respect to $\kappa$ is calculated as 
\begin{eqnarray} 
\frac{dS_1}{d\kappa} = \int d^2x \left[\xi_1' \ {\cal B}^2 +\xi_2'\ {\cal B}(
\dot\lambda-\varphi-\vartheta )
\right]. 
\end{eqnarray}
Now, the necessary condition in 
Eq. (\ref{mcond}) leads to the following equation: 
\begin{eqnarray} 
\int  [{\cal D}\phi]\ e^{i(S_{CB}^{II}+S_1)}\left [i{\cal B}^2(\xi_1^\prime - \frac{\gamma}{2}
 )+i{\cal B}(\dot\lambda-\varphi-\vartheta )(\xi_2^\prime -\gamma )
 \right]
 =0.\label{mcond2} 
\end{eqnarray}
Equating the coefficients of terms $ i{\cal B}^2$ and $ i{\cal B}(\dot\lambda-\varphi-\vartheta )$ from both sides
of above condition, 
we get following differential equations:
\begin{eqnarray}
\xi_1^\prime - \frac{\gamma}{2}&=&0,\ \
\xi_2^\prime -\gamma  = 0.
\end{eqnarray}
The solutions of above equations 
are
$
\xi_1=-\frac{1}{2}\kappa,\ \xi_2=-\kappa,$
where we have taken the parameter $\gamma =-1$.
The transformed action is obtained by adding $S_1(\kappa=1)$ to $S_{CB}^{II}$ as
\begin{eqnarray}
S_{CB}^{II}+S_1&=&\int d^2x \left[\frac{1}{2}\dot\varphi^2 -\frac{1}{2}{\varphi'}^2 +\lambda 
(\dot\varphi -\varphi' ) -\frac{1}{2}\dot\vartheta^2\right.\nonumber\\
&-&\left.\frac{1}{2}{\vartheta'}^2 +
\varphi'\vartheta' +\dot\vartheta\vartheta' -\dot\vartheta\varphi' 
-\lambda (\dot\vartheta -\vartheta') +\dot{\bar c}\dot c 
-2\bar c c\right].
\end{eqnarray}
Now the transformed generating functional becomes
\begin{eqnarray}
Z' =\int [{\cal D}\varphi {\cal D}\vartheta {\cal D}\lambda {\cal D}c {\cal D}\bar c ]\ e^{i(S_{CB}^{II}+S_1)}.
\end{eqnarray}
Performing integration over fields $\vartheta, c$ and $ \bar c$,
the above generating functional reduces to the generating functional for the self dual chiral boson 
upto some constant as
\begin{eqnarray}
Z' =\int [{\cal D}\varphi  {\cal D}\lambda] \ e^{iS_{CB}}= Z_{CB}.
\end{eqnarray}
Therefore,
\begin{equation}
Z_{CB}^{II}\left( =\int [{\cal D}\phi ]\ e^{iS_{CB}^{II}}\right)\stackrel{FFBRST}{ --\longrightarrow} 
Z_{CB}\left( =\int  [{\cal D}\varphi  {\cal D}\lambda ]\ e^{iS_{CB}}\right).
\end{equation} 
Thus, the generating functionals corresponding to the gauge invariant and gauge
non-invariant theory for self-dual chiral boson 
are connected through the FFBRST transformation given in Eq. (\ref{fanbrs}).
 
We end up this section by making conclusion that using FFBRST formulation 
the generating functional for the  theory with second-class constraint can be 
achieved by generating functional for   theory with first-class constraint.
\section{ First-class and second-class theories:  FF-anti-BRST formulation}
In this section, we consider FF-anti-BRST formulation  to show the same connection between the generating
 functionals for theories with first-class  and second-class constraints with same examples. 
The FF-anti-BRST transformation is also developed in  same fashion as FFBRST transformation,
the only key difference is the role of ghost fields are interchanged with antighost fields 
and vice-versa. 
\subsection{Relating Stueckelberg and Proca theories} 
We start with anti-BRST symmetry transformation for effective action given in Eq. (\ref{s}), as  
\begin{eqnarray}
\delta_{ab} A_\mu &=&\partial_\mu\bar c\ \ \Lambda,\ \
\delta_{ab} B  =  M\bar c\ \ \Lambda,\ \
\delta_{ab} c  =  -{\cal B}\ \ \Lambda,\nonumber\\
\delta_{ab} \bar c &=& 0,\ \
\delta_{ab} {\cal B} = 0,
\end{eqnarray}
where $\Lambda$ is infinitesimal, anticommuting and global parameter. 
The FF-anti-BRST transformation corresponding to the above anti-BRST transformation is constructed as,
\begin{eqnarray}
\delta_{ab} A_\mu &=&\partial_\mu\bar c\ \ \Theta_{ab},\ \ 
\delta_{ab} B  =  M\bar c\ \ \Theta_{ab},\ \
\delta_{ab}c  = -{\cal B}\ \ \Theta_{ab},\nonumber\\
\delta_{ab} \bar c &=& 0,\ \
\delta_{ab} {\cal B} = 0,\label{fin11} 
\end{eqnarray}
where $\Theta_{ab}$ is an arbitrary finite field dependent parameter but still anticommuting 
in nature.
To establish the connection we choose a finite field dependent parameter $\Theta_{ab}$ 
obtainable from     
\begin{equation}
\Theta_{ab}^\prime =-i\gamma\int d^4x\left[c \left(\chi MB-\frac{\chi}{2}{\cal B}+
\partial_\mu A^\mu\right)\right], 
\end{equation}
 using Eq. (\ref{81}), where $\gamma$ is an arbitrary parameter.

Using Eq. (\ref{jaceva1}) the infinitesimal change in nontrivial Jacobian can be calculated 
for this finite field dependent parameter as
\begin{eqnarray}
\frac{1}{J}\frac{dJ}{d\kappa}&=&-i\gamma\int d^4x \left[-{\cal B}
\left(\chi MB-\frac{\chi}{2}{\cal B}+
\partial_\mu A^\mu\right) 
\right].\label{j2} 
\end{eqnarray}
To Jacobian contribution can be expressed as $e^{iS_2}$. To calculate  $S_2$ we make following ansatz: 
\begin{eqnarray}
S_2&=&\int d^4x [\xi_5 (\kappa ){\cal B}^2+\xi_6 (\kappa ){\cal B} \partial_\mu A^
\mu+\xi_7
 (\kappa )\chi M{\cal B}B ],
\end{eqnarray}
where $\xi_i,\ (i=5,..,7)$ are arbitrary $\kappa$ dependent parameter and
satisfy following initial conditions: $\xi_i (\kappa =0)=0 $. 
Now, infinitesimal change in $S_2$ is calculated as 
\begin{eqnarray} 
\frac{dS_2}{d\kappa}=\int d^4x \left [ {\cal B}^2 \xi_5^\prime 
 + {\cal B}\partial_\mu A^\mu 
\xi_6^\prime  + \chi M {\cal B}B \xi_7^\prime 
\right],\label{s2} 
\end{eqnarray}
where prime denotes the differentiation  with respect to $\kappa$.

Putting the expressions (\ref{j2}) and (\ref{s2})  in the  essential condition given in 
Eq. (\ref{mcond}), we obtain
\begin{eqnarray}
 \int [{\cal D}\phi]\  e^{i(S_{ST}+S_2)}\left [ {\cal B}^2(\xi_5^\prime +\gamma \frac{\chi}{2}
 )
+ {\cal B}\partial_\mu A^\mu(\xi_6^\prime -\gamma )
+ \chi M {\cal B}B(\xi_7^\prime 
-\gamma )\right]
 =0.\label{mcond3}
\end{eqnarray}  
Equating the coefficients of terms $ i{\cal B}^2, i{\cal B}\partial_\mu A^\mu,$ and $ 
i\chi M {\cal B}B$
  from both sides
of above condition, 
we get following  differential equations: 
\begin{eqnarray}
\xi_5^\prime +\gamma \frac{\chi}{2}=0,\ \
\xi_6^\prime -\gamma =0,\ \ 
\xi_7^\prime -\gamma =0.  
\end{eqnarray}
 The solutions of the above differential equation for $\gamma =1$ are
$
\xi_5=-\frac{\chi}{2}\kappa,\ \xi_6=\kappa,\ \xi_7=\kappa.
$
The transformed action can be obtained by adding $S_2(\kappa=1)$ to $S_{ST}$ as 
\begin{eqnarray} 
S_{ST}+S_2=\int d^4x \left[-\frac{1}{4}F_{\mu \nu } F^{\mu \nu }+\frac{1}{2}M^2
\left(A_\mu  
-\frac{1}{M}\partial_\mu B\right)^2 -\bar c(\partial^2 +\chi M^2)c\right].
\end{eqnarray}
We perform integration over $B, c$ and $ \bar c$ fields
to remove the divergence of the transformed generating functional $Z'= \int [{\cal D}A_\mu {\cal D}B{\cal D}c 
{\cal D}\bar c ] e^{i(S_{ST}+S_2)}$ and hence
we get the generating functional for the Proca model as 
\begin{eqnarray} 
Z'= \int [{\cal D}A_\mu] e^{iS_P}=  Z_{P }.  
\end{eqnarray} 
Therefore, 
$
Z_{ST}\stackrel{FF-anti-BRST}{-----\longrightarrow} Z_P.
$
Thus, by constructing the appropriate finite field dependent 
parameter (given in Eq. (\ref{fin11})), we have shown that the generating functional for the Stueckelberg 
theory is related to the 
generating functional  for the Proca theory through the FF-anti-BRST transformation also. However, the finite field
dependent parameter is different from that one involved in the FFBRST transformation.  This
indicates that the Green  functions in these two theories are related through the FFBRST and the FF-anti-BRST 
transformations.
\subsection{Mapping between the gauge invariant and the gauge variant theory for chiral boson }
To connect the gauge invariant and gauge variant  theories for the chiral
 boson through the FF-anti-BRST transformation, first of all we write the anti-BRST transformation for effective action 
 (\ref{action}) as
     \begin{eqnarray}
\delta_{ab}\varphi &=&  \bar c\ \Lambda,\ \ \ \delta_{ab}\lambda =-\dot {\bar c
}\ \Lambda, \ \ \ \delta_{ab}\vartheta = \bar c\ \Lambda,\nonumber\\
\delta_{ab} c &=& -{\cal B}\ \Lambda,\ \ \ \delta_{ab} {\cal B} =0, \ \ \ \delta_{ab} \bar  c =0.
\end{eqnarray}
The corresponding  FF-anti-BRST transformation is written as
\begin{eqnarray}
\delta_{ab}\varphi &=& \bar  c\ \Theta_{ab} [\phi],\ \ \ \delta_{ab}\lambda =-\dot {\bar  c
}\ \Theta_{ab} [\phi ], \ \ \ \delta_{ab}\vartheta =\bar  c\ \Theta_{ab} [\phi],\nonumber\\
\delta_{ab}c &=& -{\cal B}\ \Theta_{ab} [\phi],\ \ \ \delta_{ab} {\cal B} =0, 
\ \ \ \delta_{ab} \bar  c =0,\label{fbrs1}
\end{eqnarray} 
where $\Theta_{ab} [\phi ]$ is the arbitrary finite field dependent parameter.
In this case, we construct the  finite field dependent anti-BRST parameter $\Theta_{ab} [\phi]$
obtainable from
\begin{equation}
\Theta_{ab}'= -i\gamma\int d^2x \left[ c (\dot \lambda -\varphi-\vartheta+\frac{1}{2}{\cal B})
\right], 
\end{equation} 
using Eq. (\ref{81}) and demand that the corresponding anti-BRST transformation will lead to the gauge variant theory 
 for the self-dual chiral boson. 
  
We make an ansatz for local functional $S_2$ in this case as 
\begin{eqnarray}
S_2&=&\int d^2x \left[\xi_5(\kappa)\ {\cal B}^2 +\xi_6 (\kappa)\ {\cal B}
(\dot\lambda-\varphi-\vartheta )
\right].
\end{eqnarray}
Now, the necessary condition in 
Eq. (\ref{mcond}) leads to the following equation: 
\begin{eqnarray} 
\int [{\cal D}\phi ]\ e^{i(S_{CB}^{II}+S_2)}\left [i{\cal B}^2(\xi_5^\prime - \frac{\gamma}{2}
 )+i{\cal B}(\dot\lambda-\varphi-\vartheta )(\xi_6^\prime -\gamma )
 \right]
 =0.\label{mcond4} 
\end{eqnarray}
Equating the coefficients of different terms on both sides
of the above equation, 
we get the following differential equations:
\begin{eqnarray}
\xi_5^\prime - \frac{\gamma}{2} = 0,\ \
\xi_6^\prime -\gamma  = 0.
\end{eqnarray}
The solutions of above equations 
are 
$
\xi_5=-\frac{1}{2}\kappa,\ \xi_6=-\kappa,
$
where the parameter $\gamma =-1$.
The transformed action can be obtained by adding $S_2(\kappa=1)$ to $S_{CB}^{II}$ as
\begin{eqnarray}
S_{CB}^{II}+S_2&=&\int d^2x \left[\frac{1}{2}\dot\varphi^2 -\frac{1}{2}{\varphi'}^2 +\lambda 
(\dot\varphi -\varphi' ) -\frac{1}{2}\dot\vartheta^2\right.\nonumber\\
&-&\left.\frac{1}{2}{\vartheta'}^2 +
\varphi'\vartheta' +\dot\vartheta\vartheta' -\dot\vartheta\varphi' 
-\lambda (\dot\vartheta -\vartheta')+\dot{\bar c}\dot c -2\bar c c\right].
\end{eqnarray}
After functional integration over fields $\vartheta, c$ and $\bar c$ in the expression of
transformed generating functional, we get the generating functional as 
\begin{eqnarray}
Z' =\int [{\cal D}\varphi {\cal D}\lambda] e^{i S_{CB}}=  Z_{CB}.
\end{eqnarray}
Therefore,
$Z_{CB}^{II} \stackrel{FF-anti-BRST}{-----\longrightarrow} Z_{CB}.$
Thus, the generating functionals corresponding to the gauge invariant and the gauge
non-invariant theory for the self-dual chiral boson 
are also connected through the FF-anti-BRST transformation given in Eq. (\ref{fbrs1}).

We end up this section by making comment that  
the generating functional for the  theory with the second-class constraint can be 
obtained from the generating functional for   theory with the first-class constraint
using both FFBRST and FF-anti-BRST transformations. The finite parameters involved in the FFBRST transformation,
which are responsible for these connection, are different from the parameters in FF-anti-BRST transformation.  
\section{Conclusions} 
The Stueckelberg theory for  massive spin  1 field and the gauge invariant 
theory for the self-dual chiral boson are  the  first-class  theories. On the other hand,
the Proca theory  for massive spin 1 field and the gauge  variant 
theory for the self-dual chiral boson are  theories with the second-class constraint. 
We have shown that the FFBRST transformation relates
the generating functionals of second-class theory  and first-class
theory. The path integral measure in the definition of generating functional
is not invariant under such FFBRST 
transformation and is responsible for such  connections. The Jacobian for 
path integral measure under such a  transformation with an appropriate finite 
parameter cancels the extra parts of
the first-class theory.
Our result is supported by two explicit examples. In the first case we have related  the 
generating functional of the Stueckelberg  
theory to the generating functional of the
Proca model and in the second case the generating functionals corresponding to the gauge 
invariant theory and the gauge variant theory for the self-dual chiral boson 
have linked through 
FFBRST transformation with appropriate choices of the finite field dependent parameter.
The same goal has been achieved by using the FF-anti-BRST transformation.
These formulations can be applied to connect the generating functionals for any first-class 
and second-class theories provided  appropriate 
finite parameters are constructed. The complicacy arises due to 
the nonlocal and field dependent Dirac brackets in the quantization of second-class theories can thus be avoided
by using FFBRST/FF-anti-BRST formulations which relate the Green functions of second-class theories 
to the first-class theories.   
These formulations can be applied to connect the generating functionals for 
any first-class (e.g. non-Abelian gauge theories)
and second-class theories provided  appropriate 
finite parameters are constructed. 

\label{Chap:Chapter7}

\chapter{Hodge-de Rham theorem in the BRST context}
In this chapter we study the different forms of BRST symmetry \cite{sud}. In particular we investigate  co-BRST and anti-co-BRST
along with usual BRST and anti-BRST symmetries. The nilpotent conserved charges for all these symmetries
are calculated and have shown to satisfy the algebra analogous to the algebra satisfied
by de Rham cohomological operators. These results are shown in a particular model namely (1+1) dimensional theory 
for a self-dual chiral boson. 
 
\vspace {-1.0cm}
\section{Self-dual chiral boson: preliminary idea}
\label{sec:1}  We start with the gauge non-invariant model \cite{pp} in (1+1) dimensions for a
 single self-dual chiral boson. The Lagrangian density for such a theory is given by
\begin{equation} 
{\cal L}=\frac{1}{2}\dot\varphi^2 -\frac{1}{2}{\varphi'}^2 +\lambda (\dot\varphi -\varphi' ),
\label{chi}
\end{equation}  
where overdot and prime 
denote time and space derivatives, respectively, and $\lambda$ is a Lagrange multiplier.
The field $\varphi$ satisfies the self-duality condition $\dot\varphi =\varphi'$ 
in this case.
We choose the Lorentz metric $g^{\mu\nu}=(1,-1)$ where $\mu, \nu =0, 1$. The associated 
momenta for the field  $\varphi$ and Lagrange multiplier are calculated as
\begin{eqnarray}
\pi_\varphi =\frac{\partial {\cal L}}{\partial \dot \varphi}=\dot\varphi +
\lambda,\ \
\pi_\lambda =\frac{\partial {\cal L}}{\partial \dot \lambda}=0,
\end{eqnarray}
which shows that the model has following primary constraint
\begin{equation}
\Omega_1\equiv \pi_\lambda \approx 0.
\end{equation}
The expression for Hamiltonian density corresponding to above Lagrangian density ${\cal L}$ 
is
\begin{eqnarray}
{\cal H}=\pi_\varphi\dot\phi +\pi_\lambda\dot\lambda -{\cal L} 
=\frac{1}{2}(\pi_\varphi -\lambda )^2 +\frac{1}{2}\varphi'^2 +\lambda \varphi'.
\end{eqnarray}
Further we write the total Hamiltonian density corresponding to this ${\cal L}$ by 
introducing  Lagrange multiplier field  $\omega$ for the primary constraint $\Omega_1$
 as
\begin{eqnarray}
{\cal H}_T = \frac{1}{2}(\pi_\varphi -\lambda )^2 +\frac{1}{2}\varphi'^2 +\lambda \varphi' +
\omega\pi_\lambda.
\end{eqnarray}
Following the Dirac prescription \cite{dir}, we obtain the secondary constraint in this case
as
\begin{equation}
\Omega_2 \equiv \dot \pi_\lambda =\{ \pi_\lambda, {\cal H}\}=\pi_\varphi -
\lambda-\varphi'\approx 0.
\end{equation}  
 The Poison bracket for primary and secondary constraint is nonvanishing,
$\{ \Omega_1, \Omega_2\} \not=0$.
 This implies the constraints $\Omega_1$ and $\Omega_2$ are of second-class,
which is an essential feature of a gauge variant theory (model).

This model is quantized by establishing the commutation relation \cite{pp}
\begin{eqnarray}
[\varphi(x), \pi_\varphi (y)]& =&[\varphi(x), \lambda(y) ]=-i\delta (x-y)\\
2[\lambda(x), \pi_\varphi(y)] &=&[\lambda(x), \lambda(y) ]= 2i\delta' (x-y),
\end{eqnarray}
and the rest of the commutators vanish.
\subsection{Wess-Zumino term and Hamiltonian formulation}
\label{sec:2}
 To construct a gauge invariant theory corresponding to this gauge non-invariant model
for chiral bosons, one generally introduces the Wess-Zumino term in the Lagrangian density
${\cal L}$.
For this purpose one has to enlarge the Hilbert space of the theory by introducing a new 
quantum field 
$\vartheta$, called as Wess-Zumino field, through the redefinition of
 fields $\varphi$ and $\lambda$ in the original Lagrangian 
density ${\cal L}$ [Eq. (\ref{chi})] as follows \cite{wz}:
\begin{equation}
\varphi\rightarrow \varphi -\vartheta, \ \ \lambda\rightarrow \lambda +\dot\vartheta.
\end{equation}
With these  redefinition of fields the modified Lagrangian density becomes
\begin{eqnarray}
{\cal L}^I&=& {\cal L} -\frac{1}{2}\dot\vartheta^2 -\frac{1}{2}{\vartheta'}^2 +
\varphi'\vartheta' +\dot\vartheta\vartheta' -\dot\vartheta\varphi' 
-\lambda (\dot\vartheta -\vartheta')\nonumber\\
&=&{\cal L} +{\cal L}^{WZ},
\end{eqnarray}
where 
\begin{equation}
{\cal L}^{WZ}=-\frac{1}{2}\dot\vartheta^2 -\frac{1}{2}{\vartheta'}^2  +
\varphi'\vartheta' +\dot\vartheta\vartheta' -\dot\vartheta\varphi' 
-\lambda (\dot\vartheta -\vartheta'),
\end{equation}
is the
Wess-Zumino part of the  ${\cal L}^I$.
It is easy to check that the above Lagrangian density is invariant under 
time-dependent chiral gauge transformation:
\begin{eqnarray}
\delta\varphi &=&\mu (x, t),\ \ \delta\vartheta =\mu (x, t),\ \ \delta\lambda =-\dot \mu (x, 
t)
\nonumber\\
\delta\pi_\varphi &=&0,\ \ \delta\pi_\vartheta =0,\ \ \delta p_\lambda =0,
\end{eqnarray} 
where $\mu(x, t)$ is an arbitrary function of the space and time.
The canonical momenta for this gauge-invariant theory are calculated as
\begin{eqnarray}
\pi_\lambda &= &\frac{\partial{\cal L}^I}{\partial\dot\lambda}=0,\ \
\pi_\varphi = \frac{\partial{\cal L}^I}{\partial\dot\varphi}=\dot\varphi +\lambda\nonumber\\
\pi_\vartheta & =&\frac{\partial{\cal L}^I}{\partial\dot\vartheta}=-\dot\vartheta 
-\varphi' +\vartheta' -\lambda.
\end{eqnarray}
This implies the theory ${\cal L}^I$ possesses a primary constraint
\begin{equation}
 \psi_1\equiv \pi_\lambda \approx 0.
\end{equation}
The Hamiltonian density corresponding to ${\cal L}^I$ is then given by
\begin{equation}
{\cal H}^I= \pi_\varphi\dot\varphi +\pi_\vartheta\dot\vartheta +\pi_\lambda\dot\lambda
-{\cal L}^I.
\end{equation}
The total Hamiltonian density after the introduction of a Lagrange multiplier field $u$ 
corresponding to the primary constraint $\Psi_1$ becomes
\begin{equation}
{\cal H}_T^I=\frac{1}{2}\pi_\varphi^2 -\frac{1}{2}\pi_\vartheta^2 +\pi_\vartheta\vartheta'-
\pi_\vartheta\varphi' -\lambda\pi_\varphi -\lambda\pi_\vartheta +u\pi_\lambda.\label{ham}
\end{equation}
Following the Dirac  method of constraint analysis we obtain the secondary
constraint 
\begin{equation}
\psi_2\equiv (\pi_\varphi +\pi_\vartheta)\approx 0.
\end{equation}
In Dirac's quantization procedure \cite{dir}, we have to change the first-class 
constraints of the theory into second-class constraints. To achieve this we impose some
additional constraints on the system in the form of gauge-fixing conditions $\partial_\mu
\vartheta=0$ ($\partial_0\vartheta=\dot\vartheta=0$ and $-\partial_1\vartheta 
=-\vartheta'=0$)
\cite{nk}.
With the above choice of gauge-fixing conditions the extra constraints of the theory are
\begin{eqnarray}
\xi_1\equiv  -\vartheta'\approx 0, \ \ \xi_2 \equiv (\pi_\vartheta -\vartheta' +\varphi' +
\lambda)\approx 0.\label{cons}
\end{eqnarray}
Now, the total set of constraints after gauge fixing are
\begin{eqnarray}
\chi_1 &=&\psi_1 \equiv \pi_\lambda\approx 0, \ \ \chi_2 = \psi_2 \equiv (\pi_\varphi 
+\pi_\vartheta)\approx 0\nonumber\\
\chi_3 &=&\xi_1\equiv  -\vartheta'\approx 0, \ \ \chi_4 =\xi_2 \equiv (\pi_\vartheta -
\vartheta' 
+\varphi' +\lambda) \approx 0.
\end{eqnarray}
The nonvanishing
commutators of gauge invariant theory are obtained as
\begin{eqnarray}
[\varphi(x), \pi_\varphi (y)]& =&[\varphi(x), \lambda(y) ]=-i\delta (x-y)\label{brac1}\\
2[\lambda(x), \pi_\varphi(y)] &=&[\lambda(x), \lambda(y) ]=+2i\delta' (x-y)\\
\ [ \vartheta (x), \pi_\vartheta(y) ] &=& 2[\varphi(x), \pi_\vartheta(y) ]=-2i\delta (x-y)\\
\ [\lambda(x), \pi_\vartheta(y)] &=&-i\delta' (x-y)\label{brac4}.
\end{eqnarray}
We end up the section with the conclusion that the above relations (\ref{brac1})-(\ref{brac4}),
together with ${\cal H}^I_T$ [Eq. (\ref{ham})],
reproduce the same quantum system described by ${\cal L}$ under the gauge condition 
(\ref{cons}). This is similar to the quantization
of a gauge invariant chiral Schwinger model (with an appropriate Wess-Zumino term) \cite{nk}.
\section{BFV formulation for model of self-dual chiral boson }
In the BFV formulation of self-dual chiral boson, we need  to introduce a pair of canonically conjugate 
ghosts $(c, p)$ with
ghost numbers 1 and $-1$ respectively, for the first-class constraint  $\pi_\lambda=0$, and 
another pair of
 ghosts $( \bar c, \bar p)$ with ghost number $-1$ and 1, respectively, for the secondary 
constraint,
$(\pi_\varphi +\pi_\vartheta) = 0 $. The effective action for this $(1+1)$ dimensional 
theory with a single self-dual 
chiral boson in this extended 
phase space then becomes
\begin{eqnarray}
S_{eff}& =& \int d^2x  \left[ \pi_\varphi\dot\varphi +\pi_\vartheta\dot\vartheta +p_u\dot u
-\pi_\lambda\dot\lambda
-\frac{1}{2}\pi_\varphi^2 +\frac{1}{2}\pi_\vartheta^2 \right.\nonumber\\
&-&\pi_\vartheta(\vartheta'-\varphi')+\left.\dot cp +\dot{\bar c}\bar p
 - \{Q_b, \Psi\}\right],
\label{hseff}
\end{eqnarray}
where $Q_b$ is the BRST charge and $\Psi$ is the gauge-fixed fermion.

The generating functional, for any gauge invariant effective theory having $\Psi$ as a
 gauge-fixed fermion, is defined as
\begin{equation}
Z_\Psi = \int {\cal{D}}\phi \exp \left[i \int d^2x \ S_{eff} \right],
\end{equation}
where $\phi$ is generic notation for all the dynamical field involved in the effective 
theory.
The BRST symmetry generator for this theory is written as
\begin{equation}
Q_b=  ic(\pi_\varphi +\pi_\vartheta )-i\bar p \pi_\lambda .\label{chr}
\end{equation}
The canonical brackets are defined for all dynamical variables as
\begin{eqnarray}
[\vartheta, \pi_\vartheta ]&=&-i,\ \ [\varphi, \pi_\varphi ]=-i, \ \ [\lambda, \pi_\lambda] 
= -i,
\nonumber\\
\ [u, p_u]&=&-i,\ \ \{ c,\dot{\bar c}\} =i, \ \ \{ \bar c,\dot{ c}\} = -i, \label{brac}
\end{eqnarray}
and the rest of the brackets are zero. The nilpotent BRST transformation, using Eqs. (\ref{anticom}) and (\ref{chr}),
is then explicitly calculated as
\begin{eqnarray}
s_b \varphi &=& -c,\ \ s_b\lambda =\bar p, \ \ s_b\bar p=0, \ \ s_b\vartheta =-c
\ \
s_b \pi_\varphi = 0,\ \ s_b u =0,\nonumber\\
s_b \pi_\vartheta &=&0,\ \ s_b p=(
\pi_\varphi +\pi_\vartheta ) \ \
s_b \bar c = \pi_\lambda,\ \ s_b \pi_\lambda =0,\ \ s_b c =0,\ \ s_b p_u =0.
\end{eqnarray}
In BFV formulation the generating functional is independent of the
gauge-fixed fermion \cite{ht,wei}, hence we have the freedom to choose it in the convenient
 way as
\begin{equation}
\Psi = {{p}}\lambda +{\bar{{c}}} \left (\vartheta +\varphi +\frac{\xi}{2}
\pi_\lambda\right ),
\end{equation}
where $\xi$ is arbitrary gauge parameter.

Putting the value of $\Psi$ in Eq. (\ref{hseff}) and using Eq. (\ref{chr}), we get
\begin{eqnarray}
S_{eff}& =& \int d^2x \left[ \pi_\varphi\dot\varphi +\pi_\vartheta\dot\vartheta +p_u\dot u
-\pi_\lambda\dot\lambda
-\frac{1}{2}\pi_\varphi^2 +\frac{1}{2}\pi_\vartheta^2 
 -\pi_\vartheta(\vartheta'-\varphi') \right.\nonumber\\
&+&\left.\dot cp +\dot{\bar c}\bar p +\lambda (\pi_\varphi +\pi_\vartheta )
+2 c\bar c -\bar pp 
+\pi_\lambda \left(\vartheta +\varphi +\frac{\xi}{2}\pi_\lambda\right )
\right]. \label{effact}
\end{eqnarray}
The generating functional for this effective theory is then  expressed as 
\begin{eqnarray}
Z_\Psi &=& \int {\cal{D}}\phi \exp \left[i \int d^2x \left\{ \pi_\varphi\dot\varphi +
\pi_\vartheta\dot\vartheta +p_u\dot u
-\pi_\lambda\dot\lambda\right.\right.\nonumber\\
&-&\left.\left.\frac{1}{2}\pi_\varphi^2 
+\frac{1}{2}\pi_\vartheta^2 
-\pi_\vartheta(\vartheta'-\varphi') 
+\dot cp +\dot{\bar c}\bar p +\lambda (\pi_\varphi +\pi_\vartheta )\right.\right.\nonumber\\
&+&\left.\left.2 c\bar c 
-\bar pp+\pi_\lambda \left(\vartheta +\varphi +\frac{\xi}{2}\pi_\lambda\right )
\right\}
\right]. 
\end{eqnarray}
Performing the integration over $p$ and $\bar p$ in the above functional integration 
we further obtain
\begin{eqnarray}
Z_\Psi &=& \int {\cal{D}}\phi' \exp \left[i \int d^2x\left\{ \pi_\varphi\dot\varphi +
\pi_\vartheta\dot\vartheta +p_u\dot u
-\pi_\lambda\dot\lambda\right.\right.\nonumber\\
&-&\left.\left.\frac{1}{2}\pi_\varphi^2 
+\frac{1}{2}\pi_\vartheta^2 -
\pi_\vartheta(\vartheta'-\varphi')
+\dot{\bar c}\dot c  +\lambda (\pi_\varphi +\pi_\vartheta )\right.\right.\nonumber\\
&+&\left.\left.2 c\bar c 
+\pi_\lambda \left(\vartheta +\varphi +\frac{\xi}{2}\pi_\lambda\right )\right\}
\right],
\end{eqnarray}
where ${\cal D}\phi'$ is the path integral measure for effective theory when integrations
 over 
fields $p$ and $\bar p$ are carried out. 
Taking the arbitrary gauge parameter $\xi =1$ and performing the integration over 
$\pi_\lambda$, we obtain an 
effective generating functional as
\begin{eqnarray}
Z_\Psi &=& \int {\cal{D}}\phi'' \exp \left[i \int d^2x \left\{ \pi_\varphi\dot\varphi +
\pi_\vartheta\dot\vartheta +p_u\dot u
-\frac{1}{2}\pi_\varphi^2  
+\frac{1}{2}\pi_\vartheta^2 \right.\right.\nonumber\\
&+&\left.\left.
\pi_\vartheta (\varphi'-\vartheta' +\lambda )
+\dot{\bar c}\dot c +\pi_\varphi\lambda  
-2\bar cc 
-\frac{ \left (\dot\lambda -\vartheta -\varphi \right )^2}{2}\right\}
\right], \label{hzfun}
\end{eqnarray}
where ${\cal D}\phi''$ denotes the measure corresponding to all the dynamical
 variable involved in this effective action. 
The expression for effective action in the above equation is exactly 
same as the BRST invariant effective action in Ref. \cite{pb}. 
The BRST symmetry transformation for this effective theory is
\begin{eqnarray}
s_b \varphi &=& -c,\ \ s_b\lambda =\dot c, \ \ s_b\vartheta =-c,\ \
s_b \pi_\varphi = 0,\ \ \ s_b u =0,\nonumber\\
 s_b \pi_\vartheta &=&0,\ \
s_b \bar c =-(\dot\lambda -\vartheta -\varphi ),\ \ s_b c =0,
\ \ s_b p_u =0,\label{brs}
\end{eqnarray}
which is {\it on-shell} nilpotent. Antighost equation of motion (i.e. $\ddot{ c}+2c=0$)
 is required
to show the nilpotency.
\section{Nilpotent symmetries: many guises}
In this section we study the different forms of the  nilpotent BRST symmetry of the system of single self-dual 
chiral boson. In particular, we discuss co-BRST anti-co-BRST, bosonic and ghost symmetries in the context of
self-dual chiral boson. 
\subsection{ Off-shell  BRST and anti-BRST Symmetry }
To study the off-shell BRST and anti-BRST transformations we  incorporate 
Nakanishi-Lautrup type auxiliary field $B$  to linearize the gauge-fixing 
part of the effective action  written as
\begin{equation}
S_{eff}=\int d^2x {\cal L}_{eff},
\end{equation}
where,
 \begin{eqnarray}
{\cal  L}_{eff}&=& \pi_\varphi\dot\varphi +\pi_\vartheta\dot\vartheta +p_u\dot u-\frac{1}{2}
\pi_\varphi^2
+\frac{1}{2}\pi_\vartheta^2 
+\pi_\vartheta (\varphi'-\vartheta' +\lambda)\nonumber\\
&+& \pi_\varphi\lambda +\frac{1}{2}B^2 +B(\dot \lambda -\varphi -\vartheta ) 
+\dot{\bar c}\dot c
-2\bar c c.\label{hlag}
\end{eqnarray}
This effective theory is invariant under the following {\it 
off-shell} nilpotent 
BRST transformation:
\begin{eqnarray}
s_b\varphi &=& - c,\ \ \ s_b\lambda =\dot { c
}, \ \ \ s_b\vartheta = -c\nonumber\\
s_b \pi_\varphi &=& 0,\ \ \ s_b u =0, \ \ \ s_b \pi_\vartheta =0
\nonumber\\
s_b\bar c &=& B,\ \ \ s_b B =0, \ \ \ s_b  c =0,\ \ s_b p_u 
=0.\label{b}
\end{eqnarray}
The corresponding anti-BRST symmetry transformation 
for this theory is written as
\begin{eqnarray}
s_{ab}\varphi &=& -\bar c,\ \ \ s_{ab}\lambda =\dot {\bar c
}, \ \ \ s_{ab}\vartheta = -\bar c\nonumber\\
s_{ab} \pi_\varphi &=& 0,\ \ \ s_{ab} u =0, \ \ \ s_{ab} \pi_\vartheta =0
\nonumber\\
s_{ab}c &=& -B,\ \ \ s_{ab} B =0, \ \ \ s_{ab} \bar c =0,\ \ s_{ab} p_u 
=0.\label{ab}
\end{eqnarray}
The conserved  BRST and anti-BRST charges
$Q_{b}$ and $Q_{ab}$, respectively, which are the generator of the 
above  BRST and anti-BRST transformations, are  
\begin{equation}
Q_b=  i(\pi_\varphi +\pi_\vartheta )c -i\pi_\lambda \dot c, 
\end{equation}
\begin{equation}
Q_{ab}=  i(\pi_\varphi +\pi_\vartheta )\bar c -i\pi_\lambda \dot{\bar c}.
\end{equation}
Further by using the equations of motion
\begin{eqnarray}
B+\dot \pi_\varphi =0,\ \ B+\dot \pi_\vartheta =0,\ \ \dot\varphi-\pi_\varphi +\lambda =0
\nonumber\\
\dot\vartheta +\pi_\vartheta +\varphi' -\vartheta' +\lambda =0,\ \ \dot B=\pi_\varphi +
\pi_\vartheta \nonumber\\
\dot u=0,\ \ \dot p_u =0, \ \ \ddot{\bar c}+2\bar c=0, \ \ \ddot{ c}+2 c=0\nonumber\\
B+\dot \lambda -\varphi -\vartheta =0,\label{eom}
\end{eqnarray}
it can be shown that these charges are constant of motion i.e. $\dot Q_b=0,\dot Q_{ab}=0$ and
  satisfy the  relations
\begin{equation}
Q_b^2=0,\ Q_{ab}^2=0,\ Q_b Q_{ab}+Q_{ab}Q_b=0.
\end{equation}
To arrive at these relations, we have used the canonical brackets 
[Eq. (\ref{brac})] of the fields and the definition
of canonical momenta, 
\begin{equation}
\pi_\lambda =B,\ \ \pi_{\bar c} =\dot c,\ \ \pi_c =-\dot {\bar c},\ \ \pi_u =p_u.
\end{equation} 
We come to the end of this section with the remark that the condition for the physical states
$Q_{b}\left|phys\right>=0$ and $Q_{ab}\left|phys\right>=0$ leads to the requirement that
\begin{equation} 
(\pi_\varphi +\pi_\vartheta)\left|phys\right>
=0
\end{equation} and 
\begin{equation}
\pi_\lambda\left|phys\right>=0.
 \end{equation}
This implies
that the operator form of the first-class constraints $\pi_\lambda\approx 0$ and 
$(\pi_\varphi +\pi_\vartheta)\approx 0$ annihilate the physical state of the theory.
Thus, the physicality criterion  is consistent with the Dirac method \cite{sund} of 
quantization.
\subsection{ Co-BRST and anti-co-BRST symmetries} 
In this subsection we investigate the nilpotent  co-BRST and anti-co-BRST (alternatively known as dual 
and anti-dual-BRST respectively)
transformations which are also the symmetry of the effective action. Further these
transformations leave the gauge-fixing term of the action invariant independently and the 
kinetic energy term (which remains invariant under BRST and anti-BRST transformations) 
transforms under it to compensate   the terms arises due to the  transformation of the ghost terms.
The gauge-fixing term has its origin in the co-exterior derivative $\delta =\pm \ast d\ast $,
where $\ast $ represents the Hodge duality operator. The $\pm$
signs is dictated by the dimensionality of the manifold \cite{egu}. Therefore, 
it is appropriate to call these transformations a  dual 
and an anti-dual-BRST transformation.

The nilpotent co-BRST transformation ($ s^2_{d}=0$) and anti-co-BRST transformation ($ 
s^2_{ad}=0$) which are absolutely
 anticommuting ($ s_d s_{ad}+s_{ad}s_d =0$) are                 
\begin{eqnarray}
s_{d}\varphi &=& -\frac{1}{2}\ \dot{\bar c},\ \ \ s_{d}\lambda =-\bar c
, \ \ \ s_{d}\vartheta = -\frac{1}{2} \ \dot{\bar c}\nonumber\\
s_{d} \pi_\varphi &=& 0,\ \ \ s_{d} u =0, \ \ \ s_{d} \pi_\vartheta =0,\ \ s_{d} p_u 
=0
\nonumber\\
s_{d}c &=& \frac{1}{2}(\pi_\varphi +\pi_\vartheta ),\ \ \ s_{d} B =0, \ \ \ s_{d} \bar c =0.
\label{d}
\end{eqnarray}
\begin{eqnarray}
s_{ad}\varphi &=& -\frac{1}{2}\ \dot{c},\ \ \ s_{ad}\lambda =-c
, \ \ \ s_{ad}\vartheta = -\frac{1}{2}\dot{ c}\nonumber\\
s_{ad} \pi_\varphi &=& 0,\ \ \ s_{ad} u =0, \ \ \ s_{ad} \pi_\vartheta =0,\ \ s_{ad} p_u 
=0
\nonumber\\
s_{ad}\bar c &=& -\frac{1}{2}(\pi_\varphi +\pi_\vartheta ),\ \ \ s_{ad} B =0, \ \ \ s_{ad} c 
=0.
\label{ad}
\end{eqnarray}
The conserved charges for the above symmetries are obtained using Noether's theorem as
\begin{equation}
Q_d = i\frac{1}{2}(\pi_\varphi +\pi_\vartheta )\dot{\bar c}+i\pi_\lambda\bar c,
\end{equation}
 \begin{equation}
Q_{ad}=i\frac{1}{2}(\pi_\varphi +\pi_\vartheta )\dot{ c}+i\pi_\lambda c.
\end{equation}
$Q_d$ and $Q_{ad}$ generate the symmetry  transformations in Eqs. (\ref{d}) and (\ref{ad}), 
respectively.
It is easy to verify the following relations satisfied by these conserved charges:
\begin{eqnarray}
s_dQ_d&=&-\{Q_d,Q_d\}=0,\nonumber\\
s_{ad}Q_{ad}&=&-\{Q_{ad},Q_{ad}\}=0,\nonumber\\
s_dQ_{ad}&=&-\{Q_{ad},Q_d\}=0,\nonumber\\
s_{ad}Q_{d}&=&-\{Q_{d},Q_{ad}\}=0.
\end{eqnarray}
These relations reflect the nilpotency and anticommutativity property of $s_{d}$ and 
$s_{ad}$ (i.e.
$ s^2_{d}=0, s^2_{ad}=0$ and $ s_d s_{ad}+s_{ad}s_d =0$).
\subsection{Bosonic symmetry}
In this subsection we construct the bosonic symmetry out of different nilpotent BRST symmetries 
of the theory.
The  BRST ($s_{b}$), anti-BRST ($s_{ab}$), co-BRST ($s_{d}$) and anti-co-BRST ($s_{ad}$) 
symmetry operators satisfy 
the following algebra: 
\begin{equation}
\{ s_d, s_{ad}\} =0,\ \ \{ s_b, s_{ab}\} =0
\end{equation}
\begin{equation}
\{ s_b, s_{ad}\} =0,\ \ \{ s_d, s_{ab}\} =0,
\end{equation}
\begin{equation}
\{ s_b, s_d\} \equiv s_w,\ \ \ \{ s_{ab}, s_{ad}\} \equiv s_{\bar w}.\label{bos}
\end{equation}
The anticommutators in Eq. (\ref{bos}) define the bosonic symmetry of the system.
Under this bosonic symmetry transformation the field variables transform  as
\begin{eqnarray}
s_{w}\varphi &=& -\frac{1}{2}(\dot B +\pi_\varphi +\pi_\vartheta ),
\ \ s_w\lambda =-\frac{1}{2}(2B-\dot\pi_\varphi -
\dot\pi_\vartheta ),\nonumber\\
 s_w\vartheta &=& -\frac{1}{2}(\dot B +\pi_\varphi +\pi_\vartheta ),\ \ 
s_w \pi_\varphi = 0,\ \ \ s_w \pi_\vartheta =0,\nonumber\\ 
 s_w u &=&0,\ \ \ s_w p_u =0, \ \ \ s_w c = 0,\ \ \ s_w B =0,\ \
s_w \bar c =0.
\end{eqnarray}
However, the symmetry operator $s_{\bar w}$ is not  an independent bosonic
symmetry transformation as shown by
\begin{eqnarray}
s_{\bar w}\varphi &=& \frac{1}{2}(\dot B +\pi_\varphi +\pi_\vartheta ),
\ \ \ s_{\bar w}\lambda =\frac{1}{2}(2B-\dot\pi_\varphi -
\dot\pi_\vartheta ),\nonumber\\
 s_{\bar w}\vartheta &=& \frac{1}{2}(\dot B +\pi_\varphi +\pi_\vartheta ),\ \ 
s_{\bar w} \pi_\varphi = 0,\ \ \ s_{\bar w} u =0, \nonumber\\
s_{\bar w} \pi_\vartheta &=&0,\ \ \
s_{\bar w} p_u =0, \ \ \ s_{\bar w} c = 0,\ \ \ s_{\bar w} B =0, \ \
 s_{\bar w} \bar c =0.
\end{eqnarray}
Now, it is easy to see that the operators $s_w$ and $s_{\bar w}$ satisfy the relation
 $s_w +s_{\bar w} =0$. This implies, from Eq. (\ref{bos}), that
\begin{equation}
\{ s_b, s_d\} =s_w=-\{ s_{ab}, s_{ad}\},
\end{equation}
It is clear from the above algebra that the operator $s_w$ is the analog  of the
Laplacian operator in the language of differential geometry and the
 conserved charge for the above symmetry transformation is calculated as
\begin{equation}
Q_w=-i\left[B^2 +\frac{1}{2}(\pi_\varphi +\pi_\vartheta)^2 \right].
\end{equation}
Using the equation of motion, it can readily be checked that
\begin{equation}
\frac{d Q_w}{dt}=-i\int dx[2B\dot B+(\pi_\varphi +\pi_\vartheta)(\dot\pi_\varphi +
\dot\pi_\vartheta)]
=0.
\end{equation}
Hence $Q_w$ is the constant of motion for this theory.
\subsection{Ghost and discrete symmetries} 
Now we would like to mention  yet another symmetry of the system namely the ghost symmetry.
The ghost number of the ghost and antighost fields  are 1 and $-1$, respectively,
the rest of the variables in the action
of the this theory have ghost number zero. 
Keeping this fact in mind we can introduce a scale transformation of the ghost field, under 
which the effective action is invariant, as
\begin{eqnarray}
\varphi &\rightarrow & \varphi,\ \ \vartheta\rightarrow \vartheta,\ \ \pi_\varphi\rightarrow 
\pi_\varphi,\ \ \pi_\vartheta\rightarrow 
\pi_\vartheta,\nonumber\\
 u&\rightarrow & u \ \ \
 p_u \rightarrow p_u,\ \ \lambda\rightarrow \lambda,\ \ B\rightarrow B,\nonumber\\
 c &\rightarrow & e^{\tau }c,\ \ \ \bar c
\rightarrow e^{-\tau }\bar c,
\end{eqnarray}
where $\tau $ is a global scale parameter.
The infinitesimal version of the ghost scale transformation can be written as
\begin{eqnarray}
s_g\varphi &=& 0, \ \ \ \ s_g\vartheta = 0,
\ \ \ \ s_g\lambda =0,\nonumber\\
s_g \pi_\varphi &=& 0,\ \ \ \ s_g u =0, \ \ \ \ s_g \pi_\vartheta =0,\nonumber\\
s_g p_u &=&0, \ \ \ \ s_g c = c,\ \ \ \ s_g B =0,\nonumber\\
s_g \bar c &=& -\bar c.
\end{eqnarray}
The Noether conserved charge for the above symmetry transformations is
\begin{equation}
Q_g =i[\dot{\bar c}c+\dot c\bar c].
\end{equation}
In addition to the above continuous symmetry transformation, the ghost
sector respects the  discrete symmetry transformations
\begin{equation}
c \rightarrow  \pm  i \bar c,\ \ \ \bar c \rightarrow  \pm  i c.
\end{equation}
The above discrete symmetry transformation is useful  to obtain
the anti-BRST symmetry transformation from the BRST symmetry transformation and vice versa.

\section{Geometrical cohomology} 
In this section we study the de Rham 
cohomological operators and their realization in terms of
 conserved charges which generate the symmetries for the
theory of self-dual chiral boson. 
In particular we point out the similarity between the algebra obeyed by de Rham 
cohomological operators and that by different BRST conserved charges.
\subsection{Hodge-de Rham decomposition theorem and differential operators}
The de Rham cohomological operators in
differential geometry obey the following algebra:
\begin{eqnarray}
 d^2&=&\delta^2 =0, \ \ \Delta =(d +\delta )^2 =d\delta +\delta d\equiv \{d,\delta\}
\nonumber\\
\ \ [\Delta, \delta ]&=&0, \ \ [\Delta,d ]=0, \label{alg}
\end{eqnarray}
where $d, \delta$ and $\Delta$ are the exterior, co-exterior and Laplace-Beltrami operator, 
respectively.
The operators $d$ and $\delta$ are adjoint or dual to each other and  $\Delta$ 
is self-adjoint operator \cite{brac}.
It is well known that the exterior derivative raises the degree of a form by
one when it operates on it (i.e. $df_n\sim f_{n+1}$). On the other hand, the dual-exterior 
derivative lowers the degree of a form by one when it operates on forms 
(i.e. $\delta f_n\sim f_{n-1}$). However, $\Delta$ does not change the degree of form 
(i.e. $df_n\sim f_n$).
Here $f_n$ denotes an arbitrary n-form  object.

Let $M$ be a compact, orientable Riemannian manifold; then an inner product on the vector 
space 
$E^n (M)$ of $n$-forms on $M$ can be defined as \cite{gold}
\begin{equation}
(\alpha, \beta ) =\int_M \alpha\wedge \ast \beta,
\end{equation}
for $\alpha, \beta\in E^n (M)$ and $\ast $ represents the Hodge duality operator \cite{mor}.
Suppose that $\alpha$ and $\beta$ are forms of degree $n$ and $n+1$, respectively,
then the following relation for inner product will be satisfied
\begin{equation}
(d\alpha, \beta )=(\alpha, \delta\beta ).
\end{equation}
Similarly, if $\beta$ is a form of degree $n-1$, then we have the relation 
$(\alpha, d\beta )=(\delta\alpha, \beta )$. Thus, the necessary and 
sufficient condition for $\alpha$ 
to be closed is that it should be orthogonal to all co-exact forms of degree $n$.
The form $\omega\in E^n (M)$ is called harmonic if $\Delta \omega =0$. Now let $\beta$ be a 
$n$-form  on $M$ and if there exists another $n$-form $\alpha$ such that $\Delta\alpha=\beta$,
then for a harmonic form $\gamma\in H^n$, 
\begin{equation}
(\beta, \gamma)=(\Delta\alpha, \gamma)=(\alpha, \Delta\gamma)=0,\label{Del}
\end{equation}
where $H^n(M)$ denote the subspace of $E^n(M)$ of harmonic forms on $M$.
Therefore, if a form $\alpha$ exists with the property that $\Delta\alpha =\beta$,
then Eq. (\ref{Del}) is a necessary and sufficient condition for $\beta$ to be orthogonal
 to the subspace $H^n$.
This reasoning leads to the idea that $E^n(M)$ can be partitioned into three 
distinct subspaces $\Lambda^n_d$, $\Lambda^n_\delta$ and $H^n$ which are consistent
with exact, co-exact  and harmonic forms, respectively.
Therefore, the Hodge-de Rham decomposition theorem can be stated \cite{wan}. 

{\textit{ A regular differential form of degree n ($\alpha$) may be uniquely decomposed into 
a 
sum of  the harmonic form ($\alpha_H$), exact form ($\alpha_d$) and co-exact form 
($\alpha_\delta$) i.e.
\begin{equation}
\alpha =\alpha_H +\alpha_d +\alpha_\delta,
\end{equation}
where $\alpha_H\in H^n, \alpha_\delta\in \Lambda^n_\delta$ and $\alpha_d\in \Lambda^n_d$.}}
\subsection{Hodge-de Rham decomposition theorem and conserved charges} 
 The conserved charges for all the 
symmetry transformations
  satisfy the following algebra: 
\begin{eqnarray}
Q_{b}^2&=&0,\ \ Q_{ab}^2=0,\ \ Q_{d}^2=0, \ \ Q_{ad}^2=0,\nonumber\\
 \{ Q_b, Q_{ab}\} &=&0,\ \ \{ Q_d, Q_{ad}\} =0,\ \{ Q_b, Q_{ad}\} =0, 
\nonumber\\
\{ Q_d,Q_{ab}\} &=&0,\ [ Q_g, Q_b] = Q_b,\ \ [ Q_g, Q_{ad}]= Q_{ad},\nonumber\\
 \left[ Q_g, Q_d\right]&=&- Q_d,  \ \ [ Q_g, Q_{ab}]= -Q_{ab},\nonumber\\
 \{Q_b, Q_d\} &=&  Q_w = - \{ Q_{ad}, Q_{ab}\},\ \left[ Q_w, Q_r\right] = 0. 
\label{cgs}
\end{eqnarray}
This can be checked easily by using the canonical brackets in Eq. (\ref{brac}), 
where $Q_r$ generically represents the charges for BRST symmetry ($Q_b$), anti-BRST 
symmetry ($Q_{ab}$), dual-BRST symmetry ($Q_d$), anti-dual-BRST symmetry ($Q_{ad}$) and
ghost symmetry ($Q_g$). We note that the 
relations between the conserved charges $Q_b$ and $Q_d$ as well 
as $Q_{ab}$ and $Q_{ad}$  mentioned in the last line of (\ref{cgs}) can be established  
by using the equation of motions only.  

This algebra is reminiscent of the algebra satisfied by the de Rham cohomological operators
of differential geometry given in Eq. (\ref{alg}). Comparing (\ref{alg}) and (\ref{cgs}) we 
obtain the following analogies:   
\begin{eqnarray}
(Q_b, Q_{ad})\rightarrow d, \ \ (Q_d, Q_{ab})\rightarrow \delta, \ \ Q_w\rightarrow \Delta.
\end{eqnarray}
Let $n$ be the ghost number associated with a particular state $\left|\psi\right>_n$ 
defined in the total Hilbert space of states, i.e.,
\begin{equation}
iQ_g\left|\psi\right>_n = n\left|\psi\right>_n
\end{equation}
Then it is easy to verify the relations 
\begin{eqnarray}
 Q_g Q_b\left|\psi\right>_n &=& (n+1)Q_b\left|\psi\right>_n,\ \
 Q_g Q_{ad}\left|\psi\right>_n = (n+1)Q_{ad}\left|\psi\right>_n,\nonumber\\
 Q_g Q_d\left|\psi\right>_n &=& (n-1)Q_b\left|\psi\right>_n,\ \
 Q_g Q_{ab}\left|\psi\right>_n  = (n-1)Q_b\left|\psi\right>_n,\nonumber\\
 Q_g Q_w\left|\psi\right>_n &=& nQ_w\left|\psi\right>_n,\label{cgh}
\end{eqnarray}
which imply that the ghost numbers of the states 
$Q_b\left|\psi\right>_n$, $Q_d\left|\psi\right>_n $ and $Q_w\left|\psi\right>_n $
are $(n + 1), (n - 1)$ and $n $, respectively.
The states $ Q_{ab}\left|\psi\right>_n$ and $ Q_{ad}\left|\psi\right>_n $
have ghost numbers $ (n - 1)$ and $(n + 1)$, respectively. 
The properties of $d$ and $\delta$ are mimicked by the sets ($Q_b,Q_{ad}$) and ($Q_d,Q_{ab}$), 
respectively. It is evident from Eq. (\ref{cgh}) that  the set ($Q_b,Q_{ad}$) raises 
the ghost number of a
state by one and on the other hand the set ($Q_d, Q_{ab}$) lowers the ghost number of 
the same state
by one.
 These observations, keeping the analogy with the Hodge-de Rham 
decomposition theorem, enable us to express any arbitrary state 
$\left|\psi\right>_n$ in terms of the sets
 ($Q_b, Q_d, Q_w$) and ($Q_{ad}, Q_{ab}, Q_w$) as
\begin{equation}
\left|\psi\right>_n = \left|w\right>_n +Q_b\left|\chi\right>_{n-1}+Q_d\left|\phi\right>_
{n+1},
\end{equation}
\begin{equation}
\left|\psi\right>_n = \left|w\right>_n +Q_{ad}\left|\chi\right>_{n-1}+Q_{ab}
\left|\phi\right>_
{n+1},
\end{equation}
where the most symmetric state is the harmonic state $\left|w\right>_n$ that satisfies 
\begin{eqnarray}
Q_w\left|w\right>_n &=&0, \ \ Q_{b}\left|w\right>_n =0, \ \ Q_{d}
\left|w\right>_n=0,\nonumber\\
Q_{ab}\left|w\right>_n &=&0,\ \ Q_{ad}\left|w\right>_n =0,
\end{eqnarray}
analogous to Eq. (\ref{Del}).
It is quite interesting to point out that the physicality criterion  of  all the fermionic 
charges $Q_{b},  Q_{ab}, Q_{d}$ and $ Q_{ad}$,
i.e.,
\begin{eqnarray}
Q_{b}\left|phys\right> &=&0, \ \ Q_{ab}\left|phys\right> =0,\nonumber \\ 
Q_{d}\left|phys\right> &=&0, \ \ Q_{ad}\left|phys\right> =0,
\end{eqnarray}
leads to the following conditions:
\begin{eqnarray}
 \pi_\lambda\left|phys\right> =0, \ \ \
 (\pi_\varphi +\pi_\vartheta )\left|phys\right> =0.
\end{eqnarray}
This is the operator form
of the first-class constraint which annihilates the physical state as a consequence of
the above physicality criterion, which is consistent with the Dirac 
method of quantization of a system with first-class constraints.
\section{Conclusions}
 The BFV technique plays an important role in the gauge theory to analyze the 
 constraint and the symmetry of the system. We have considered such a
powerful technique to study the theory of a self-dual chiral boson. In particular we have 
derived the nilpotent BRST symmetry transformation for this theory using the BFV technique.
Further we have studied the dual-BRST transformation which is also the symmetry 
of the effective
action and which leaves the gauge-fixing part of effective action invariant separately. 
Interchanging the role of ghost and antighost fields the anti-BRST and anti-dual-BRST 
symmetry  transformations are also constructed.
We have shown that the nilpotent BRST and anti-dual-BRST symmetry transformations are 
analogous to the exterior derivative as the ghost number of  the state
 $\left|\psi\right>_n$ on the 
total Hilbert space is  increased by one when these operate on $\left|\psi\right>_n$
and the algebra followed by these are the same as the algebra obeyed by the
de Rham cohomological operators.
 In a similar
 fashion the dual-BRST and anti-BRST symmetry transformations are
 linked to the co-exterior derivative.
 The anticommutator of either the BRST and the dual-BRST transformations or
 anti-BRST and anti-dual-BRST transformations
 leads to a bosonic symmetry  which turns out to be
the analog  of the Laplacian operator. Further, the effective theory has a non-nilpotent
 ghost symmetry transformation which leaves the  ghost terms of the effective 
action invariant independently. Further we have noted that the   
 Hodge duality operator $(\ast )$ does not exist for the theory of a self-dual chiral boson 
in  (1+1) dimensions
because effectively this theory reduces to a theory in (0+1) dimensions
due to the self-duality condition of fields ($\dot\varphi=\varphi'$ as well as 
$\dot\vartheta=\vartheta'$). 
 The algebra satisfied by the conserved charges is exactly the same in appearance as the 
algebra of the de Rham cohomological operators of differential geometry.  These lead to the conclusion 
 that the theory for self-dual
chiral boson is a Hodge theory.
\label{Chap:Chapter8}


\chapter{Concluding Remarks}
In this thesis we have considered the extension of the finite field dependent BRST transformation (FFBRST)
and its applications on gauge field theories.   

We have started with the brief introduction of the BRST formalism and its importance in the study of gauge field theories.
We have not only discussed the BRST formalism for simple gauge theories where the gauge algebra
is closed or irreducible but have also discussed the BRST quantization for the more general class
of the gauge theories when the gauge group is allowed to be open and/or irreducible. In particular,
 BV (field/antifield)
formulation and BFV technique have been discussed using the BRST transformation. 
There are various generalizations of the BRST transformation in literature. The finite field dependent BRST 
transformation is one of the most important generalizations of the BRST transformation. 
The FFBRST transformation is also the symmetry of the effective action and nilpotent. 
However, such a finite transformation does not leave the generating functional invariant as the Jacobian
in the definition of generating functional is not invariant due to finiteness  of the parameter.
But this nontrivial Jacobian under certain condition can be replaced by $e^{iS_1}$, where $S_1$ 
is local functional of the fields. Thus, the FFBRST transformation is extremely useful in connecting
two different effective theories. 

In this thesis we mainly focused on further generalization of such transformation with new applications.
We have started with some basic techniques and mathematical tools
relevant for this thesis in chapter two.

The generalization of BRST transformation considered earlier was on-shell nilpotent as equation
of motion for some of the fields were used to show the nilpotency. In chapter three, we have constructed the
 off-shell nilpotent FFBRST transformation in 1-form gauge theories by introducing a Nakanishi-Lautrup
type auxiliary field. Further, we have developed
both the on-shell and the off-shell FF-anti-BRST transformations, by making the infinitesimal
anti-BRST parameter finite and field dependent, for 1-form gauge theories. The FF-anti-BRST transformation
also plays the same role as the FFBRST transformation and connects the two different effective theories. 
Thus, same pair of the theories are shown to be connected through both FFBRST and FF-anti-BRST
transformations but with the different finite parameters.  We have shown 
these by connecting the YM theories in the
different covariant and noncovariant gauges (like the Lorentz gauge, the axial gauge, the Coulomb gauge
and the quadratic gauge). The connection between generating functional for the YM theory  to that of
the most general Faddeev-popov effective theory have also been established. 
We have noted that the nontrivial Jacobians of the path integral measures which arise due to the FFBRST and 
the FF-anti-BRST transformations are responsible for such connection. 
We have further observed that the off-shell FFBRST and FF-anti-BRST formulations are more simpler and
  elegant to that of the on-shell transformations. 

In chapter four, we have constructed the FFBRST transformation for an Abelian rank-2 tensor field theory 
 by making the infinitesimal BRST parameter finite and field dependent.  It has been shown that 
such finite transformation plays a crucial role in studying the Abelian 2-form gauge theory in noncovariant
 gauges. We have considered the axial gauge and the Coulomb gauge as the possible candidates 
for the noncovariant gauges.
The generating functional of the 2-form gauge theories in different gauges have been connected through such
FFBRST transformation with the different choices of finite field dependent parameters. We have further
derived the FF-anti-BRST transformation  and have shown that the
FF-anti-BRST transformation with different finite parameters  also relates the generating
functional for the different effective theories. The BV formulation for this Abelian 2-form gauge theory 
has also been studied in the context of the FFBRST transformation. The effective theories of Abelian rank-2 tensor 
field in different gauges are considered as the different solutions of the quantum master equation. We
 have shown that the FFBRST transformation with appropriate finite parameters
connects the generating functional corresponding to the different solutions of the quantum master equation in BV 
formulation.
 
In the YM theories, even after fixing the gauge, the redundancy of the gauge fields is not completely removed in 
certain covariant gauges for the large gauge fields. This Gribov problem  has been resolved in the GZ theory 
by adding an extra term, known as horizon term, to the YM effective action. The Kugo-Ojima (KO)
 criterion for color confinement in manifestly covariant gauge is not satisfied
in the GZ theory due to the presence of the horizon term which breaks the usual BRST symmetry.
However, this theory has been extended to restore the BRST symmetry and hence to satisfy the KO criterion
for color confinement. In {chapter five}, we have developed the FFBRST transformation
for the GZ theory with appropriate horizon term exhibiting the exact BRST invariance. By constructing
appropriate finite field dependent parameter we have mapped the generating functional of the GZ theory, 
which is free from the
Gribov copies to that of the Yang-Mills theory through FFBRST transformation.
Thus, the theory with Gribov copies are related through a field transformation to the theory without 
Gribov copies.  We have shown that same results also holds in the BV formulation of the GZ theory.

In {chapter six}, we have  introduced a new finite nilpotent symmetry, namely finite field dependent 
mixed BRST (FFMBRST) symmetry. Mixed BRST transformation is the combination of usual BRST and anti-BRST symmetry. 
Infinitesimal mixed BRST transformation  has been integrated to construct the FFMBRST transformation. 
These finite transformations are an exact nilpotent symmetry of  both the effective action as well as  the 
generating functional for the certain choices of the finite parameters. 
Further, it has been shown that the Jacobian contributions for the path 
integral measure arising from BRST and anti-BRST part compensate each 
other. Such symmetry transformation has also been considered in the field/antifield formulation to show that 
the solutions of the quantum master equation remain invariant under this.

We have shown yet another 
application of the FFBRST transformation by establishing the connection between the generating functional for the 
first-class  theories and the second-class theories. The Stueckelberg theory for the  massive spin  1 field and 
the gauge 
invariant 
theory for a self-dual chiral boson have considered as the first-class  theories. However,
the Proca theory  for massive spin 1 field and gauge  variant theory for self-dual chiral boson are considered 
as the second-class constraint theories. 
The connection  have been made with the help of explicit calculations 
in the two above mentioned models. In the first example, the generating functional of the Proca model has been 
obtained 
from the generating functional of the Stueckelberg theory for massive spin 1 vector field. 
In the other example, we have related the generating functionals for the gauge invariant and the 
gauge variant theory for 
self-dual chiral boson. Thus, we can obtain the Green  functions of the second-class theories from the Green 
functions of the first-class theories in our formulation. Complicated  nonlocal Dirac bracket analysis
for the study of the second-class theory is thus avoided in our FFBRST formulation. 

In {chapter eight}, the (1+1) dimensional theory for a single
self-dual chiral boson has been considered as a classical model for gauge theory. Using the  BFV technique, the 
nilpotent BRST and  anti-BRST symmetry transformations
for this theory have been explored. In this model
other forms of the nilpotent symmetry transformations like co-BRST and anti-co-BRST, which leave the gauge-fixing part
of the action invariant, have also been investigated. We have shown that the
nilpotent charges for these symmetry transformations satisfy
the algebra of the de Rham cohomological operators in the differential
geometry. The Hodge decomposition theorem on 
compact manifold  has also been studied in the context of the conserved
charges. Further the theory for single
self-dual chiral boson has been realized as a field theoretic model for the Hodge theory.

In this thesis we have made an attempt to extend the FFBRST formulation by incorporating it in
different field theoretic models. We do believe  that our formulation will find many more 
new applications in future.  In particular, it will be helpful in removing the
discrepancy of the anomalous dimension calculation for a gauge invariant operators \cite{hamvan}. Exploring 
our formulation in the context of the different field theoretic models having spontaneous symmetry breaking will 
also be very
exciting.

\label{Chap:Chapter9}

\appendix
\numberwithin{equation}{chapter} 
\chapter{Mathematical details of 2-form gauge theories}
\section{FFBRST in Axial gauge}
Under the FFBRST transformation with finite parameter given in Eq. (\ref{2afin}), 
the path integral measure for the generating functional in Eq. (\ref{2zfun}) transforms as 
\begin{equation}
\int {\cal D}\phi^\prime =\int {\cal D}\phi\; J(\kappa),
\end{equation}
$J(\kappa)$ can be replaced by $e^{iS_1^A}$ if the condition in Eq. (\ref{mcond}) satisfies.
We start with an ansatz for $S_1^A$ to connect the theory in Lorentz to axial gauge as
\begin{eqnarray}
S_1^A&=&\int d^4x \left[\xi_1\beta_\nu\partial_\mu B^{\mu\nu} +\xi_2\beta_\nu\eta_\mu B^{\mu\nu} +
\xi_3\beta_\nu\beta^\nu +i\xi_4\tilde\rho_\nu\partial_{\mu}(\partial^\mu\rho^\nu -
\partial^\nu\rho^\mu)\right.\nonumber\\
&+&\left. i\xi_5\tilde\rho_\nu\eta_{\mu}(\partial^\mu\rho^\nu -\partial^\nu\rho^\mu )+i
\xi_6\chi\tilde\chi +i\xi_7 \tilde\chi\partial_\mu\rho^\mu +i\xi_8 
\tilde\chi\eta_\mu\rho^\mu\right.\nonumber\\
&+&\left. \xi_9\tilde\sigma\partial_\mu\partial^\mu\sigma +\xi_{10}
\tilde\sigma\eta_\mu\partial^\mu\sigma +\xi_{11} \partial_\mu\beta^\mu\varphi +\xi_{12} 
\eta_\mu\beta^\mu\varphi\right.\nonumber\\
&+&i\left.\xi_{13}\chi\partial_\mu\tilde\rho^\mu +i\xi_{14}\chi\eta_\mu\tilde\rho^\mu \right] 
\label{2ans1}
\end{eqnarray}
where $\xi_i(i=1, 2,...,14)$ are explicit $\kappa$ dependent parameters to be determined by 
using Eq. (\ref{mcond}).
The infinitesimal change in Jacobian, using Eq. (\ref{jaceva}) with finite parameter in Eq. (\ref
{2afin}), is calculated as
\begin{eqnarray}
\frac{1}{J}\frac{dJ}{d\kappa}=-\int &d^4x&\left[-i\gamma_1\beta_\nu (\partial_\mu B^{\mu\nu}-
\eta_\mu B^{\mu\nu})+\gamma_1\tilde\rho_\nu\partial_\mu (\partial^\mu\rho^\nu -
\partial^\nu\rho^\mu )\right.\nonumber\\
&-&\left.\gamma_1\tilde\rho_\nu\eta_\mu (\partial^\mu\rho^\nu -\partial^\nu\rho^\mu )-i
\gamma_2 \lambda_1\beta_\nu\beta^\nu -\gamma_1\tilde\chi\partial_\mu\rho^\mu \right.\nonumber\\
&+&\left.\gamma_1\tilde\chi\eta_\mu\rho^\mu +\gamma_2 \lambda_2\chi\tilde\chi +i
\gamma_1\tilde\sigma\partial_\mu\partial^\mu\sigma -i
\gamma_1\tilde\sigma\eta_\mu\partial^\mu\sigma\right.\nonumber\\
&+&\left. i\gamma_1\beta_\mu(\partial^\mu\varphi +\eta^\mu\varphi) -
\gamma_1\chi\partial_\mu\tilde\rho^\mu +\gamma_1\chi\eta_\mu\tilde\rho^\mu\right].
\end{eqnarray}

The condition (\ref{mcond}) will be satisfied iff
\begin{eqnarray}
\int &[{\cal D}\phi ]&e^{i(S_{eff}+S_1^A)}\left [i\beta_\nu\partial_\mu B^{\mu\nu}(\xi_1^\prime -\gamma_1
 )+i\beta_\nu\eta_\mu B^{\mu\nu}(\xi_2^\prime +\gamma_1 )+i\beta_\nu\beta^\nu (\xi_3^\prime -
\gamma_2 \lambda_1)\right.\nonumber\\
&-&\left.\tilde\rho_\nu\partial_\mu (\partial^\mu\rho^\nu-\partial^\nu\rho^\mu )(\xi_4^\prime 
-\gamma_1 )-\tilde\rho_\nu\eta_\mu (\partial^\mu\rho^\nu-\partial^\nu\rho^\mu )(\xi_5^\prime +
\gamma_1 )\right.\nonumber\\
&-&\left.\chi\tilde\chi (\xi_6^\prime -\gamma_2 \lambda_2) -\tilde\chi\partial_\mu\rho^\mu (\xi_7^\prime +\gamma_1 )-
\tilde\chi\eta_\mu\rho^\mu (
\xi_8^\prime -\gamma_1 ) +i\tilde\sigma\partial_\mu\partial^\mu\sigma (\xi_9^\prime +\gamma_1 
)\right.\nonumber\\
&+&\left.\tilde\sigma\eta_\mu\partial^\mu\sigma (\xi_{10}^\prime -\gamma_1 )+i
\partial_\mu\beta^\mu\varphi (\xi_{11}^\prime -\gamma_1 )+i\eta_\mu\beta^\mu\varphi (\xi_{12}^
\prime +\gamma_1 )\right.\nonumber\\
&-&\left.\chi\partial_\mu\tilde\rho^\mu (\xi_{13}^\prime +\gamma_1 )-
\chi\eta_\mu\tilde\rho^\mu (\xi_{14}^\prime -\gamma_1 )+i\partial_\mu (
\partial^\mu\rho^\nu-\partial^\nu\rho^\mu )\Theta^\prime_b [
\beta_\nu(\xi_4-\xi_1)]\right.\nonumber\\
&+&\left.i\eta_\mu (\partial^\mu\rho^\nu-\partial^\nu\rho^\mu )\Theta^\prime_b [
\beta_\nu(\xi_5-\xi_2)] -i\partial_\mu\partial^\mu\sigma\Theta^\prime_b [\tilde\chi(\xi_7 -
\xi_9)]\right.\nonumber\\
&-&\left.i\eta_\mu\partial^\mu\sigma\Theta^\prime_b [\tilde\chi (\xi_8 -\xi_{10})]-i
\partial^\mu\chi\Theta^\prime_b [\beta_\mu (\xi_{11}+\xi_{13})]\right.\nonumber\\
&-&\left. i\eta^\mu\chi\Theta^\prime_b [\beta_\mu (\xi_{12}+\xi_{14})]\right] =0\label{2mcond1}
\end{eqnarray}

The contribution of antighost and ghost of antighost can possibly vanish by using the 
equations of motion of the $\tilde\rho_\mu$ and $\tilde\sigma$. It will happen if the ratio of 
the coefficient of terms in the above equation and the similar terms in $S^L_{eff} +S_1^A$ is 
identical \cite{sdj4}.
This requires that
\begin{eqnarray}
\frac{\xi_4^\prime -\gamma_1}{\xi_4 +1}&=&\frac{\xi_5^\prime +\gamma_1}{\xi_5}\nonumber\\
\frac{\xi_9^\prime +\gamma_1}{\xi_9 -1}&=&\frac{\xi_{10}^\prime -\gamma_1}{\xi_{10}}
\label{2req1}
\end{eqnarray}
The $\Theta^\prime$ dependent terms can be converted into local terms by the equation of 
motion of different fields. This can only work if the following conditions are satisfied
\begin{eqnarray}
\frac{\xi_4-\xi_1}{\xi_4+1}&=&\frac{\xi_5 -\xi_2}{\xi_5}\nonumber\\
\frac{\xi_7-\xi_9}{\xi_9 -1}&=&\frac{\xi_8 -\xi_{10}}{\xi_{10}}\nonumber\\
\frac{\xi_{11}+\xi_{13}}{\xi_{13} -1}&=&\frac{\xi_{12} +\xi_{14}}{\xi_{14}}
\label{2req2}
\end{eqnarray}
Further by comparing the coefficients of different terms $ 
i\beta_\nu\partial_\mu B^{\mu\nu}$, $i\beta_\nu\eta_\mu B^{\mu\nu}$, $i\beta_\nu\beta^\nu$, $
\tilde\rho_\nu\partial_\mu (\partial^\mu\rho^\nu-\partial^\nu\rho^\mu )$, $
\tilde\rho_\nu\eta_\mu (\partial^\mu\rho^\nu-\partial^\nu\rho^\mu )$, $
\chi\tilde\chi$, $\tilde\chi\partial_\mu\rho^\mu$, $\tilde\chi\eta_\mu\rho^\mu$, $ 
i\tilde\sigma\partial_\mu\partial^\mu\sigma$, $i\tilde\sigma\eta_\mu\partial^\mu\sigma$,
 $i\partial_\mu\beta^\mu\varphi$, $i\eta_\mu\beta^\mu\varphi$, $\chi\partial_\mu\tilde\rho^\mu$
and $\chi\eta_\mu\tilde\rho^\mu$ in both sides of Eq. (\ref{2mcond1}), we obtain the following conditions
\begin{eqnarray}
&&\xi_1^\prime  -\gamma_1 +\gamma_1 (\xi_4 -\xi_1 )+\gamma_1 (\xi_5 -\xi_2 )=0\nonumber\\
&&\xi_2^\prime +\gamma_1 -\gamma_1 (\xi_4 -\xi_1 )-\gamma_1 (\xi_5 -\xi_2 )=0\nonumber\\
&&\xi_3^\prime -\gamma_2 \lambda_1 +\gamma_2 \lambda_1(\xi_4 -\xi_1 )+\gamma_2 \lambda_1 (
\xi_5 -\xi_2 )=0\nonumber\\
&&\xi_4^\prime -\gamma_1 =0\nonumber\\
&&\xi_5^\prime +\gamma_1 =0\nonumber\\
&&\xi_6^\prime -\gamma_2 \lambda_2 -\gamma_2 \lambda_2(\xi_7 -\xi_9 )-\gamma_2 \lambda_2 (
\xi_8 -\xi_{10} )=0\nonumber\\
&&\xi_7^\prime +\gamma_1 -\gamma_1 (\xi_7 -\xi_9 )-\gamma_1 (\xi_8 -\xi_{10})=0\nonumber\\
&&\xi_8^\prime -\gamma_1 +\gamma_1 (\xi_7 -\xi_9 )+\gamma_1 (\xi_8 -\xi_{10})=0\nonumber\\
&&\xi_9^\prime +\gamma_1 =0\nonumber\\
&&\xi_{10}^\prime -\gamma_1 =0\nonumber\\
&&\xi_{11}^\prime -\gamma_1 -\gamma_1 (\xi_{11} +\xi_{13} )-\gamma_1 (\xi_{12} +\xi_{14})=0
\nonumber\\
&&\xi_{12}^\prime +\gamma_1 +\gamma_1 (\xi_{11} +\xi_{13} )+\gamma_1 (\xi_{12} +\xi_{14})=0
\nonumber\\
&&\xi_{13}^\prime +\gamma_1 =0\nonumber\\
&&\xi_{14}^\prime -\gamma_1 =0.
\end{eqnarray}
A particular solution of the above differential equations subjected to the conditions in
Eqs. (\ref{2req1}) and (\ref{2req2}) with initial condition $\xi_i(\kappa=0)=0$ 
can be written as 
\begin{eqnarray}
\xi_1 &=&-\kappa,\ \ \xi_2 =\kappa,\ \ \
\xi_3 =\gamma_2 \lambda_1\kappa,\  \ \xi_4 =-\kappa\ \
\xi_5 =\kappa,\nonumber\\
 \xi_6 &=&\gamma_2 \lambda_2\kappa,\ \ \
\xi_7 =\kappa,\ \ \xi_8 =-\kappa\ \ \
\xi_9 =\kappa,\ \ \xi_{10} =-\kappa\nonumber\\
\xi_{11} &=&-\kappa,\ \ \ \ \xi_{12} =\kappa,\ \ \ \
\xi_{13} =\kappa,\ \ \ \ \xi_{14} =-\kappa, \label{2soln}
\end{eqnarray}
where we have chosen the arbitrary parameter $\gamma_1 =-1$.
Putting the values of $\xi_i$ in Eq. (\ref{2ans1}) we obtain $S_1^A$ at $\kappa =1$ as
\begin{eqnarray}
S_1^A&=&\int d^4x \left[-\beta_\nu\partial_\mu B^{\mu\nu} +\beta_\nu\eta_\mu B^{\mu\nu} +\gamma_2
\lambda_1\beta_\nu\beta^\nu -i\tilde\rho_\nu\partial_{\mu}(\partial^\mu\rho^\nu -
\partial^\nu\rho^\mu)\right.\nonumber\\
&+&\left. i\tilde\rho_\nu\eta_{\mu}(\partial^\mu\rho^\nu -\partial^\nu\rho^\mu )+i\gamma_2 
\lambda_2\chi\tilde\chi +i\tilde\chi\partial_\mu\rho^\mu -i
\tilde\chi\eta_\mu\rho^\mu\right.\nonumber\\
&+&\left. \tilde\sigma\partial_\mu\partial^\mu\sigma -\tilde\sigma\eta_\mu\partial^\mu\sigma - 
\partial_\mu\beta^\mu\varphi + \eta_\mu\beta^\mu\varphi +i\chi\partial_\mu\tilde\rho^\mu 
 - i\chi\eta_\mu\tilde\rho^\mu \right].
\end{eqnarray}

\section{FFBRST in Coulomb gauge}
For the finite parameter given in Eq. (\ref{2par})  we make following ansatz for $S_1^C$ 

\begin{eqnarray}
S_1^C&=&\int d^4x \left[\xi_1\beta_\nu\partial_\mu B^{\mu\nu} +\xi_2\beta_\nu\partial_i B^{i\nu} +
\xi_3\beta_\nu\beta^\nu +i\xi_4\tilde\rho_\nu\partial_{\mu}(\partial^\mu\rho^\nu -
\partial^\nu\rho^\mu)\right.\nonumber\\
&+&\left. i\xi_5\tilde\rho_\nu\partial_i(\partial^i\rho^\nu -\partial^\nu\rho^i )+i
\xi_6\chi\tilde\chi +i\xi_7 \tilde\chi\partial_\mu\rho^\mu +i\xi_8 
\tilde\chi\partial_i\rho^i\right.\nonumber\\
&+&\left. \xi_9\tilde\sigma\partial_\mu\partial^\mu\sigma +\xi_{10}
\tilde\sigma\partial_i\partial^i\sigma +\xi_{11} \partial_\mu\beta^\mu\varphi +\xi_{12} 
\partial_i\beta^i\varphi\right.\nonumber\\
&+&\left. i\xi_{13}\chi\partial_\mu\tilde\rho^\mu +i\xi_{14}\chi\partial_i\tilde\rho^i \right].
\end{eqnarray}

The infinitesimal change in Jacobian for Coulomb gauge is calculated as
\begin{eqnarray}
\frac{1}{J}\frac{dJ}{d\kappa}=-\int &d^4x&\left[-i\gamma_1\beta_\nu (\partial_\mu B^{\mu\nu}-
\partial_i B^{i\nu})+\gamma_1\tilde\rho_\nu\partial_\mu (\partial^\mu\rho^\nu -
\partial^\nu\rho^\mu )\right.\nonumber\\
&-&\left.\gamma_1\tilde\rho_\nu\partial_i (\partial^i\rho^\nu -\partial^\nu\rho^i )-i\gamma_2
\lambda_1\beta_\nu\beta^\nu -\gamma_1\tilde\chi\partial_\mu\rho^\mu \right.\nonumber\\
&+&\left.\gamma_1\tilde\chi\partial_i\rho^i +\gamma_2 \lambda_2\chi\tilde\chi +i
\gamma_1\tilde\sigma\partial_\mu\partial^\mu\sigma -i
\gamma_1\tilde\sigma\partial_i\partial^i\sigma\right.\nonumber\\
&+&\left. i\gamma_1\beta_\mu\partial^\mu\varphi -i\gamma_1\beta_i\partial^i\varphi -
\gamma_1\chi\partial_\mu\tilde\rho^\mu +\gamma_1\chi\partial_i\tilde\rho^i\right].
\end{eqnarray}
The condition will be satisfied iff
\begin{eqnarray}
\int &[{\cal D}\phi ]&e^{i(S_{eff}+S_1^C)}\left [i\beta_\nu\partial_\mu B^{\mu\nu}(\xi_1^\prime -\gamma_1
 )+i\beta_\nu\partial_i B^{i\nu}(\xi_2^\prime +\gamma_1 )+i\beta_\nu\beta^\nu (\xi_3^\prime -
\gamma_2 \lambda_1)\right.\nonumber\\
&-&\left.\tilde\rho_\nu\partial_\mu (\partial^\mu\rho^\nu-\partial^\nu\rho^\mu )(\xi_4^\prime 
-\gamma_1 )-\tilde\rho_\nu\partial_i (\partial^i\rho^\nu -\partial^\nu\rho^i )(\xi_5^\prime +
\gamma_1 )\right.\nonumber\\
&-&\left.\chi\tilde\chi (\xi_6^\prime -\gamma_2 \lambda_2) -\tilde\chi\partial_\mu\rho^\mu (\xi_7^\prime +\gamma_1 )-
\tilde\chi\partial_i\rho^i (
\xi_8^\prime -\gamma_1 )+i\tilde\sigma\partial_\mu\partial^\mu\sigma (\xi_9^\prime +\gamma_1 )
\right.\nonumber\\
&+&\left.\tilde\sigma\partial_i\partial^i\sigma (\xi_{10}^\prime -\gamma_1 )+i
\partial_\mu\beta^\mu\varphi (\xi_{11}^\prime -\gamma_1 )+i\partial_i\beta^i\varphi (\xi_{12}^
\prime +\gamma_1 )\right.\nonumber\\
&-&\left.\chi\partial_\mu\tilde\rho^\mu (\xi_{13}^\prime +\gamma_1 )-
\chi\partial_i\tilde\rho^i (\xi_{14}^\prime -\gamma_1 )+i\partial_\mu (
\partial^\mu\rho^\nu-\partial^\nu\rho^\mu )\Theta^\prime_b [
\beta_\nu(\xi_4-\xi_1)]\right.\nonumber\\
&+&\left.i\partial_i(\partial^i\rho^\nu-\partial^\nu\rho^i )\Theta^\prime_b [
\beta_\nu(\xi_5-\xi_2)] -i\partial_\mu\partial^\mu\sigma\Theta^\prime_b [\tilde\chi(\xi_7 -
\xi_9)]\right.\nonumber\\
&-&\left.i\partial_i\partial^i\sigma\Theta^\prime_b [\tilde\chi (\xi_8 -\xi_{10})]-
\partial^\mu\chi\Theta^\prime_b [\beta_\mu (\xi_{11}+\xi_{13})]\right.\nonumber\\
&-&\left. i\partial^i\chi\Theta^\prime_b [\beta_i (\xi_{12}+\xi_{14})]\right] =0.
\end{eqnarray}
Following the method exactly similar to Appendix A, we obtain the solution for the parameters $\xi_i$ which is 
exactly same as in Eq. (\ref{2soln}). 

Thus we obtain $S_1^C$ at $\kappa =1$ as
\begin{eqnarray}
S^C_1&=&\int d^4x \left[-\beta_\nu\partial_\mu B^{\mu\nu} +\beta_\nu\partial_i B^{i\nu} +\gamma_2
\lambda_1\beta_\nu\beta^\nu -i\tilde\rho_\nu\partial_{\mu}(\partial^\mu\rho^\nu -
\partial^\nu\rho^\mu)\right.\nonumber\\
&+&\left. i\tilde\rho_\nu\partial_i(\partial^i\rho^\nu -\partial^\nu\rho^i )+i\gamma_2 
\lambda_2\chi\tilde\chi +i\tilde\chi\partial_\mu\rho^\mu -i
\tilde\chi\partial_i\rho^i\right.\nonumber\\
&+&\left. \tilde\sigma\partial_\mu\partial^\mu\sigma -\tilde\sigma\partial_i\partial^i\sigma - 
\partial_\mu\beta^\mu\varphi + \partial_i\beta^i\varphi +i\chi\partial_\mu\tilde\rho^\mu 
\right.\nonumber\\
&-&\left. i\chi\partial_i\tilde\rho^i \right].
\end{eqnarray}



\begin{thebibliography}{99} 
\bibitem{ym} C. N. Yang and R. L. Mills, {\it Conservation of isotopic spin and
isotopic gauge invariance}, Phys. Rev. {\bf 96}, 191  (1954). 
\bibitem{fp}  L. D. Faddeev  and V. N. Popov,  {\it Feynman diagrams for the Yang-Mills
field}, {Phys. Lett. } {\bf B25}, {29} (1967).
 \bibitem{brst} C. Becchi, A. Rouet and R. Stora, {\it Renormalization of gauge theories}, 
Annals Phys. {\bf{98}}, 287 (1976). 
 \bibitem {tyu}I. V. Tyutin, {\it Gauge Invariance in Field Theory and Statistical Physics in Operator Formalism}, 
Lebedev Physics Institute preprint {\bf 39} (1975).
 \bibitem{ht} M. Henneaux and C. Teitelboim, {\it{ Quantization of gauge
systems}}, Princeton, USA: Univ. Press (1992).
\bibitem{wei} S. Weinberg, {\it{ The quantum theory of fields, Vol-II: Modern
applications}}, Cambridge, UK Univ. Press (1996).
\bibitem{bv} I. A. Batalin and G. A. Vilkovisky, {\it Gauge algebra and quantization}, 
{ Phys. Lett.}{\bf{ B102}}, 27 (1981).
\bibitem{bv1} I. A. Batalin and G. A. Vilkovisky, {\it Quantization of gauge theories with
linearly dependent generators}, { {Phys. Rev.}} {\bf{D28}}, 2567 (1983); 
Erratum ibid {\bf{D30}}, 508 (1984).
\bibitem{bv2} I. A. Batalin and G. A. Vilkovisky, {\it Feynman rules for reducible gauge theories}, 
{Phys. Lett.} {\bf{ B120}} 166 (1983).
\bibitem{c} M. Sato and S. Yahikozawa, {\it ``Topological" formulation of effective vortex strings},
 {Nucl. Phys.} {\bf{B436}}, 100 (1995).
\bibitem{frad} E. S. Fradkin and G. A. Vilkovisky, {\it Quantization of relativistic systems with
constraints},  Phys. Lett. {\bf B55}, (1975) {224}.
  \bibitem{ku}  T. Kugo,  and I.  Ojima, {\it Local Covariant Operator Formalism of Non-Abelian Gauge 
Theories and Quark Confinement Problem}, Suppl.  Prog. Theor. Phys.  {\bf 66},(1979) {14}.
 \bibitem{dir} P. A. M. Dirac, \textit{Lectures on Quantum Mechanics}, (Yeshiva Univ. Press, 
New York, 1964).
\bibitem{ste} A. Restuccia,   and J. Stephany, {\it Gauge fixing in extended phase space and path integral quantization of systems with second class constraints}, Phys. Lett.  {\bf B305}, 
 {348} (1993). 
\bibitem{bt}   I. A. Batalin and  I. V. Tyutin, {\it Existence theorem for the effective gauge algebra in the
generalized canonical formalism with Abelian conversion of second-class constraints},
 Int. J. Mod. Phys.   {\bf A6}, 
 {3255} (1991).
\bibitem{bf}   I. A. Batalin and E. S. Fradkin, {\it Operational quantization of dynamical systems subject to second 
class constraints},  Nucl. Phys.  {\bf B279}, {514} (1987);  {\it Operator quantization of dynamical systems with irreducible first- and second-class constraints}, 
Phys. Lett.  {\bf B180},  {157} (1986).
\bibitem{fik} T. Fujiwara, Y. Igarashi and  J.   Kubo, {\it Anomalous gauge theories and subcritical strings based on the Batalin-Fradkin formalism}, Nucl. Phys.   {\bf B341}, 
 {695} (1990).
\bibitem{kkk}   Y. W. Kim, S. K. Kim,  W. T. Kim, Y. J. Park,  K. Y. Kim and Y. Kim, {\it Chiral Schwinger model 
based on the Batalin-Fradkin-Vilkovisky formalism},
Phys. Rev.   {\bf D46},  {4574} (1992).
\bibitem{fs}  L. D. Faddeev and S. L. Shatashivili, {\it Realization of the Schwinger term in the Gauss law and the possibility of correct quantization of a theory with anomalies},  Phys. Lett.  {\bf B167}, 
 {225} (1986).
\bibitem{bs}  O. Babelon, F. A. Shaposnik and C. M. Vialett, {\it Quantization of gauge theories with Weyl fermions},
  Phys. Lett.  {\bf B177}, {385} (1986).
\bibitem{wz} J. Wess and B. Zumino, {\it Consequences of anomalous ward identities}, Phys. Lett. {\textbf B37},  95 (1971).
\bibitem{bv0} E. S. Fradkin, G. A. Vilkovisky, {\it Quantization of relativistic systems with constraints},
  Phys. Lett. {\bf B55},  224 (1975).
\bibitem{hen} M. Henneaux, {\it Hamiltonian form of the path integral for theories with a gauge freedom},
 Phys. Rep. {\bf 126}, 1  (1985).
 \bibitem{liv} M. Lavelle and D. McMullan, {\it Nonlocal Symmetry for QED}, Phys. Rev.
Lett. {\bf 71}, 3758 (1993).
\bibitem{tan} Z. Tang and D. Finkelstein, {\it Relativistically Covariant Symmetry in QED}, 
Phys. Rev. Lett. {\bf 73}, 3055 (1994).
\bibitem{yan} H. S. Yang and B. H. Lee, {\it Noncovariant Local Symmetry in Abelian
Gauge Theories}, J. Korean Phys. Soc. {\bf 28}, 572 (1995).
\bibitem{jm} S. D. Joglekar and B. P. Mandal, {\it Finite field-dependent BRS transformations}, Phys. Rev. 
{\bf{D51}}, 1919 (1995).
\bibitem{sdj0} S. D. Joglekar and A. Misra, {\it Relating Green's functions in axial and Lorentz gauges using finite 
field-dependent BRS transformations},
J. Math. Phys. {\bf{41}}, 1755 (2000).
\bibitem{sdj2} S. D. Joglekar and A. Misra, {\it Correct treatment of $1/( \eta\cdot\kappa)^p$ singularities in the 
axial gauge propagator},  Int. J. Mod. Phys. {\bf{A15}}, 1453 (2000).
\bibitem{sdj3} S. D. Joglekar and A. Misra, {\it 	
A Derivation of the correct treatment of $1/(\eta\cdot\kappa)^p$ singularities in axial gauges}, Mod. Phys. Lett. 
{\bf{A14}}, 2083 (1999).
\bibitem{sdj4} S. D. Joglekar and A. Misra, {\it Wilson loop and the treatment of axial gauge poles},
 Mod. Phys. Lett. {\bf{A15}}, 541 (2000).
\bibitem{sd} S. D. Joglekar, {\it Green's functions in Axial- and Lorentz-type gauges and BRS transformations}, Mod. Phys. Lett {\bf{A15}}, 245 (2000).
\bibitem{sdbp} S. D. Joglekar and B. P. Mandal, {\it Application of finite field-dependent BRS transformations
to problems of the Coulomb gauge}, Int. J. Mod. Phys. {\bf A17}, 1279 (2002).
\bibitem{subm} B. P. Mandal,  S. K. Rai and S. Upadhyay, {\it Finite nilpotent symmetry in Batalin-Vilkovisky 
formalism},  Euro. Phys. Lett. {\bf 92}, 21001 (2010).
\bibitem{sdj} S. D. Joglekar, {\it Connecting Green functions in an arbitrary pair of gauges and an application
to planar gauges}, Int. J. Mod. Phys {\bf{A16}}, 5043 (2000). 
\bibitem{rb} R. Banerjee and B. P. Mandal, {\it Quantum gauge symmetry from finite field dependent BRST
 transformations}, Phys. Lett. {\bf{B488}}, 27 (2000).
\bibitem{sdj1}R. S. Bandhu and S. D. Joglekar, {\it Finite field-dependent BRS transformation and axial gauges},
 J. Phys. {\bf{A31}}, 4217 (1998).
\bibitem{etc} S. D. Joglekar, {\it Additional considerations in the definition and renormalization of
non-covariant gauges}, Mod. Phys. lett. {\bf{A18}}, 843 (2003).
\bibitem{sudha1} M. Faizal, B. P. Mandal and S. Upadhyay, {\it Finite BRST Transformations for the Bagger-Lambert-Gustavasson Theory},  Phys. Lett. {\bf B721}, 159, (2013). 
\bibitem{sudha2} S. Upadhyay, M. K. Dwivedi and B. P. Mandal, {\it The noncovariant gauges in 3-form theories},
Int. J. Mod. Phys. {\bf A28}, 1350033   (2013).
\bibitem{sudha3} R. Banerjee, B. Paul and S. Upadhyay, {\it BRST symmetry and $W$-algebra in higher derivative models}, arXiv:1306.0744.

 

\bibitem{susk} S. Upadhyay, S. K. Rai and B. P. Mandal, {\it Off-shell nilpotent finite Becchi-Rouet-Stora-Tyutin
(BRST)/anti-BRST transformations}, J. Math. Phys. {\bf 52}, 022301 (2011). 
\bibitem{boto} L. Bonora and M. Tonin, {\it Superfield formulation of extended BRS symmetry}, Phys. Lett. {\bf 
B98}, 48 (1981).
\bibitem{gaba} L. Alvarez-Gaume and L. Baulieu, {\it The two quantum symmetries associated with a classical 
symmetry}, 
Nucl. Phys. {\bf B212}, 255 (1982).
\bibitem{sm1} S. Upadhyay and B. P. Mandal, {\it Generalized BRST transformation in Abelian rank-2
antisymmetric tensor field theory},   Mod. Phys. Lett. {\bf A40}, 3347 (2010).
\bibitem{a} M. Kalb and P. Ramond, {\it Classical direct interstring action}, {Phys. Rev.} {\bf{D9}}, 2273 (1974).
\bibitem{b} F. Lund and T. Regge, {\it Unified approach to strings and vortices with soliton solutions}, {Phys. Rev.} 
{\bf{D14}}, 1524 (1976).
\bibitem{d} A. Sugamoto, {\it Dual transformation in Abelian gauge theories}, { {Phys. Rev.}} {\bf{D19}}, 1820 (1979). 
\bibitem{e} R. L. Davis and E. P. S. Shellard, {\it Antisymmetric tensors and spontaneous symmetry breaking}, {Phys. Lett.}, {\bf{B 214}}, 219 (1988).
\bibitem{g} A. Salam and E. Sezgin, {\it{Supergravities in Diverse Dimensions}}, 
(North-Hplland and World Scientific, 1989).
\bibitem{h} M. B. Green, J. H. Schwarz and E. Witten, {\it{Superstring Theory}}, (Cambridge 
Univ. Press, 1987).
\bibitem{i} J. Polchinski, {\it{String Theory}}, (Cambridge Univ. Press, 1998).
\bibitem{gri}  V. N. Gribov, {\it Quantization of non-Abelian gauge theories},  Nucl. Phys. {\bf B139}, 1 (1978).
\bibitem{zwan}  D. Zwanziger, {\it Local and renormalizable action from the gribov horizon},  Nucl. Phys. {\bf B323},
 {513} (1989).
\bibitem{zwan2}  D. Zwanziger, {\it Renormalizability of the critical limit of lattice gauge theory by BRS 
invariance},    Nucl. Phys. {\bf B399}, 477 (1993).
\bibitem{kon0}    K. I. Kondo, {\it Gluon and ghost propagators from the viewpoint of general principles of quantized 
gauge field theories},  Nucl. Phys. {\bf B129}, 715 (2004).
\bibitem{zwan1}   D. Zwanziger, {\it Critical limit of lattice gauge theory},   Nucl. Phys. {\bf B378}, 525 (1992).
\bibitem{sore1}  D. Dudal, J. A. Gracey, S. P. Sorella, N.  Vandersickel  and H. Verschelde, {\it Refinement of the 
Gribov-Zwanziger approach in the Landau gauge: Infrared propagators in harmony with the lattice results}
 Phys. Rev. {\bf D78}, 065047 (2008).
\bibitem{sor} {D. Dudal, S. P. Sorella  and N. Vandersickel } {\it More on the renormalization of the horizon function
 of the Gribov–Zwanziger action and the Kugo–Ojima Green function(s)},
  Eur. Phys. J.   {\bf C68}, 283 (2010).
 \bibitem{dud}   D. Dudal, S. P. Sorella, N. Vandersickel   and H. Verschelde, {\it Gribov no-pole condition,
 horizon function, Kugo-Ojima confinement criterion, boundary conditions, BRST breaking and all that}, 	Phys. Rev. {
\bf D79}, 121701 (2009).
\bibitem{fuj}   K. Fujikawa, {\it Dynamical stability of the BRS supersymmetry and the Gribov problem},
   Nucl. Phys. {\bf B223}, 218 (1983).
\bibitem{sor1} S. P. Sorella, {\it Gribov horizon and BRST symmetry: a few remarks}, Phys. Rev. {\bf D80}, 025013 
(2009).
\bibitem{sudbm}  S. Upadhyay and B. P. Mandal, {\it Relating Gribov-Zwanziger theory to Yang-Mills theory},  Euro.
 Phys. Lett. {\bf 93}, 31001 (2011). 
 \bibitem{sudbpm}  S. Upadhyay and B. P. Mandal, {\it Relating Gribov-Zwanziger theory and Yang-Mills theory in 
Batalin-Vilkovisky formalism}, AIP Conf. Proc. {\bf 1444}, 213 (2012).
 \bibitem{epjc}  S. Upadhyay and B. P. Mandal, {\it Field dependent nilpotent symmetry for gauge theories},
 Eur. Phys. J. {\bf C72}, 2065 (2012). 
\bibitem{pp} P. P. Srivastva, {\it Quantization of self-dual field revisited},  Phys. Rev. Lett. {\bf  63}, 2791 
(1989).
\bibitem{pp1} P. P. Srivastva, {\it On a gauge theory of the self-dual field and its quantization}, 
Phys. Lett. {\bf B234}, 93  (1990).
\bibitem{kk} D. S. Kulshreshtha and Muller-Kirsten, {\it Faddeev-Jackiw quantization of self-dual fields},
 Phys. Rev. {\bf D45}  R 393 (1992).
\bibitem{fj} R. Floreanini and R. Jackiw, {\it Self-dual fields as charge-density solitons},
 Phys. Rev. Lett. {\bf 59},  1873 (1987).
\bibitem{vo} H. O. Girotti, M. Gomes and  V. O. Rivelles, {\it Chiral bosons through linear constraints}, 
Phys. Rev. {\bf D45},  R 3329  (1992).
\bibitem{cg}  M. E. V. Costa and H. O. Girotti, {\it Comment on ``Self-dual fields as charge-density solitons"}, Phys. Rev. Lett. {\bf   60}, {1771}  (1988).
\bibitem{ggk}  H. O.  Girotti, M. Gomes, V. Kurak, V. O. Rivelles and 
A. J. da Silva, {\it Chiral Bosonization}, Phys. Rev. Lett. {\bf 60}, {1913} (1988).
\bibitem{sud} S Upadhyay and B. P. Mandal, {\it The model for self-dual chiral bosons as a Hodge theory}, Eur. Phys. J. {\bf C71}, 1759 (2011). 
\bibitem{sub}  S. Ghosh, {\it Quantization of a $U(1)$ gauged chiral boson in the Batalin-Fradkin-Vilkovisky scheme}, 
Phys. Rev. {\bf D49}, {2990} (1994).
\bibitem{ms} N. Marcus and J. Schwarz, {\it Field theories that have no manifestly Lorentz-invariant formulation},
 Phys. Lett. {\bf B115}, 111 (1982).
\bibitem{wen} X.G. Wen, {\it Electrodynamical properties of gapless edge excitations in the fractional quantum Hall 
states},  Phys. Rev. Lett. {\bf 64}, 2206 (1990).
\bibitem{stuc1}   E. C. G. Stueckelberg, {\it The interaction forces in electrodynamics and in the field theory of nuclear forces (I)},  Helv. Phys. Acta  {\bf 11}, {225} (1938).
\bibitem{stuc2}  E. C. G. Stueckelberg, {\it The interaction forces in electrodynamics and in the field theory of nuclear forces (II)}, Helv. Phys. Acta   {\bf 11}, {299} (1938).
\bibitem{stuc3}   E. C. G. Stueckelberg, {\it The interaction forces in electrodynamics and in the field theory of nuclear forces (III)},  Helv. Phys. Acta   {\bf 11}, {312} (1938).
\bibitem{al}   G. Aldazabal, L. E. Ibanez and F. Quevedo, {\it A D-brane alternative to the MSSM},
 JHEP  {\bf 02}, {015} (2000).
\bibitem{mr}  C.  Marshall and P. Ramond, {\it Field theory of the interacting string: The closed string},  Nucl. 
Phys.  {\bf B85}, {375} (1975).
\bibitem{r}  P. Ramond, {\it A Pedestrian Approach to Covariant String Theory}, Prog. Theor. Phys. Suppl.  {\bf 86}, {126} (1986). 
\bibitem{aop}  S. Upadhyay and B. P. Mandal, {\it Finite BRST transformation and constrained systems},
  Annals of Physics, {\bf 327},  2885 (2012).
\bibitem{sund} K. Sundermeyer, {\it  Constrained Dynamics}, Lecture notes in Physics,
vol. 169 (Springer, Berlin, 1982). 

\bibitem{egu} T. Equchi, P. B. Gilkey  and A. Hanson, {\it Gravitation, gauge theories and differential geometry},
 Phys. Rep. {\bf 66},  213 (1980).
\bibitem{nis} K. Nishijima, {\it The Casimir Operator in the Representations of BRS Algebra},
 Prog. Theor. Phys. {\bf 80},  897 (1988).
\bibitem{nis1} K. Nishijima, {\it Observable States in the Representations of BRS Algebra},
 Prog. Theor. Phys. {\bf 80},  905  (1988).
\bibitem{kala} W. Kalau and J. W. van Holten, {\it 	BRST cohomology and BRST gauge fixing},  Nucl. Phys. {\bf 
B361},  233 (1991).
\bibitem{hol} J. W. van Holten, {\it Becchi-Rouet-Stora-Tyutin cohomology of compact gauge algebras},
  Phys. Rev. Lett. {\bf 64},   2863 (1990).
\bibitem{hol1} J. W. van Holten, {\it The BRST complex and the cohomology of compact lie algebras}, Nucl. Phys. {\bf 
 B339},   158 (1990).
\bibitem{arn} H. Aratyn, {\it BRS cohomology in string theory: Geometry of Abelization and the quartet mechanism},
 J. Math. Phys. {\bf 31}, 1240 (1990).
\bibitem{hari} E. Harikumar, R. P. Malik and M. Sivakumar, {\it Hodge decomposition theorem for Abelian two form 
gauge theory}, J. Phys. A: Math. Gen. {\bf 33},  7149 (2000).
\bibitem{sr}S. Gupta and R. P. Malik, {\it A field-theoretic model for Hodge theory},  Eur. Phys. J. {\bf C58}, 517 
(2008).
\bibitem{sr1}S. Gupta and R. P. Malik, {\it Rigid Rotor as a Toy Model for Hodge Theory}, Eur. Phys. J. {\bf C68},  
325 (2010).
 



\bibitem{cf} G. Curci and R. Ferrari, {\it On a class of Lagrangian models for massive and massless Yang-mills 
fields}, {  Nuovo Cimento Soc. Ital. Fis.} {\bf A32}, 151 (1976).
\bibitem{cf1} G. Curci and R. Ferrari, {\it The unitarity problem and the zero-mass limit for a model of massive 
yang-mills theory}, { Nuovo Cimento Soc. Ital. Fis.} {\bf A35}, 1 (1976); 47, 555 (1978).
\bibitem{del} R. Delbourgo and P. D. Jarvis, {\it Extended BRS invariance and OSp(4/2) supersymmetry},
 {  J. Phys.} {\bf A15}, 611 (1982).


\bibitem{thoo} G. 't Hooft and M. Veltman, {\it Combinatorics of gauge fields},  Nucl. Phys.   {\bf B50}, 
 318 (1972).
 \bibitem{bath} L. Baulieu and J. Thierry-Mieg, {\it The principle of BRS symmetry: An alternative approach to 
Yang-Mills theories}, Nucl. Phys. {\bf B197}, 477 (1982).
\bibitem{deg} S. Deguchi and Y. Kokubo, {\it Quantization of massive Abelian antisymmetric tensor field and linear
potential}, Mod. Phys. Lett. {\bf A17}, 503  (2002).
\bibitem{cfdj} R. Delbourgo and P. D. Jarvis, {\it Extended BRS invariance and OSp (4/2) supersymmetry},
 J. Phys. {\bf A 15},  611 (1982).
\bibitem{nk} N. K. Falck and G. Kramer, {\it Gauge invariance, anomalies, and the chiral Schwinger model},
 Ann. Phys. {\bf 176},  330 (1987).
\bibitem{pb} P. Bracken, {\it Quantization of Two Classical Models by Means of the BRST Quantization Method},
 Int. J. Theor. Phys. {\bf 47}, 3321 (2008) .
\bibitem{brac} P. Bracken, {\it  The Hodge-de Rham Decomposition Theorem And Some Applications Pertaining to Partial 
Differential Equations}, { ArXiv}: {1001.2348} [math.DG].
\bibitem{gold} S. Goldberg, {\it curvature and Homology}, Dover, NY, (1970).
\bibitem{mor} S. Morita, {\it Geometry of Differential Forms, AMS Translations 
of Mathematical Monographs}, Vol. 209, Providence, RI, 2001.
 \bibitem{wan} Warner  and W. Frank, {\it Foundations of Differentiable Manifolds and Lie Groups},
 Berlin, New York (1983) Springer-Verlag, ISBN 978-0-387-90894-6.
\bibitem{hamvan} R. Hamberg and W. Van Neerven, {\it The correct renormalization of the gluon operator in a covariant gauge}, Nucl. Phys. {\bf B379}, 143 (1992).
\end{thebibliography}
\end{document}